%% file: grsk-chargediffuse-arxiv-v4.tex
\documentclass[final, 11pt]{article}
\pdfoutput=1
\usepackage{MRnotes}

\input{gravsk-macros}

\title{An effective description of momentum diffusion in a charged plasma from holography}
\author[a]{Temple He,}
\author[b]{R. Loganayagam,}
\author[a]{Mukund Rangamani,}
\author[a]{Julio Virrueta}

\affiliation[a]{
	Center for Quantum Mathematics and Physics (QMAP)\\
	Department of Physics \& Astronomy, University of California, Davis, CA 95616 USA}
\affiliation[b]{
	International Centre for Theoretical Sciences (ICTS-TIFR), \\ 
	Tata Institute of Fundamental Research, Shivakote, Hesaraghatta, Bangalore 560089, India.}

\emailAdd{tmhe@ucdavis.edu}
\emailAdd{nayagam@icts.res.in}
\emailAdd{mukund@physics.ucdavis.edu}
\emailAdd{jvirrueta@ucdavis.edu}

\abstract{
We discuss the physics of momentum diffusion in a charged plasma. Following the holographic strategy outlined in  \cite{Ghosh:2020lel} we construct an open effective field theory for the low-lying modes of the conserved currents. The charged plasma is modeled holographically in terms of a \RNAdS{d+1} black hole. We analyze  graviton and photon fluctuations about this background, decoupling in the process the long-lived momentum diffusion mode from the short-lived  charged transport mode. Furthermore, as in the aforementioned reference, we argue that the dynamics of these modes are captured by a set of designer scalars in the background geometry. These scalars have  their gravitational coupling modulated by an auxiliary dilaton with long-lived modes being weakly coupled near the spacetime asymptopia. Aided by these observations,  we obtain the quadratic effective action that governs the fluctuating hydrodynamics of the charge current and stress tensor,  reproducing in the process transport data computed  previously.  
We also point out an interesting length scale lying between the inner and outer horizon radii  of the charged black hole associated with Ohmic conductivity.
}

\begin{document}
\maketitle
 

\section{Introduction}
\label{sec:intro}

The dissipative dynamics of planar AdS black holes is encoded in their quasinormal spectrum, while the associated quantum and stochastic fluctuations are captured by  Hawking quanta. Recently, a useful picture for the unified description of these two facets of black holes has started to emerge \cite{Chakrabarty:2019aeu,Jana:2020vyx,Ghosh:2020lel}, inspired in part by a proposal for computing real-time  observables in holography  \cite{Glorioso:2018mmw}.\footnote{Various aspects of the problem have been discussed extensively in the literature, cf.,  \cite{Son:2002sd,Herzog:2002pc,Skenderis:2008dg,vanRees:2009rw} for computing  real-time observables in holography,
 \cite{Horowitz:1999jd,Policastro:2001yc,Policastro:2002se,Kovtun:2005ev,Morgan:2009pn} for analysis of quasinormal modes, and  \cite{Kovtun:2012rj,Grozdanov:2013dba,Kovtun:2014hpa,Haehl:2015pja,Crossley:2015evo,Haehl:2015uoc,Jensen:2017kzi,Haehl:2018lcu,Jensen:2018hse,Chen-Lin:2018kfl} for  construction of effective actions for hydrodynamics in field theory. We refer the reader to the introduction of \cite{Ghosh:2020lel}  (which we build on) for an overview of the salient developments. Applications of the real-time holographic techniques to diverse settings are discussed in the recent works \cite{Loganayagam:2020eue,Loganayagam:2020iol,Chakrabarty:2020ohe,Bu:2020jfo,Bu:2021clf}.} The upshot of these developments is that one can analyze the problem of computing Schwinger-Keldysh observables in gravity in a complex two-sheeted geometry obtained from the eternal black hole solution, by gluing two copies of the future-half of the domain of outer-communication, the so called grSK geometry reviewed in \cite{Jana:2020vyx}. Such a  stochastic effective action capturing both  dissipation and fluctuation is highly desirable from the viewpoint of understanding an open effective field theoretic description of strongly correlated quantum systems. 
We will describe here another application of these techniques -- the effective description of dynamics of momentum diffusion in a  charged plasma. 

To set the stage, we recall from \cite{Jana:2020vyx} that one can conveniently formulate the stochastic dynamics of a thermal plasma as that of an open quantum system. Imagine coupling the plasma to an external quantum system (a read-out device) and integrating out the plasma degrees of freedom -- the resulting open effective dynamics of our external probe system is what we seek to understand. The examples studied in \cite{Jana:2020vyx} comprised of plasma-system couplings where the plasma operators had short thermal relaxation times, or in gravitational parlance, short-lived quasinormal modes. On the other hand, should one couple the system to conserved current operators of the plasma, then one encounters long-lived hydrodynamic modes. In this case it is a-priori unclear whether there is a useful  local description of the physics.

This question was explored in detail in \cite{Ghosh:2020lel} who considered  a neutral (conformal) thermal plasma, modeled holographically in terms of a \SAdS{d+1} black hole. They argued there that one should distinguish the two classes of dynamical modes described above: the short-lived Markovian modes and the long-lived non-Markovian modes. The latter includes the modes that drive momentum and charge diffusion, as well as sound modes in the plasma. The authors of \cite{Ghosh:2020lel} analyzed the effective description of a probe conserved current in a neutral plasma and the physics of momentum diffusion. We will summarize their main observations below and show that the general lessons proposed there continue to extend to other settings.  

The conserved current operator in a thermal system comprises of an admixture of both Markovian and non-Markovian degrees of freedom. For example, the charge current in a $d$ dimensional plasma has a $d-2$ Markovian degrees of freedom corresponding to the (short-lived) physical charge waves in the plasma, and a single non-Markovian degree of freedom corresponding to the long-lived charge diffusion mode. Likewise the energy-momentum tensor has $\frac{d(d-3)}{2}$ short-lived Markovian modes corresponding to momentum waves, $d-2$ momentum diffusion modes, and a single sound mode. If one talks about the currents en masse one does not disentangle the long-time and short-time physics, a problem from an effective field theory perspective. 

There are related issues in the gravitational description: conserved currents in holographic field theories are dual to gauge fields in the bulk; charge currents map to bulk Maxwell fields, and energy-momentum tensor to gravitational dynamics. Thus, in the AdS black hole description one has to account for the bulk gauge invariance, which leads to two issues. Firstly, canonical gauge fixing choices (eg., radial gauge in AdS) results naively in singular solutions on the grSK geometry (cf., \cite[Appendix B.2]{Ghosh:2020lel}). Secondly, the radial gauge (Gauss or momentum) constraint forces the difference Schwinger-Keldysh current to be on-shell. If we were to attempt computing a generating function of current correlators we would be forced to confront the fact that we are missing degrees of freedom. Earlier works  \cite{Glorioso:2018mmw,deBoer:2018qqm} and the more recent \cite{Bu:2021clf} attempt to take the difference current off-shell by postulating some new sources on the horizon. This is a somewhat ad-hoc procedure as explained in \cite{Ghosh:2020lel}. In any event this doesn't fully help: the issue of  locality for the effective description of non-Markovian modes remains.

The primary thesis of \cite{Ghosh:2020lel} was that the gravitational description provides a clean resolution to all of the aforementioned problems. Clearly, one should consider a parameterization of the physics in terms of the explicit Markovian and non-Markovian degrees of freedom. While the currents themselves in a strongly correlated system may not offer insight into how to do this, the dual holographic description naturally does! This is  achieved by working in terms of gauge invariant combinations which immediately allows for disambiguating Markovian and non-Markovian degrees of freedom. Elements of this were already present in earlier analysis of gravitational perturbations  \cite{Kodama:2003jz,Kodama:2003kk} and studies of quasinormal modes \cite{Kovtun:2005ev}. 

The choice of suitable parameterization  also explains how to deal with the locality issue. The trick is to not compute the generating function of the correlators (which would involve integrating out the long-lived modes leading to non-locality), but rather to parameterize the effective action in terms of the long-lived moduli fields. This choice is naturally forced upon one from the bulk gravity: the gauge invariant combinations are required to be quantized with alternate (Neumann) boundary conditions to ensure that the parent gauge or gravitational perturbations satisfy the standard (Dirichlet) boundary conditions.\footnote{We emphasize here that the boundary conditions on the gauge invariant variables are induced from the canonical choice and not put in by hand. We explain elements of this in our current set-up in \cref{sec:vectorsAct}.}

\begin{figure}[htbp]
\centering
\begin{tikzpicture}
\draw[thick, black,->] (-3,0) -- (6,0) node[below]{$r$};
\draw[thick, black,->] (-3,0) -- (-3,7) node[left] {$e^\chi$};
\draw[thick,dotted,gray] (0,0) node[below]{$r_+$} -- (0,7);
\draw[thick,rust] (-2.5,6.5) .. controls (-1,1) and (2,0.5) .. (6,0.2);
\draw[thick,blue] (-2.5,0.2) .. controls (-1,0.5) and (4,2) .. (6,6);
\node at (3,2.5) [right] {\color{blue}{Markovian}};
\node at (-1.5,4) [right] {\color{rust}{non-Markovian}};
\end{tikzpicture}

\caption{A cartoon of the effective modulation of the gravitational coupling, parameterized as an effective dilaton $e^{\chi}$ for Markovian and non-Markovian modes in a black hole geometry with a horizon at $r=r_+$. }
\label{fig:cartoon}
\end{figure}
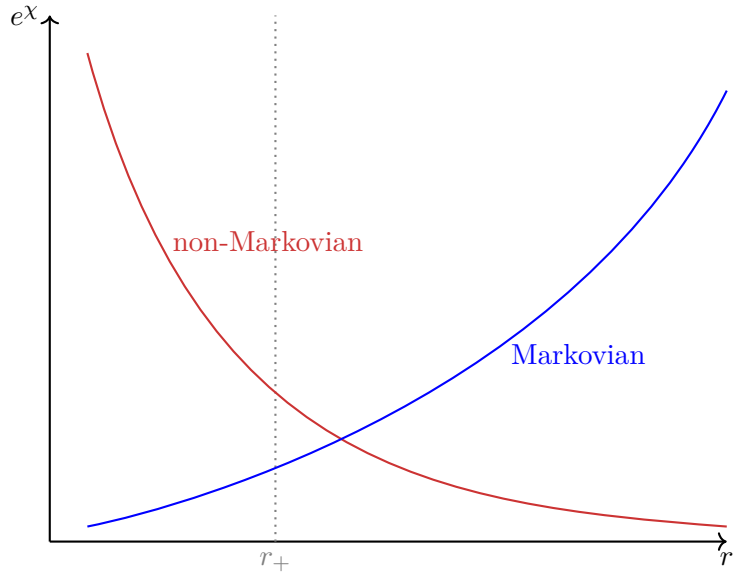

Motivated by these observations \cite{Ghosh:2020lel} proposed that the natural quantity to compute is a Wilsonian gadget (called the Wilsonian influence functional) parameterized by the sources for the Markovian data and field configurations (the Wilsonian effective fields) for the non-Markovian data. Moreover, the bulk dynamics of these modes, it was argued, could be repackaged into effective scalar degrees of freedom, which at the Gaussian order obey decoupled wave equations, with a gravitational coupling that is modulated between the horizon and the boundary. For Markovian fields one finds that the boundary is strongly repulsive and the horizon strongly attractive: these modes therefore fall into the black hole and dissipate on short time scales. The non-Markovian modes on the other hand are floppy near the boundary and mildly repulsed by the horizon, see \cref{fig:cartoon}.\footnote{As we shall see in detail later the non-Markovian modes are repulsed from some region well inside the horizon (which is not technically part of the grSK geometry). The details appear to depend on the nature of the mode in question; the core locus of repulsion is picked out by some physical scales.}
 It is this floppiness that leads to the non-Markovian modes being quantized with alternate boundary conditions.  

 As mentioned above these general ideas were illustrated in \cite{Ghosh:2020lel} for the case of diffusive modes in a neutral plasma, demonstrating that the corresponding non-Markovian modes had rather simple radial modulation of their gravitational coupling. The dilatonic modulation was a simple power law parameterized by a Markovianity index $\ann$, viz.,  $e^{\chi} = r^{\ann + 1-d}$, with the zero point chosen so as to have familiar dynamics for a minimally coupled scalar (which would have $\ann =d-1$). It was also empirically observed in \cite{Ghosh:2020lel} that order by order in a low energy, long wavelength, gradient expansion, the holographic solution for non-Markovian modes which have $\ann \leq -1$ can be obtained from those for Markovian modes with $\ann > -1$ by analytically continuing $\ann \to -\ann$. 

One natural question is whether this picture continues to hold in more intricate examples. In the current work we argue definitively that the basic principles espoused in \cite{Ghosh:2020lel} continue to hold in the case of a charged plasma. Our focus will again be on the physics of  momentum diffusion. The key element of novelty in the problem is an issue of mode coupling. A charged plasma has both a charge current and an energy-momentum current. Holographically, the dual geometry is the  planar \RNAdS{d+1} black hole, a solution to Einstein-Maxwell equations (with negative  cosmological constant). The short-lived energy-momentum modes correspond to the transverse tensor 
polarizations of the gravitons whose behaviour is qualitatively similar to the neutral plasma case.
In contrast the momentum diffusion non-Markovian mode  corresponds to transverse  vector polarizations of gravitons and it mixes non-trivially with transverse polarizations of the photons, which are, however,  Markovian. Once again by passing to suitable combination of gauge invariant variables, we are able to decouple the long-lived and short-lived modes, and show that they can be broadly understood in the scheme of designer scalar dynamics introduced in \cite{Ghosh:2020lel}. 

The main novelty in the present discussion will be two-fold: the dilatonic modulation of the gravitational coupling of the effective modes is no longer a simple power law and the resulting scalar wave equation has a non-trivial momentum dependent potential. Nevertheless, the primary thesis of \cite{Ghosh:2020lel} illustrated in \cref{fig:cartoon} remains: the asymptotic fall-off (UV of the plasma) of the modes is still parameterized by a simple Markovianity index. We will see below that the momentum diffusion is characterized by a non-Markovian mode with index $\ann = 1-d$, while the admixed charge wave is  Markovian with index $\ann = d-3$. We will use the gravitational description to argue for a suitable parameterization of the CFT currents which decouples the Markovian and non-Markovian sectors. It should become clear during the course of our discussion that such should always be possible purely in field theoretic terms (i.e., no assumptions of holographic duals). We find this to be a useful lesson from the holographic modeling, suggesting a valuable general lesson for constructing open effective field theories, cf., \cref{sec:discuss}. 

The outline of the paper is as follows. We will begin with a quick overview of our set-up, reviewing the \RNAdS{d+1} geometry and introduce some physical parameterization inspired by problem (some of these details are to our knowledge not discussed elsewhere). We then describe in \cref{sec:pertGP} the basic perturbation equations we need to solve in the grSK geometry for the graviton and photon fluctuations.  The resulting solution and the parameterization of the Schwinger-Keldysh effective action at the Gaussian level are explained in \cref{sec:cdiff}, where we translate the bulk analysis directly into field theoretic terms. Much of the analysis will be for general $d$-dimensional plasma, though we do comment on some special cases.\footnote{The results for $d=4$ do not fully capture the dynamics of R-charged  $\mathcal{N}=4$ SYM plasma, as we eschew the Chern-Simons term in the bulk. The parity-odd part of the current which arises from R-charge `t Hooft anomaly is thus excluded from our discussion. It should be straightforward to extend our analysis to include this, but for sake of simplicity we refrained from doing so in the current work.} We end with some general lessons in \cref{sec:discuss} where we also outline some open questions and also comment on the sound mode which couples to the charge diffusion mode.

Since a large part of our analysis follows the set-up of  \cite{Jana:2020vyx} and \cite{Ghosh:2020lel} we will be brief in providing some of the details relating to the gravitational Schwinger-Keldysh analysis, focusing instead on the novelties of the charged plasma system. Even so, there are several intricate pieces of calculation that we relegate to appendices to keep the main text streamlined. The derivation of the effective description of perturbations of the \RNAdS{d+1} black hole is explained in \cref{sec:bulkaction}. The reader interested in understanding how to decouple the bulk degrees of freedom and the resulting variational principle is invited to consult \cref{sec:vectorsAct}. Details of how the bulk equations of motion are solved order by order in a boundary gradient expansion and the derivation of various Green's functions are given in \cref{sec:probeM,sec:probenM} for the toy problem of probe Markovian and non-Markovian scalar fields which are employed with modifications to the physical problem of parameterizing the solutions for the transverse vector  perturbations in \cref{sec:XYsolns}. 
 The relation between bulk and boundary observables is described in detail in \cref{sec:bdyobs}, which we employ extensively in our analysis.

\section{The background geometry and setup}
\label{sec:setup}

The background \RNAdS{d+1} geometry we work with is a solution to the Einstein-Maxwell theory:\footnote{A useful reference for the background solution is \cite{Azeyanagi:2013xea} though we have chosen to fix the coupling of the Maxwell field slightly differently to simplify expressions.}
\begin{equation}\label{eq:SEMax}
\begin{split}
S_\text{EM} 
&= 
	\frac{1}{16\pi G_N}\, \int d^{d+1} x\, \sqrt{-g} \, 
		\left[ R + d(d-1) - \frac{1}{2} \,F_{AB}\, F^{AB}\right] + 
		  S_\text{bdy} + S_\text{ct} \,,\\
S_\text{bdy}
&=  \frac{1}{8\pi G_N}\, \int d^d x\, \sqrt{-\gamma} \, K		  \,.
\end{split}
\end{equation}
 Here $g_{AB}$ is the bulk metric, $\gamma_{\mu\nu}$ the induced metric on the timelike asymptotic boundary, and $K$ is the extrinsic curvature of the boundary.\footnote{We work in units  where the  AdS length scale is set to unity $\lads =1$. Dimensions of physical quantities can be restored using it, eg., the cosmological constant is given by   $-\frac{d(d-1)}{2\lads^2}$.\label{fn:lads}} Here $G_N$ is the $(d+1)$ dimensional Newton's constant.\footnote{We will use uppercase Latin alphabet ($A,B,\cdots$) to indicate bulk spacetime indices,  Greek alphabets ($\mu,\nu, \cdots$) will refer to boundary spacetime indices, while lowercase Latin alphabets ($i,j, \cdots$) will be used to refer to the spatial directions along the boundary. } The counterterm action $S_\text{ct}$ is necessary to obtain finite physical answers for correlation functions and can be found in \cref{sec:bulkaction}. We will often refer to the bulk Einstein-Maxwell action as $S_\text{EM,bulk}$ for brevity.

The equations of motion  from \eqref{eq:SEMax} are
\begin{equation}\label{eq:EMeqns}
\begin{split}
\EEin_{AB}&\equiv
R_{AB} - \frac{1}{2}\, R\, g_{AB} - \frac{d(d-1)}{2}\, g_{AB} 
=
	 g^{CD}\, F_{AC} \, F_{BD} - \frac{1}{4}\, g_{AB}\, F_{CD}\, F^{CD} \,,\\
\EMax_B&\equiv
\nabla^A F_{AB} = 0	\,.
\end{split}
\end{equation}
The \RNAdS{d+1} geometry solves these equations with line element and gauge potential given in ingoing Eddington-Finkelstein coordinates by
\begin{equation}\label{eq:RNAdS}
\begin{aligned}
ds^2 =
	2 dv dr - r^2\, f(r)\, dv^2 + r^2\, d\vb{x}^2 \,,  \qquad   \vb{A} =  -a(r)\, dv \,.
\end{aligned}
\end{equation}
The background geometry functions are themselves parameterized by two parameters $r_+$ and $Q$, and are
\begin{equation}\label{eq:RNAdSfns}
f(r) = 1 - (1+Q^2)\,\left(\frac{r_+}{r}\right)^d + Q^2\, \left(\frac{r_+}{r}\right)^{2(d-1)}\,, \qquad
a(r) = \sqrt{\frac{d-1}{d-2}}\, Q\, \frac{r_+^{d-1}}{r^{d-2}}\,.
\end{equation}	
The dimensionless parameter $Q$ is our proxy for the charge while $r_+$ is the radius of the outer horizon. 

The solution can be viewed as a charged thermal plasma of the dual CFT with intensive thermodynamic parameters temperature and chemical potential   being
\begin{equation}\label{eq:TRN}
\begin{split}
T 
&=  
	\frac{d -(d-2)\, Q^2}{4\pi}\, r_+  \,, \qquad \mu =\sqrt{\frac{d-1}{d-2}} \, Q \, r_+  \,. 
\end{split}
\end{equation}	
The physical energy and charge density can be read off from the stress tensor and charge current which take the ideal fluid form:\footnote{We define the effective central charge of the boundary theory as $c_\text{eff} = \frac{\lads^{d-1}}{16\pi G_N}$. \label{fn:centralcharge} }
\begin{equation}\label{eq:idealfluid}
\begin{split}
T^\text{Ideal}_{\mu\nu} 
&=  
	c_\text{eff}\, (1+  Q^2)\, r_+^d  \left(\eta_{\mu\nu} + d\, u_\mu\, u_\nu\right) \,, \\
J^\text{Ideal}_\mu 
&= 
	c_\text{eff}\,  \sqrt{(d-1)(d-2)} \, Q \, r_+^{d-1} \, u_\mu\,,\end{split}
\end{equation}	
with $u^\mu \,u_\mu =-1$ and $u^\mu = \left(\pdv{v}\right)^\mu$ on the boundary. 

The parameter $Q$ lives in a bounded domain 
\begin{equation}\label{eq:Qbound}
0 \leq Q \leq  \sqrt{\frac{d}{d-2}}
\end{equation}	
with $Q=0$ being the neutral \SAdS{d+1} solution and the upper limit corresponding to the extremal solution $T=0$ in  \eqref{eq:TRN}.

It will prove convenient to introduce a new length scale, the \emph{Ohmic radius} of the charged black hole which is related to the ratio  of energy density to charge squared up to a normalization factor. The rationale behind this terminology will become clear below. We let
\begin{equation}\label{eq:RQdef}
\RQ^{d-2} = \frac{d-1}{d}\, \frac{2\,Q^2}{1+Q^2}   \, r_+^{d-2}  \,.
\end{equation}	
Note that $ \RQ \in [0, r_+]$ with the upper limit corresponding to the extremal solution and the lower limit to the neutral one, cf.,  \eqref{eq:Qbound}. Moreover, the locus $r=\RQ$ lies outside the inner horizon $r=r_-$, satisfying the constraint $r_- \leq \RQ \leq r_+$ with strict equality only being attained at extremality. In fact,  as we illustrate  in \cref{fig:RQrm} these length scales satisfy  $r_- \leq \RQ \leq \frac{r_++r_-}{2}$ with the upper inequality being saturated in the near-extremal limit.

We will find it useful to work with the length scales $r_+$ and $\RQ$ rather than the temperature and chemical potential. Since physical data of the conformal plasma can only depend on the ratio of scales once we have used scalar invariance to measure quantities in units of $r_+$ (the entropy/horizon scale), it will be helpful to define a dimensionless quantity, the \emph{Ohmic parameter} $\sdc$. We define it in a dimension dependent fashion as
\begin{equation}\label{eq:sdcdef}
\sdc \equiv  \frac{\RQ^{d-2}}{r_+^{d-2}} \,.
\end{equation}	
$\sdc \in [0,1]$ with $\sdc =0$ for \SAdS{d+1} and $\sdc=1$ for the extremal solution.

Not only will this parameter play a crucial role, it also controls the dynamical equations we encounter through the \emph{Ohmic function}
\begin{equation}\label{eq:hfndef}
h(r) =  1-\frac{\RQ^{d-2}}{r^{d-2}} = 1-\sdc\, \frac{r_+^{d-2}}{r^{d-2}} \,.
\end{equation}	
This function  will appear repeatedly in our analysis below.\footnote{While its appearance here may seem a bit ad hoc, we note that $h(r)$ is closely related to the derivative of the metric function $f'(r)$, in fact
$\dv{r} f(r) = \frac{d \,(1+Q^2) r_+^d}{r^{d+1}}\, h(r)$. }
Interestingly, the Ohmic radius determines the DC conductivity of the black hole. We will verify below in \eqref{eq:sigma2} that 
\begin{equation}\label{eq:sigmadc}
\sigma_\text{dc} = r_+^{d-3} \, h(r_+)^2 = r_+^{d-3} \, (1-\sdc)^2  \,,
\end{equation}	
consistent with the earlier derivation in \cite{Hartnoll:2007ip}.\footnote{ For simplicity, we  define the charge conductivity  by stripping off an overall factor of $c_\text{eff}$.}  This relation justifies our terminology.

\begin{figure}[h!]
\begin{center}
\begin{tikzpicture}[scale=0.6]
\draw[thick,color=rust,fill=rust] (-5,0) circle (0.45ex);
\draw[thick,color=black,fill=black] (5,1) circle (0.45ex);
\draw[thick,color=black,fill=black] (5,-1) circle (0.45ex);
\draw[very thick,snake it, color=orange] (-5,0) node [below] {$\scriptstyle{r_+}$} -- (5,0) node [right] {$\scriptstyle{r_c}$};
\draw[thick,color=black, ->-] (5,1)  node [right] {$\scriptstyle{r_c+i\varepsilon}$} -- (0,1) node [above] {$\scriptstyle{\Re(\ctor) =0}$} -- (-4,1);
\draw[thick,color=black,->-] (-4,-1) -- (0,-1) node [below] {$\scriptstyle{\Re(\ctor) =1}$} -- (5,-1) node [right] {$\scriptstyle{r_c-i\varepsilon}$};
\draw[thick,color=black,->-] (-4,1) arc (45:315:1.414);
\draw[thin, color=black,  ->] (9,-0.5) -- (9,0.5) node [above] {$\scriptstyle{\Im(r)}$};
\draw[thin, color=black,  ->] (9,-0.5) -- (10,-0.5) node [right] {$\scriptstyle{\Re(r)}$};  
\draw[thick,color=orange,fill=rust] (-10,0) circle (0.45ex);
\draw[thick,color=red,fill=rust] (-8,0) circle (0.45ex) node[below] {$\RQ$};
\draw[thick,snake it, color=orange] (-10,0) node [below] {$\scriptstyle{r_-}$} -- (-15,0) ;
\end{tikzpicture}
\caption{ The complex $r$ plane with the locations of the two regulated boundaries (with cut-off $r_c$), the outer and inner horizons at $r_\pm$ and the Ohmic radius $\RQ$ marked. The grSK contour is a codimension-1 surface in this plane (drawn at fixed $v$). As indicated the direction of the contour is counter-clockwise and it encircles the branch point at the outer horizon with the cut running out to the boundary. The cut emanating from the inner horizon, and the locus at the Ohmic radius are not encountered by the contour.}
\label{fig:mockt}
\end{center}
\end{figure}
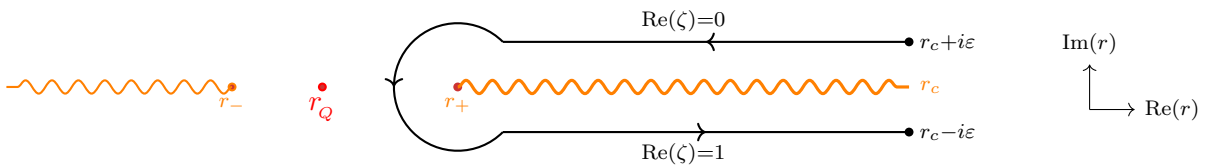

We are interested in computing real-time correlation functions of the energy-momentum tensor  and charge currents, and will employ the grSK geometry to extract these.  For the most part we will follow the conventions outlined in \cite{Jana:2020vyx,Ghosh:2020lel}. The grSK geometry is given by a complex two-sheeted metric
\begin{equation}\label{eq:grSKgeom}
ds^2 = -r^2\, f\, dv^2 + i\,\beta r^2\, f\, dv \,d \ctor + r^2 \, d\mathbf{x}^2 \,, \qquad \dv{r}{\ctor} = \frac{i\,\beta}{2}\, r^2f \,.
\end{equation}	
The coordinate $\ctor$ is the mock tortoise coordinate and $\beta = T^{-1}$, the inverse temperature.  The former is defined on the complex $r$ plane along a contour that encircles the cut emanating from the horizon at $r=r_+$, cf., \cref{fig:mockt}. Note that the Ohmic radius $r=\RQ$ is not in the part of the grSK geometry.

We work with the grSK derivative operators introduced in \cite{Ghosh:2020lel}
\begin{equation}\label{eq:Dz}
\Dz_\pm = r^2f \, \pdv{}{r} \pm \, \pdv{}{v} \,, \qquad \Dz_\pm = r^2f \, \pdv{}{r} \mp \, i \, \omega \,, 
\end{equation}	
in the time and frequency domain, respectively.   
 
While the choice of ingoing coordinates breaks the explicit time-reversal invariance, the geometry retains a time-reversal $\mathbb{Z}_2$ isometry: the transformation $v\mapsto i\beta\ctor-v$ preserves the form of the metric. The  operator $ \Dz_+$ is  naturally covariant under this isometry \cite{Ghosh:2020lel}. 
The 1-forms $\{\frac{dr}{r^2 f} , dv - \frac{dr}{r^2 f}, dx^i\}$ furnish a basis of cotangent space that is covariant under the time-reversal $\mathbb{Z}_2$. Likewise the dual derivative operators $\{\Dz_+,\partial_v, \partial_i\}$ furnish a natural basis of the bulk tangent space covariant under time-reversal. For further details on the grSK geometry and computations therein see \cite{Jana:2020vyx,Ghosh:2020lel}. The reader can also find a quick overview of probe Markovian and non-Markovian fields in \cref{sec:probeM,sec:probenM}.

\section{Linearized perturbations}
\label{sec:pertGP}

We will now consider linearized perturbations of the geometry focusing on the dynamics of gravitons and photons. We let 
\begin{equation}\label{eq:perturb}
\begin{split}
ds^2
& = 
	\left(g^{bg}_{AB} + h_{AB}  \right)dx^A dx^B \,, \\
\vb{A}
&=  A_A\, dx^A = 
	-a(r)\, dv + \Vm_A(v,r,\bx)\, dx^A \,, 
\end{split}
\end{equation}
and expand the perturbations in harmonics along $\mathbb{R}^{d-1,1}$. We discuss the different polarizations in turn in the subsections below. It will be important to understand the diffeomorphism and gauge invariant combinations. We note for now that  under a diffeomorphism along $\xi^A$ and with a gauge parameter $\lambda$ we have the following transformations:  
\begin{equation}\label{eq:diffgauge}
\begin{split}
\delta g_{AB} 
&=
	 \lieD_\xi \,g_{AB} = 2\, \nabla_{(A}\, \xi_{B)} \,, \\
\delta A_A 
&=
	\lieD_\xi A_A + \partial_A \lambda\,.
\end{split}
\end{equation}	
%

\subsection{Tensor perturbations}
\label{sec:tensors}

There are no tensor perturbations of the gauge potential as it is a vector. So the only tensor perturbations are those of the gravitons which are furthermore polarized transverse to the momentum vector $\bk$.
We expand the metric fluctuations in terms of mode components as
\begin{equation}\label{eq:hAtensor}
(h_{AB})^{\text{\tiny{Tens}}} \, dx^A\, dx^B  =
	r^2 \, \int_k \; \sum_{\bi=1}^{N_T} \, \tGR_\bi(r,\omega,\bk)\,\TT^\bi_{ij}(\omega,\bk|v,\bx)\ dx^i dx^j \,,\\
\end{equation}	
noting that there are $N_T = \frac{d(d-3)}{2} $ transverse tensor polarizations of the gravitons indexed by $\ai$. These are trivially diffeomorphism invariant  (there are no tensor diffeos). Our conventions for the harmonics follow those described in \cite[Appendix F]{Ghosh:2020lel}. We will use a short-hand for the momentum space integrals by writing  
\begin{equation}\label{eq:intk}
\int_k \;\equiv\;\int\frac{d\omega}{2\pi} \int \frac{d^{d-1} k}{(2\pi)^{d-1}} \,.
\end{equation}	

It is straightforward to check that the tensor perturbations satisfy a minimally coupled massless scalar wave equation
\begin{equation}\label{eq:tensorKG}
\nabla_A \nabla^A \, \tGR_\bi = \frac{1}{r^{d-1}}\, \Dz_+(r^{d-1}\, \Dz_+ \tGR_\bi) + r_+^2(\bwt^2 - \bqt^2\, f) \tGR_\bi =0 \,,
\end{equation}	
in the \RNAdS{d+1} background \eqref{eq:RNAdS}.  The field $\tGR_\bi$ is time-reversal even and we have  written the Klein-Gordon equation in  an explicitly time-reversal invariant form. In the language of \cite{Ghosh:2020lel} $\tGR_\bi$ is a Markovian field of Markovianity index $\ann = d-1$. The equation is written in terms of dimensionless frequencies and momenta (cf., \cref{fn:lads})
\begin{equation}\label{eq:dimlesswk}
\bwt = \frac{\omega}{r_+} \,, \qquad \bqt = \frac{k}{r_+} \,.
\end{equation}	
%

\subsection{Vector perturbations}
\label{sec:vectors}
The vector polarizations are present for both the gauge potential and the metric and we have 
$N_V = d-2$  degrees of freedom for every transverse vector in $\mathbb{R}^{d-1,1}$. The components can be expanded as 
\begin{equation}\label{eq:hAvector}
\begin{split}
 \Vm_i (v, r, \bx) 
 &=
 	\int_k  \sum_{\ai=1}^{N_V}\, \vMax_\ai(r,\omega,\bk) \, \VV^\ai_i(\omega,\bk|v,\bx)  \,, \\
 (h_{AB})^{\text{\tiny{Vec}}} \, dx^A\, dx^B 
 &= 
 	r^2\, \int_k\,  
  	\sum_{\ai=1}^{N_V} \bigg(2\ (\vGR_r^\ai(r,\omega,\bk) \, dr +\vGR_v^\ai (r,\omega,\bk) dv)\,
  	\VV^\ai_i(\omega,\bk|v,\bx) dx^i \\
  & \qquad \qquad \quad 
  	+ i\, \vGR_x^\ai(r,\omega,\bk)\ \VV_{ij}^\ai\,  dx^i dx^j\bigg), 		
\end{split}
\end{equation}

Under a vector diffeomorphism 
\begin{equation}
\begin{split}
 x^i
 &\mapsto 
 	x^i + \int_k\,\sum_{\ai=1}^{N_V}\Lambda_\ai(r,\omega,\bk) \, \VV_i^\ai (\omega,\bk|v,\bx)\,, \\
dx^i &\mapsto 
		dx^i + \int_k\,\sum_{\ai=1}^{N_V}\,\left[
	 	\dv{\Lambda_\ai}{r} \,    dr 
		+ \Lambda_\ai(r,\omega,\bk)\   ( dv\, \partial_v+ dx^j\ \partial_j)  \right]\VV_i^\ai (\omega,\bk|v,\bx) \,.
 \end{split}
 \end{equation}
Upto linear order in $\Lambda_\ai$ this gives a shift 
\begin{equation}\label{eq:gaugediffeo}
\begin{split}{}
\vGR_r^\ai 
&	\mapsto 
	\vGR_r^\ai + \dv{\Lambda_\ai}{r}\ ,
	\quad 
	\vGR_v^\ai \mapsto \vGR_v^\ai -i\omega\  \Lambda_\ai \ ,\quad 
	\vGR_x^\ai \mapsto \vGR_x^\ai  -ik\,\Lambda_\ai\,, \\
\vMax_\ai
&\mapsto
	\vMax_\ai  \,.
\end{split}
\end{equation}
Thus $\vMax_\ai$ are separately gauge-invariant (there is no vector gauge transformation) and diffeomorphism invariant.  On the other hand the triple $\{\vGR_v^\ai,\vGR_v^\ai, \vGR_x^\ai\}$ form an auxiliary diffusive gauge system as described in \cite[\S8]{Ghosh:2020lel}. This can be checked by working out the gauge transformations using \eqref{eq:gaugediffeo} (see  \cref{sec:Vcoupled}). 

While there appear to be four dynamical variables per polarization index $\ai$ in \eqref{eq:hAvector} there are only two dynamical degrees of freedom because of the underlying gauge invariance. Working with gauge invariant variables introduced in \cite{Kodama:2003kk} (as we explain in \cref{sec:vectorEom}) these can be shown to satisfy two decoupled equations for a non-Markovian and a Markovian designer field, $\MX_\ai$ and $\MY_\ai$, respectively. 

To motivate this, one examines the Einstein's equations and realizes that they can be solved identically by cleverly parameterizing $\{\vGR_v^\ai,\vGR_v^\ai, \vGR_x^\ai, \vMax_\ai\}$ in terms of two fields $X_\ai$ and $Y_\ai$ as follows
\begin{equation}\label{eq:XYpar1}
\begin{aligned}
&
\dv{\vGR_v^\ai}{r} + i\omega  \, \vGR_r^\ai  
=
	\frac{k^2}{r^{d+1}}  (X_\ai +2\,Y_\ai)  \,, && \quad 
\dv{\vGR_x^\ai}{r}+ik \,\vGR_r^\ai 
 = -
 	\frac{ik}{r^{d-1}} \, \dv{X_\ai}{r} \,,\\ 
&k\, \vGR_v^\ai -\omega \, \vGR_x^\ai 
= 
	\frac{k}{r^{d-1}}\, \Dz_+ X_\ai \,, && \quad 
\vMax_\ai 
=
	 -\frac{k^2}{(d-2)\, \mu\,r_+^{d-2} } \, Y_\ai \,.\\	
\end{aligned}
\end{equation}	
Self-consistency of the parameterization and Maxwell's equation result in a pair of second order coupled differential equations for $X_\ai$ and $Y_\ai$, see \eqref{eq:XYcoupled}. These may in turn be decoupled by the  functional linear combination
\begin{equation}\label{eq:CXYdef}
\begin{split}
X_\ai 
&=
	 -(\BQT^2+2) \, \MX_\ai + \BQT^2\, \frac{h}{1-h} \, \MY_\ai\,,\\
Y_\ai 
&=
	 (1-h) \,\MX_\ai + h\, \MY_\ai \,,
\end{split}
\end{equation}
leading to a remarkably simple decoupled dynamical system:
\begin{equation}\label{eq:XYfinal}
\begin{split}
\frac{1}{r^{d-3} \, (1-h)^2} \Dz_+ \left((1-h)^2\, r^{d-3}\, \Dz_+ \MX_\ai\right) 
+ r_+^2 \left(\bwt^2 - \bqt^2 f\,+  \bRQ^2\, (1-h) \, \BQT^2 \, f\right) \MX_\alpha 
&=0 \,,\\
\frac{1}{r^{d-3} \,h^2} \Dz_+ \left(h^2\, r^{d-3}\, \Dz_+ \MY_\ai\right) 
+ r_+^2 \left(\bwt^2 - \bqt^2 f\, - \bRQ^2\, (1-h) \, \BQT^2\, f\right) \MY_\alpha 
&=0 \,.
\end{split}
\end{equation}
We have introduced here a deformed momentum parameter $\BQT$ which appears courtesy the basis rotation coefficients when we decouple the Einstein-Maxwell system:
\begin{equation}\label{eq:BQTdef}
\begin{split}
\BQT^2 
&=
	 \sqrt{1+2\frac{\bqt^2}{\bRQ^2}} - 1  = 
	\frac{\bqt^2}{\bRQ^2} \left(1- \frac{1}{2}\, \frac{\bqt^2}{\bRQ^2} +\cdots \right) \,,
	\qquad 
	\bRQ \equiv \frac{(d-2)\, \mu}{r_+\, \sdc}\,.
\end{split}
\end{equation}	

The effective fields $\MX_\ai$ and $\MY_\ai$ are indeed `designer scalars' as introduced in \cite{Ghosh:2020lel}, albeit with a non-trivial dilaton:
\begin{equation}\label{eq:dilXY}
e^{\dilX}= \frac{1}{r^{2(d-1)}} \,, \qquad e^{\dilY} = \frac{h^2}{r^2} \,.
\end{equation}	
The field $\MX_\ai$ is a non-Markovian field with index $\ann = -(d-1)$, since the dilaton factor simplifies to a simple monomial,  $(1-h)^2\, r^{d-3} \propto r^{1-d}$. The field $\MY_\ai$ has a bit more complicated dilaton; its asymptotics is that of a Markovian scalar of index $\ann = d-3$, but this behaviour is modulated by the Ohmic function $h(r)$ as we probe the interior of the spacetime.  Accounting for the measure factor $\sqrt{-g} = r^{d-1}$ we arrive at the expressions quoted above. These dilatonic modulations, one can check, realize the paradigm depicted in \cref{fig:cartoon}. The main difference is that the momentum dependence is more complicated. Instead of $\bqt^2\, f$ for a designer field as the potential, we have an additional contribution in $\pm \,\bRQ^2 \, \BQT^2\, (1-h)$. The solution to these equations up to the quartic order in a low frequency and momentum expansion is presented in \cref{sec:XYsolns}. 

To get some intuition for the dynamical system, let us consider switching off the charge; in the $Q \to 0$ limit, $\MX_\ai$ is the non-Markovian momentum diffusion graviton mode, while $\MY_\ai$ is the Markovian transverse photon mode which decays away quickly, as explained in \cite{Ghosh:2020lel}. In a charged plasma, the mixing between vector polarizations of gravitons and photons implies that while the vector component charge current wants to decay away quickly it is dragged by the momentum flux. The gravitational description of the system indicates that this coupled dynamics can be decoupled at the quadratic level and packaged neatly into the general paradigm suggested in \cite{Ghosh:2020lel} (modulo more complicated designer dilaton potentials). We believe this is a general phenomenon (valid also
for the scalar polarizations not discussed herein); for further comments see \cref{sec:discuss}.

\section{Diffusion in a charged plasma}
\label{sec:cdiff}

We would like to compute the effective action governing the dynamics of the conserved currents, the energy-momentum tensor $(\Tcft)^{\mu\nu}$  and charge current $(\Jcft)^\mu$, focusing only on the transverse tensor and vector polarizations. These are effectively encoded in our designer scalar fields: tensor modes are captured by $\tGR_\bi$ which are minimally coupled, massless scalars \eqref{eq:tensorKG}, while  the vector modes packaged into $\MX_\ai$ and $\MY_\ai$ satisfy a more complicated dynamics as described in \eqref{eq:XYfinal}. Two of these, $\tGR_\bi$ and $\MY_\ai$,  are Markovian,  while $\MX_\ai$ is a non-Markovian field capturing the physics of momentum diffusion in the charged plasma.

The general scheme for solving the problem of Markovian and non-Markovian fluctuations of the \SAdS{d+1} background was described in \cite{Ghosh:2020lel} and can be immediately applied to the problem at hand. 
We briefly review salient elements necessary for our discussion, generalizing the results to an arbitrary planar black hole geometry. Some of the technical details are collected in  \cref{sec:probeM,sec:probenM}. The reader is encouraged to consult \cite{Ghosh:2020lel} for further details. 

\subsection{Review of Wilsonian influence functionals}
\label{sec:wifc}

First, let us focus on the simpler case of  Markovian dynamics. For purposes of illustration consider a designer scalar $\sen{\ann}$ with a simple dilatonic coupling $e^{\dils} = r^{\ann+1-d}$, with bulk action \eqref{eq:probeMAct}. To obtain its boundary effective action, one  solves  for the ingoing bulk-boundary propagator $\Gin{\ann}$ with unit source on the boundary, order by order in a gradient expansion, imposing Dirichlet boundary conditions with sources $J_\skR$ and $J_\skL$ on the two asymptotic boundaries of the grSK geometry,
\begin{equation}\label{eq:Msources}
\lim_{r\to \infty + i0} \sen{\ann}^\text{SK} = J_L \,, \qquad \lim_{r\to \infty - i0} \sen{\ann}^\text{SK} = J_R\,.
\end{equation}  
The general expression for $\Gin{\ann}$ and functions entering it are given in \cref{sec:probenM}. They can be readily specialized to the tensor modes by setting $\ann = d-1$. For the field $\MY_\ai$ the corresponding Green's function will be given below and is obtained using the data given in \cref{sec:XYsolns}. 

The solution for the designer scalar on the grSK geometry with the aforementioned boundary conditions is \cite{Jana:2020vyx}
\begin{equation}\label{eq:senM}
\sen{\ann}^\text{SK}(\ctor,\omega,\bk) 
= \Gin{\ann}\, \JMar_a + \left[\left(\nB + \frac{1}{2}\right) \Gin{\ann} -  \nB\, e^{\beta \omega (1-\ctor)} \, \Grev{\ann}\right] \JMar_d \,,
\end{equation}  
with $\Grev{\ann}(r,\omega,\bk) = \Gin{\ann}(r,-\omega,\bk)$ is the time-reversed propagator and  $\JMar_a, \JMar_d$ are the average and difference sources, respectively, defined as 
\begin{equation}\label{eq:Jad}
\JMar_a = \frac{1}{2} (\JMar_\skR + \JMar_\skL) \,, \qquad \JMar_d = \JMar_\skR - \JMar_\skL \,,
\end{equation}  
and $\nB$ is the Bose-Einstein distribution function
\begin{equation}\label{eq:BEdist}
\nB = \frac{1}{e^{\beta \omega} -1} \,.
\end{equation}  
As argued in \cite{Jana:2020vyx} the above form ensures that a gradient expansion exists for the SK
solution provided $\Gin{\ann}$ has a gradient expansion. 

The boundary correlation function we seek is captured by taking an appropriate limit of the bulk to boundary propagator with suitable regulators to obtain the function,  $\Kin{\ann}(\omega, \bk)$, which feeds into the on-shell action directly. On the grSK geometry one finds
\begin{equation}\label{eq:sosM}
S[\sen{\ann}]\bigg|_\text{on-shell}
 = - \int_k\, \JMar_d^\dagger\, \Kin{\ann} \left[ \JMar_a + \left(\nB+\frac{1}{2}\right) \JMar_d\right] .
\end{equation}  

We also note that the one-point function of the  dual boundary operators $\mathcal{O}$ in the presence of an external source is given by
\begin{equation}\label{eq:MarkJO}
\begin{split}
\expval{\mathcal{O}_a(\omega,\bk)} 
&= 
  -\Kin{\ann}\, J_a - \left(\nB + \frac{1}{2}\right) \left[ \Kin{\ann} -\Krev{\ann}\right] \, J_d \,, \\ 
\expval{\mathcal{O}_d(\omega,\bk)} 
&= 
  -\Krev{\ann}\, J_d\,.
 \end{split}
\end{equation}

The quantity $\Kin{\ann}$ is the retarded Green's function which is obtained directly from the regularized asymptotic value of the momentum conjugate $\cpen{\ann}$ to the field $\sen{\ann}$. An explicit expression for a general Markovian field $\sen{\ann}$ can be found in  \eqref{eq:KinMark}. The function $\Krev{\ann}$ is its time-reversed counterpart and is obtained as 
\begin{equation}\label{eq:tRKin}
\Krev{\ann}(\omega,\bk) = \Kin{\ann}(-\omega,\bk) \,.
\end{equation}  
The structure of the Gaussian part of the on-shell action guarantees that the fluctuation dissipation relation is satisfied, since the $J_d^\dag J_a$ coefficient, the retarded Green's function, is related to that of $J_d^\dag \, J_d$, the quantum fluctuations in the thermal state. It will be helpful to record the explicit form for the two point functions which we will use later: 
\begin{equation}\label{eq:M2ptfns}
\begin{split}
\expval{\mathcal{O}(-\omega,-\bk) \, \mathcal{O}(\omega,\bk)}^\text{Ret} 
&= 
  i\, \Kin{\ann}(\omega,\bk)\,, \\
\expval{\mathcal{O}(-\omega,-\bk) \, \mathcal{O}(\omega,\bk)}^\text{Kel} 
&= 
  -\frac{1}{2}\, \coth\left(\frac{\beta\omega}{2}\right) \, \Im\left[\Kin{\ann}(\omega,\bk)\right] ,
\end{split}
\end{equation}  
where we have used \eqref{eq:tRKin}. For Markovian modes we thus recover the expected picture at the quadratic order, consistent with thermal field theory expectations, as we see explicitly the fluctuation dissipation relation:
\begin{equation}\label{eq:FDT}
\expval{\mathcal{O}(-\omega,-\bk) \, \mathcal{O}(\omega,\bk)}^\text{Kel} =
    \frac{1}{2}\,\coth\left(\frac{\beta\omega}{2}\right) \Re\left[\expval{\mathcal{O}(-\omega,-\bk) \, \mathcal{O}(\omega,\bk)}^\text{Ret} \right] .  
\end{equation}  

Turning next to non-Markovian fields, the fact that the field is free to fluctuate near the boundary of the spacetime means that we have mode functions that grow in a non-normalizable fashion. The main premise of \cite{Ghosh:2020lel} was that these fields should be quantized in the bulk using  Neumann boundary conditions. Specifically, rather than computing the generating function of Schwinger-Keldysh correlators as we did above for the Markovian fields, the idea was to parameterize the hydrodynamic moduli space by some boundary field configurations and obtain the Wilsonian influence functional in terms of long-distance moduli fields, $\snMar_\skR$ and $\snMar_\skL$ instead. They can be viewed as parameterizing the expectation value of the non-Markovian operator $\OnMar$, and in particular, we can take
\begin{equation}\label{eq:OnMar}
\expval{\OnMar_\skR}  = \snMar_\skR \,, \qquad 
\expval{\OnMar_\skL}  = \snMar_\skL \,.
\end{equation}  

While for a probe non-Markovian field this would be a natural choice, it is a remarkable fact that for the non-Markovian components of the conserved currents, the bulk gravitational dynamics, when distilled into a gauge invariant modes dictates  by itself that the particular non-Markovian degrees of freedom should be quantized with Neumann boundary conditions. For bulk gravitons and gauge fields the standard Dirichlet boundary conditions transmutes into Neumann boundary conditions for the non-Markovian sector. This was explained for probe gauge bosons and graviton fluctuations in  \cite{Ghosh:2020lel}. This continues to hold for the charged plasma modes $\MX_\ai$ and $\MY_\ai$, even though we have to diagonalize quadratic action to decouple these degrees of freedom. We demonstrate this explicitly in \cref{sec:vectorsAct}.

Sticking to a general non-Markovian field $\sen{-\ann}$ we obtain the effective action by analytic continuation of $\ann$ to $-\ann$, which effectively converts the sources for the Markovian problem to the moduli fields of the non-Markovian problem. The bulk solution on the grSK geometry is similar:
\begin{equation}\label{eq:sennM}
\sen{-\ann}^\text{SK}(\ctor,\omega,\bk) 
= \Gin{-\ann}\, \snMar_a + \left[\left(\nB + \frac{1}{2}\right) \Gin{-\ann} -  \nB\, e^{\beta \omega (1-\ctor)} \, \Grev{-\ann}\right] \snMar_d \,.
\end{equation}  
Furthermore, the one-point functions are computed modulo counterterm contributions by the asymptotic field value
\begin{equation}\label{eq:}
\expval{\snMar_{\skL,\skR}} = \lim_{r\to \infty \pm i0}\, \left[\sen{-\ann} + \text{counterterms} \right] .
\end{equation}  

Since we are parameterizing the solution in terms of the asymptotic normalizable mode, it is helpful to also write down the relation to the non-normalizable sources $\JnMar$ for the non-Markovian scalar which owing to the Neumann boundary conditions is given by the asymptotic value of the conjugate momentum $\cpen{-\ann} = -r^{-\ann}\, \Dz_+ \sen{-\ann}$. The result is simply given in the boundary dispersion function $\Kin{-\ann}$:
\begin{equation}\label{eq:JOnMrel}
\begin{split}
\JnMar_a &= 
  \Kin{-\ann} \, \snMar_a + \left(\nB+ \frac{1}{2}\right) \left[ \Kin{-\ann} - \Krev{-\ann}\right] \snMar_d \,,\\
\JnMar_d &= 
  \Krev{-\ann}\,\snMar_d \,.
\end{split}
\end{equation}  
These equations have to be interpreted as the dynamical equations for the long-lived non-Markovian modes. Equivalently they can also be thought of as real-time Schwinger-Dyson equations within the 
open quantum field theory describing these modes.

These results can be immediately derived by computing the on-shell action with fixed boundary field configurations, which leads to the Wilsonian influence functional for the non-Markovian fields \cite{Ghosh:2020lel} parameterized as indicated in terms of the Schwinger-Keldysh moduli fields $\snMar{a,d}$.
\begin{equation}\label{eq:sosnM}
S[\sen{-\ann}] \bigg|_\text{on-shell}
 = - \int_k\, \snMar_d^\dagger\, \Kin{-\ann} \left[ \snMar_a + \left(\nB+\frac{1}{2}\right) \snMar_d\right] .
\end{equation}  
Legendre transforming this expression with respect to the non-Markovian moduli $\snMar_{a,d}$ leads to the aforementioned relation between the sources and moduli, \eqref{eq:JOnMrel}. 

The advantage of working with the moduli fields is that the effective action is completely local, since the dispersion function has a nice gradient expansion. The Legendre transformation to compute the generating function of correlators leads to the expected two-point function with the hydrodynamic poles. 

To obtain Green's functions for the non-Markovian boundary operator $\OnMar$ one can simply invert  the relation  \eqref{eq:JOnMrel}  solving for moduli field $\snMar_{a,d}$ as a functional of the background source. Using \eqref{eq:OnMar} we obtain the Schwinger-Keldysh correlation functions consistent with fluctuation dissipation relations:
\begin{equation}\label{eq:nM2ptfns}
\begin{split}
\expval{\OnMar(-\omega,-\bk) \, \OnMar(\omega,\bk)}^\text{Ret} 
&= 
   \frac{1}{i\, \Kin{-\ann}(\omega,\bk)} \,,\\
\expval{\OnMar(-\omega,-\bk) \, \OnMar(\omega,\bk)}^\text{Kel} 
&= 
 -  \frac{1}{2}\,\coth\left(\frac{\beta\omega}{2}\right) \frac{\Im\left[\Kin{-\ann}(\omega,\bk)\right]}{\abs{\Kin{-\ann}(\omega,\bk)}^2} . 
\end{split}
\end{equation}
%

\subsection{Wilsonian influence functional for the  charged plasma}
\label{sec:cplasmadata}

For the charged plasma, we have following boundary data parameterizing the boundary Wilsonian influence functional:
\begin{itemize}[wide,left=0pt]
\item The sources for the tensor polarizations of the graviton, which are the transverse traceless components of the boundary metric, denoted in the average difference basis $\JMarP_a^\bi$ and $\JMarP_d^\bi$. Note that we have traded the boundary spacetime indices for the polarization label $\bi$. These are the sources for the transverse tensor polarizations of the stress tensor in the grSK geometry and are defined in the R, L basis  as 
\begin{equation}\label{eq:TTsource}
\JMarP_{\skL}^\bi = \lim_{r\to \infty+i0}\, \Phi_\bi \,, \qquad 
\JMarP_{\skR}^\bi = \lim_{r\to \infty-i0}\, \Phi_\bi\,.
\end{equation}  
\item The sources for the transverse vector charge mode which is an admixture of the transverse vector polarizations of the bulk metric and Maxwell potential. We will identify this mode with the field $\MY_\ai$ in the bulk. The corresponding boundary sources will be denoted $\YsQ_a^\ai$ and $\YsQ_d^\ai$, respectively and are obtained from the L, R sources 
\begin{equation}\label{eq:MVsource}
\YsQ_{\skL}^\ai = \lim_{r\to \infty+i0}\, \MY_\ai \,, \qquad 
\YsQ_{\skR}^\ai = \lim_{r\to \infty-i0}\, \MY_\ai\,.
\end{equation}  
The boundary operator that couples to these sources will be labeled $\YOp^\ai$.  
\item The momentum flux vectors capturing shear modes which arise from a linearly independent admixture of the transverse vector polarizations of the bulk metric and Maxwell potential are captured by the field $\MX_\ai$ in the bulk geometry. The corresponding boundary moduli fields will be denoted as  $\XP_a^\ai$ and $\XP_d^\ai$. These are related to  the expectation values of the dual operator $(\XOp^\ai)_a$ and $(\XOp^\ai)_d$. We define 
\begin{equation}\label{eq:nMVop}
\begin{split}
\XP_\skL^\ai 
&= 
  \lim_{r\to \infty + i 0} \left[\MX_\ai + \text{counterterms}\right] ,\\
\XP_\skR^\ai 
&= 
  \lim_{r\to \infty - i 0} \left[\MX_\ai + \text{counterterms}\right] ,
\end{split}
\end{equation}  
with 
\begin{equation}\label{eq:OXP}
\expval{(\XOp^\ai)_{\skL}} = \XP_{\skL}^\ai \,, \qquad
\expval{(\XOp^\ai)_{\skR}} = \XP_{\skR}^\ai \,.
\end{equation}	
The boundary sources for these non-Markovian modes will be denoted as $\XsJ^\ai$ and are defined in \eqref{eq:BC}.
\end{itemize}

\subsubsection{Dynamics of the decoupled vector modes}
\label{sec:XYwif}

Let us first discuss the dynamics of the fields $\MX_\ai$ and $\MY_\ai$ as probe fields in the \RNAdS{d+1} black hole background subject to the equations of motion \eqref{eq:XYfinal}. One can show that the Wilsonian influence functional for these probe fields in the grSK geometry takes the general form described above:\footnote{ The probe action we report below is the minimal action which is compatible with the equations of motion in \eqref{eq:XYfinal}. The Einstein-Maxwell dynamics itself does simplify to a similar effective action, albeit one with non-canonical kinetic terms, see \eqref{eq:WIFEin} and \cref{sec:osaction}. \label{fn:XYminimal}}
\begin{equation}\label{eq:SXYprobe}
\begin{split}
S_\text{probe}[\MX_\ai] 
&\propto 
   -\, \int_k   \sum_{\ai=1}^{N_V}   (\XP^\ai_d)^\dag \,    \Kin{\MX} \left[\XP^\ai_{a}+\left(\nB+\frac{1}{2}\right)\: \XP^\ai_d \right]  ,  \\ 
S_\text{probe}[\MY_\ai] 
&\propto
 -\, \int_k   \sum_{\ai=1}^{N_V}   (\YsQ_d^\ai)^\dag \,
  \Kin{\MY} \left[\YsQ_a^\ai+\left(\nB+\frac{1}{2}\right)\: \YsQ_d^\ai \right]    .
\end{split}
\end{equation}  
We note there that this expression is to be viewed as a heuristic explaining the structure of the Wilsonian influence function. As we footnote above this is useful mnemonic, but one that does not account for the correct dimensions of the sources and operators, which for the physical Einstein-Maxwell problem are induced directly from \eqref{eq:SEMax}. We will present the correct influence functional for the Einstein-Maxwell system in \eqref{eq:WIFEin} with dimensional factors and normalizations completely fixed.

\begin{figure}[t]
\centering
\subfloat{
 \begin{minipage}[t]{0.5\textwidth}
 \vspace{0.5cm}
\hspace{0pt}
\begin{tikzpicture}
  \node (img)  {\includegraphics[scale=0.275]{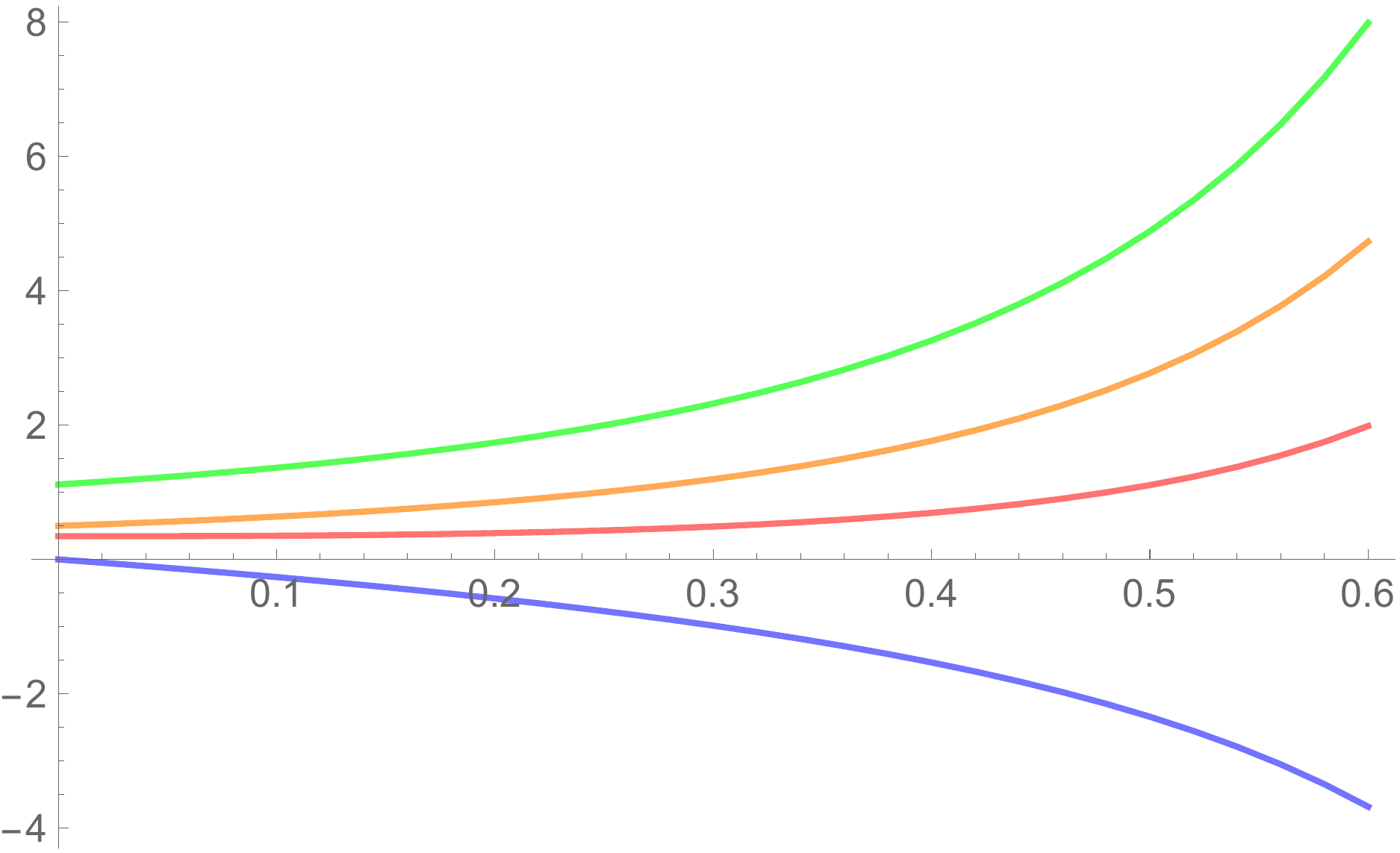}};
  \node[right=of img, node distance=0cm,font=\small,xshift=-1.15cm,yshift=-0.5cm] {$\sdc$};
  \node[left=of img, node distance=0cm, anchor=center,font=\footnotesize,xshift=0.5cm] {$\Dfn{\MY}{2,0}(r_+)$};
  \node[right=of img, node distance=0cm,font=\scriptsize,xshift=-1.25cm,yshift=1.5cm] {$d=5$};
  \node[right=of img, node distance=0cm,font=\scriptsize,xshift=-1.25cm,yshift=0.8cm] {$d=6$};
  \node[right=of img, node distance=0cm,font=\scriptsize,xshift=-1.25cm,yshift=0.15cm] {$d=4$};
   \node[right=of img, node distance=0cm,font=\scriptsize,xshift=-1.25cm,yshift=-1.5cm] {$d=3$};
 \end{tikzpicture}
 \end{minipage}
}\\
\subfloat{
\begin{minipage}[t]{0.5\textwidth}
\vspace{0pt}
\hspace{0pt}
\begin{tikzpicture}
  \node (img)  {\includegraphics[scale=0.275]{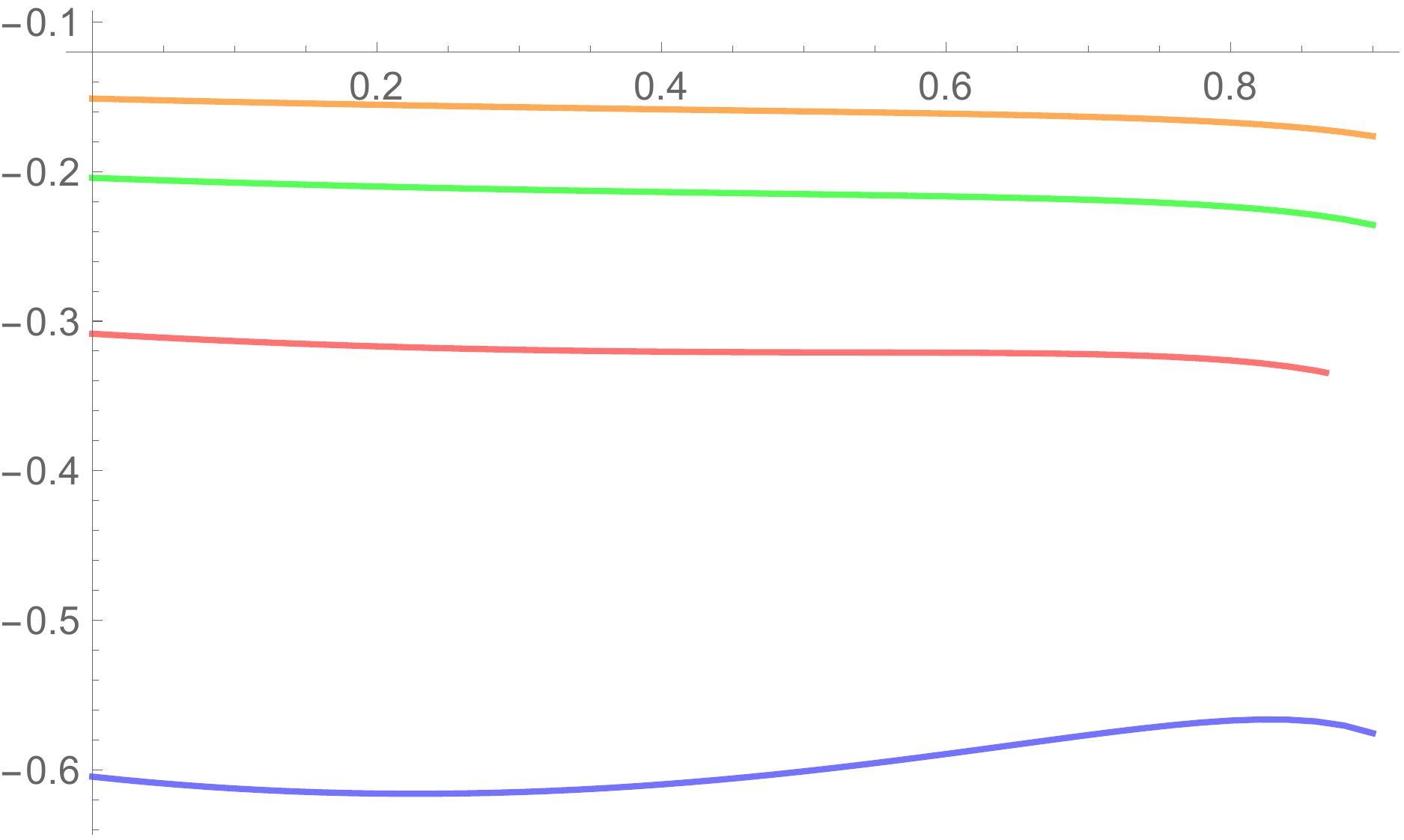}};
  \node[above=of img, node distance=0cm,font=\small,yshift=-1.25cm] {$\sdc$};
  \node[left=of img, node distance=0cm, anchor=center,font=\footnotesize,xshift=0.5cm] {$\yser{0,2}(r_+)$};
  \node[right=of img, node distance=0cm,font=\scriptsize,xshift=-1.25cm,yshift=1.15cm] {$d=6$};
  \node[right=of img, node distance=0cm,font=\scriptsize,xshift=-1.25cm,yshift=0.8cm] {$d=5$};
  \node[right=of img, node distance=0cm,font=\scriptsize,xshift=-1.25cm,yshift=0.25cm] {$d=4$};
   \node[right=of img, node distance=0cm,font=\scriptsize,xshift=-1.25cm,yshift=-1.25cm] {$d=3$};
 \end{tikzpicture}
 \end{minipage}}
\subfloat{
 \begin{minipage}[t]{0.5\textwidth}
 \vspace{0pt}
 \hspace{0pt}
\begin{tikzpicture}
  \node (img)  {\includegraphics[scale=0.275]{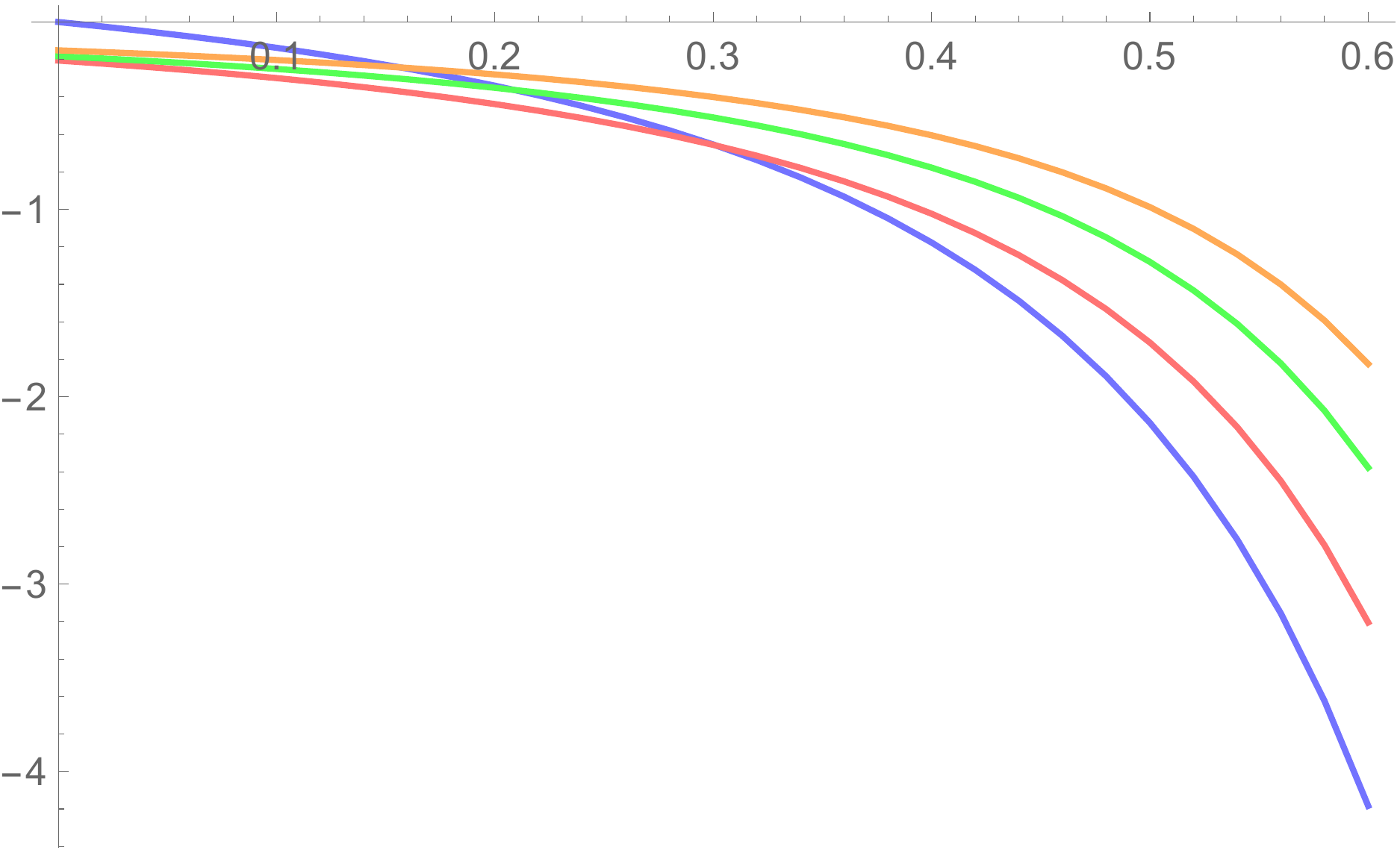}};
  \node[above=of img, node distance=0cm, font=\small,yshift=-1.25cm] {$\sdc$};
  \node[left=of img, node distance=0cm, anchor=center,font=\footnotesize,xshift=0.5cm] {$\yser{0,2}(r_+)$};
  \node[right=of img, node distance=0cm,font=\scriptsize,xshift=-1.25cm,yshift=0.25cm] {$d=6$};
  \node[right=of img, node distance=0cm,font=\scriptsize,xshift=-1.25cm,yshift=-0.25cm] {$d=5$};
  \node[right=of img, node distance=0cm,font=\scriptsize,xshift=-1.25cm,yshift=-0.75cm] {$d=4$};
   \node[right=of img, node distance=0cm,font=\scriptsize,xshift=-1.25cm,yshift=-1.5cm] {$d=3$};
 \end{tikzpicture}
 \end{minipage}
}
\caption{ The charge dependence of the coefficients in Green's function of the Markovian vector polarizations encoded in $\MY_\ai$ in  dimensions $d=3,\ldots, 6$. We use the Ohmic radius $\sdc$ defined in Eq.~(\ref{eq:sdcdef}) as the proxy for the charge reminding the reader that $\sdc=0$ corresponds to the neutral \SAdS{d+1} black hole while $\sdc\to1$ is the extremal limit (where the various functions diverge). }
\label{fig:KinMarY}
\end{figure}
The two pieces of data entering the above are the boundary Green's function for the Markovian field $\Kin{\MY}(\omega,\bk)$ and the inverse propagator $\Kin{\MX}(\omega,\bk)$ for the non-Markovian component.  The former is given by
\begin{equation}\label{eq:KinY}
\begin{split}
\Kin{\MY}(\omega,\bk) 
&=
   r_+^{d-2} 
   \bigg\{ -i\,   (1-\sdc)^2\, \bwt 
  - \left[\frac{1}{d-4} + \frac{\sdc}{2} + \frac{\sdc^2}{d} - \frac{\sdc^3}{2(d-1)}\right] \bqt^2  \\
&\qquad 
  +   (1-\sdc)^2 \left[\Dfn{\MY}{2,0}(r_+) \, \bwt^2  - 2i\, \yser{0,2}(r_+)\,  \bwt \bqt^2 + 2i \,\yser{2,0}(r_+)\, \bwt^3 + \cdots \right] \bigg\} \,.
\end{split}
\end{equation}  
The functions whose horizon values\footnote{ In evaluating the functions appearing the gradient expansion we work with dimensionless variables rescaling out a length scale set by the horizon radius $r_+$.  These functions are non-trivial functions of $\RQ/r_+$, or equivalently $\sdc$ introduced in \eqref{eq:sdcdef},  whose dependence we leave implicit. We also alert the reader that we work with the dimensionless coordinate $\ri = r_+/r$ in \cref{sec:probeM,sec:probenM,sec:XYsolns}. As a consequence horizon values of the gradient expansion functions will be evaluated at unity (i.e., at $\ri =1$).}  enter the expression above are defined in \eqref{eq:Ygradfns} and \eqref{eq:DYfns}.\footnote{Some of these expression are valid  only for $d>4$. As we discuss in \cref{sec:XYsolns} in $d=4$ we have to be careful to take care of logarithmic divergences.}  

The charge dependence of the coefficients appearing in $\Kin{\MY}$  until the cubic order is plotted in \cref{fig:KinMarY}.  We have obtained the functions $\Kin{\MY}$ and analogous expressions for the other modes up to the quartic order in gradients. However, in the main text we will only give expressions to cubic order in gradients to avoid writing complicated formulae.  For $\Kin{\MY}$ the expression accurate to quartic order in derivatives can be found in \eqref{eq:KinYA} (these are plotted in \cref{fig:KinMarY4th}).

\begin{figure}[t]
\centering
\subfloat{
 \begin{minipage}[t]{0.5\textwidth}
 \vspace{0.5cm}
\hspace{0pt}
\begin{tikzpicture}
  \node (img)  {\includegraphics[scale=0.275]{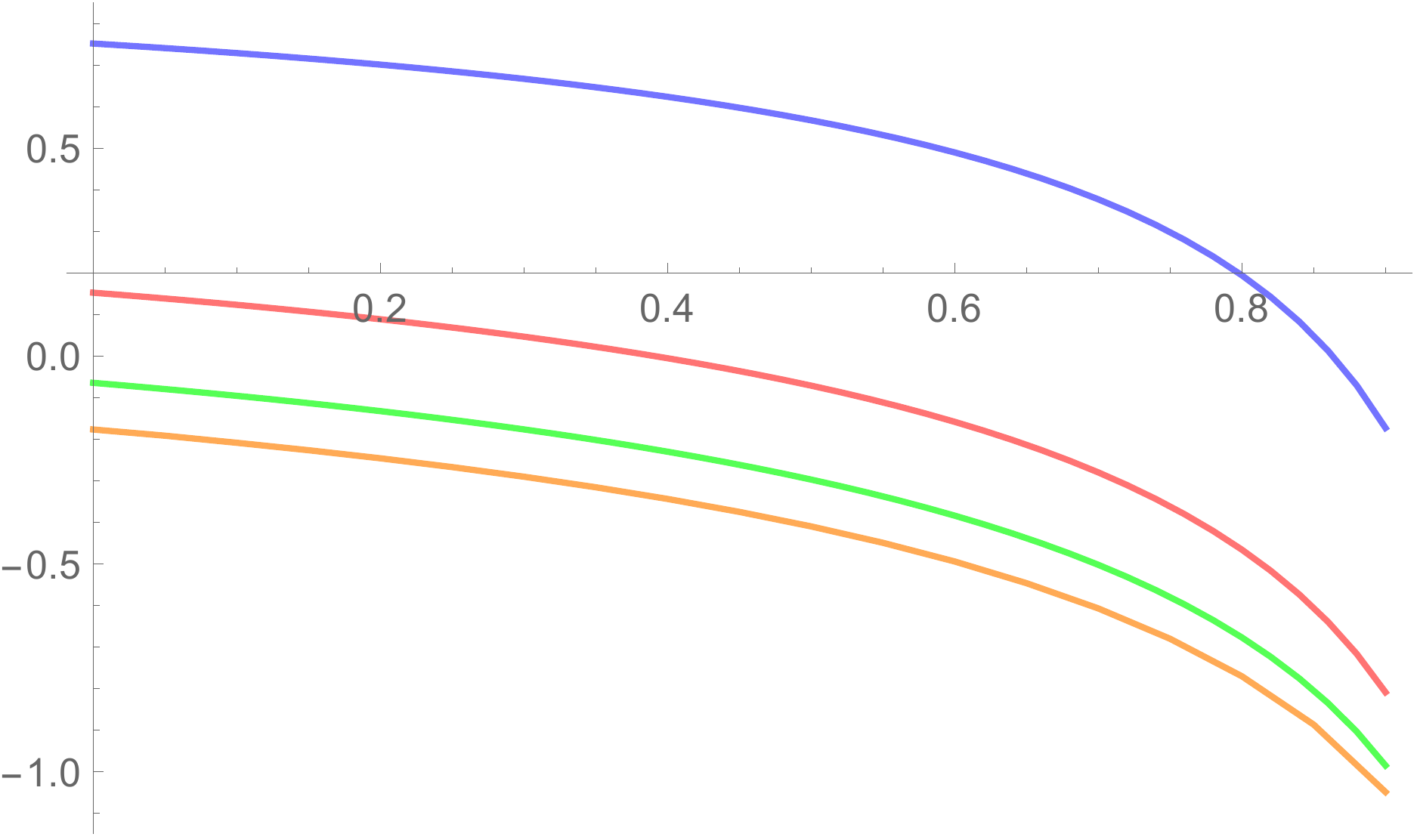}};
  \node[right=of img, node distance=0cm,font=\small,xshift=-1.15cm,yshift=0.625cm] {$\sdc$};
  \node[left=of img, node distance=0cm, anchor=center,font=\footnotesize,xshift=0.5cm] {$\Dfn{d-1}{2,0}(r_+)$};
  \node[right=of img, node distance=0cm,font=\scriptsize,xshift=-1.25cm,yshift=0.1cm] {$d=3$};
  \node[right=of img, node distance=0cm,font=\scriptsize,xshift=-1.25cm,yshift=-0.75cm] {$d=4$};
  \node[right=of img, node distance=0cm,font=\scriptsize,xshift=-1.25cm,yshift=-1.25cm] {$d=5$};
   \node[right=of img, node distance=0cm,font=\scriptsize,xshift=-1.25cm,yshift=-1.5cm] {$d=6$};
 \end{tikzpicture}
 \end{minipage}
}\\
\subfloat{
\begin{minipage}[t]{0.5\textwidth}
\vspace{0pt}
\hspace{0pt}
\begin{tikzpicture}
  \node (img)  {\includegraphics[scale=0.275]{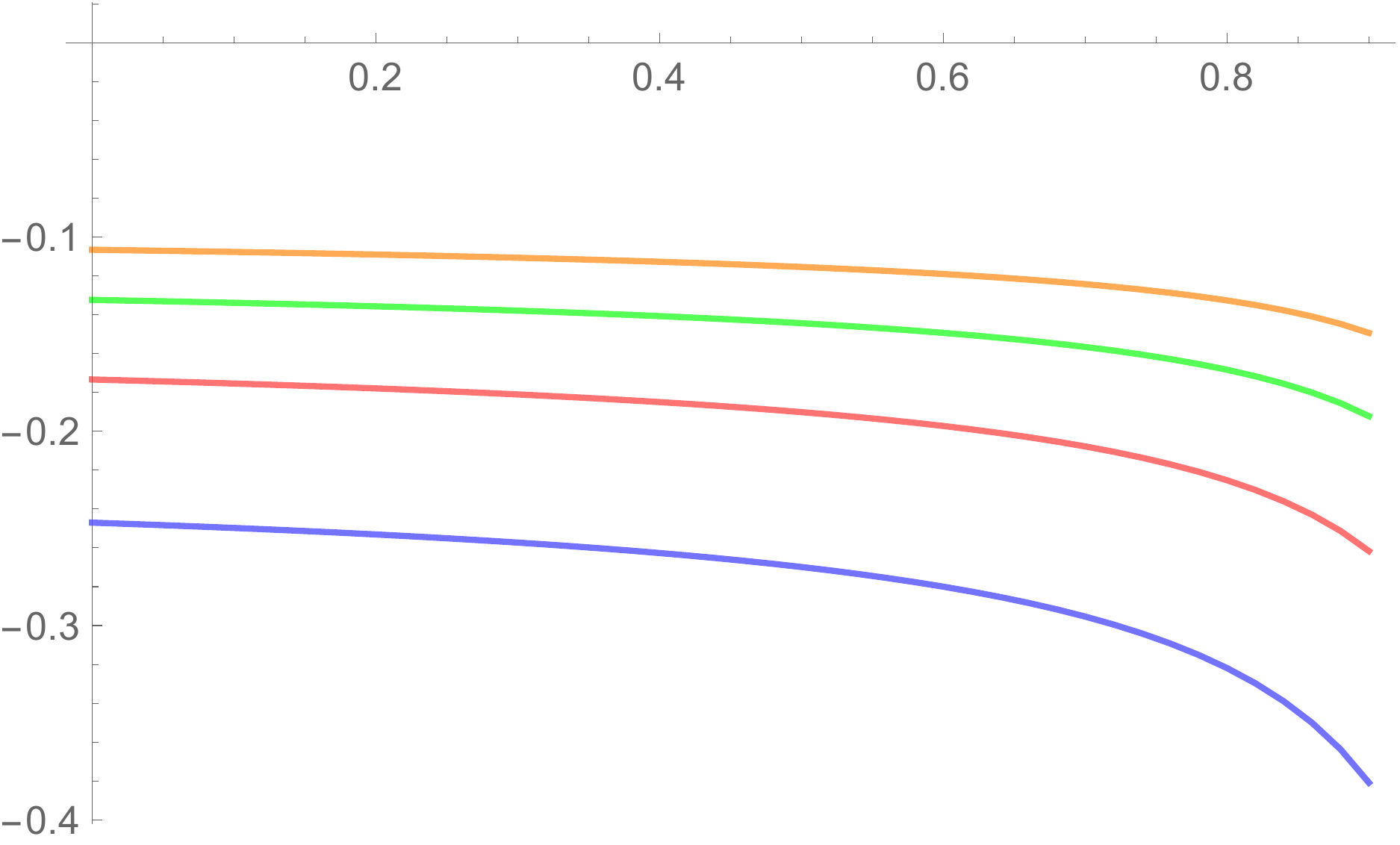}};
  \node[above=of img, node distance=0cm,font=\small,yshift=-1.25cm] {$\sdc$};
  \node[left=of img, node distance=0cm, anchor=center,font=\footnotesize,xshift=0.5cm] {$\Mser{d-1}{0,2}(r_+)$};
  \node[right=of img, node distance=0cm,font=\scriptsize,xshift=-1.25cm,yshift=0.45cm] {$d=6$};
  \node[right=of img, node distance=0cm,font=\scriptsize,xshift=-1.25cm,yshift=-0.0cm] {$d=5$};
  \node[right=of img, node distance=0cm,font=\scriptsize,xshift=-1.25cm,yshift=-0.5cm] {$d=4$};
   \node[right=of img, node distance=0cm,font=\scriptsize,xshift=-1.25cm,yshift=-1.5cm] {$d=3$};
 \end{tikzpicture}
 \end{minipage}
}
\subfloat{ 
 \begin{minipage}[t]{0.5\textwidth}
 \vspace{0pt}
 \hspace{0pt}
\begin{tikzpicture}
  \node (img)  {\includegraphics[scale=0.275]{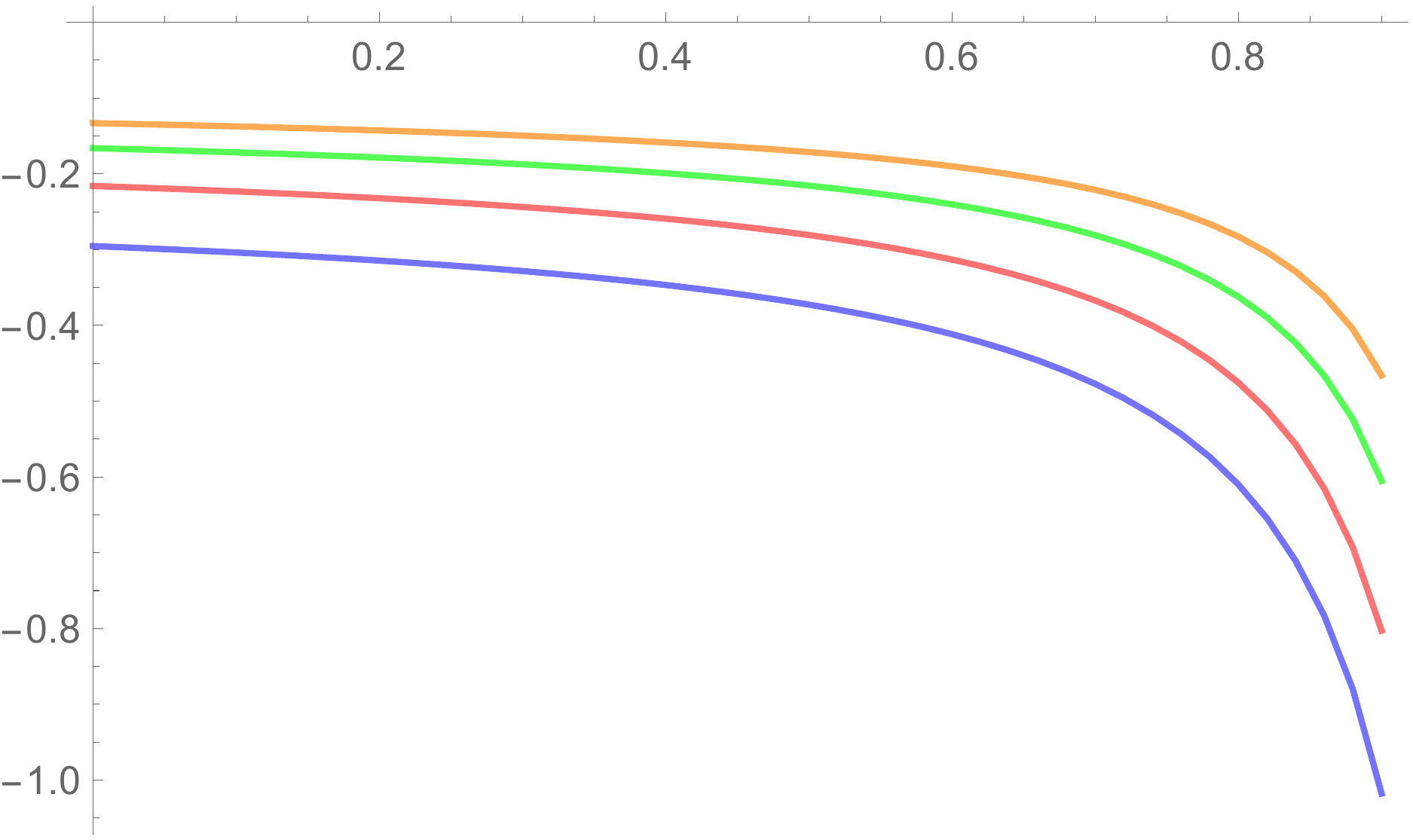}};
  \node[above=of img, node distance=0cm, font=\small,yshift=-1.25cm] {$\sdc$};
  \node[left=of img, node distance=0cm, anchor=center,font=\footnotesize,xshift=0.5cm] {$\Mser{d-1}{0,2}(r_+)$};
  \node[right=of img, node distance=0cm,font=\scriptsize,xshift=-1.25cm,yshift=0.25cm] {$d=6$};
  \node[right=of img, node distance=0cm,font=\scriptsize,xshift=-1.25cm,yshift=-0.25cm] {$d=5$};
  \node[right=of img, node distance=0cm,font=\scriptsize,xshift=-1.25cm,yshift=-0.75cm] {$d=4$};
   \node[right=of img, node distance=0cm,font=\scriptsize,xshift=-1.25cm,yshift=-1.5cm] {$d=3$};
 \end{tikzpicture}
 \end{minipage} 
}
\caption{ The charge dependence of the coefficients in Green's function of the Markovian vector polarizations $\MX_\ai$ in dimensions $d=3,\ldots,6$. Since $\MX_\ai$ is non-Markovian of index $\ann = 1-d$, the functions entering it are obtained by analytically continuing those for a Markovian field of index $\ann=d-1$. The latter happily happens to coincide with the  tensor graviton polarizations, so the data above enters both $\Kin{\MX}$ and $\Kin{d-1}$. Our conventions are as described in \cref{fig:KinMarY}.}
\label{fig:KinMarT}
\end{figure}
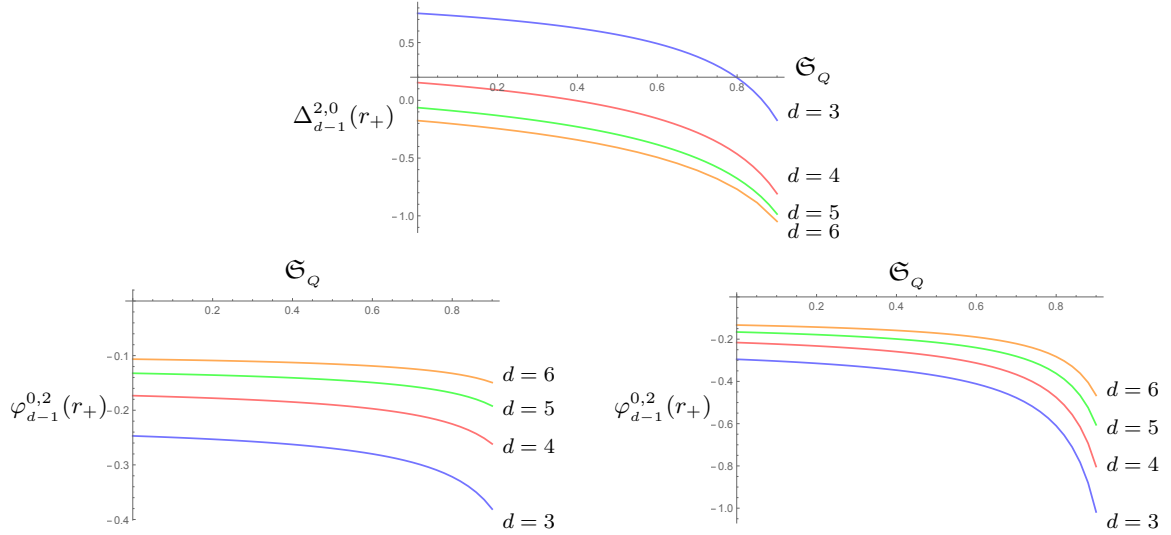

The inverse Green's function for $\MX_\ai$ is obtained up to some small changes from that for a non-Markovian probe field\footnote{ The changes are related to the presence of an additional potential term which kicks in at higher orders in spatial momenta.}
\begin{equation}\label{eq:KinX}
\begin{split}
\Kin{\MX}(\omega,\bk) 
&=
 r_+^{2-d}\bigg\{  -i\, \bwt 
  + \left[\frac{1}{d} - \frac{\sdc}{2(d-1)} \right] \bqt^2  - \Dfn{d-1}{2,0}(r_+) \, \bwt^2
    + i \left[\Dfn{d-1}{2,0}(r_+)^2  -2\Mser{d-1}{2,0}(r_+) \right]  \bwt^3 \\
&\qquad 
  + i \left[ \frac{2(d-2)}{d}\, \Mser{d-1}{0,2}(r_+) 
  + \left(\frac{\sdc}{d-1}-\frac{2}{d}\right) \Dfn{d-1}{2,0}(r_+)
  \right]  \bwt\,\bqt^2+ \cdots \bigg\}\,.
\end{split}
\end{equation}
The coefficients appearing in the above expression  are linear combinations of the  horizon values of functions plotted in \cref{fig:KinMarT}.  Until the cubic order in gradients we only encounter functions that already appear in the tensor sector. This is because $\MX_\ai$ satisfies a non-Markovian equation with index $\ann =1-d$ whose solution can be obtained from that for a Markovian mode with $\ann = d-1$ by analytically continuing $\ann \to - \ann$. At higher orders in gradients we encounter new functions resulting from the $\bRQ^2\, \BQT^2\, (1-h) $ term in \eqref{eq:XYfinal}. The expression for $\Kin{\MX}$ accurate to quartic order in gradients can be found in \eqref{eq:KinXA}. The functions entering in the dispersion function are collected in \eqref{eq:Xgradfns} and \eqref{eq:DXfns}, respectively, and plotted in \cref{fig:KinMarT4th}. 

From the Wilsonian influence functional one can read off  the expectation values of  dual operators in the presence of a boundary source. For the Markovian mode $\MY_\ai$ we obtain  the analog of \eqref{eq:MarkJO} 
\begin{equation}\label{eq:Yop1pt}
\begin{split}
\expval{(\YOp)_a} 
&= 
  - \left[\Kin{\MY}\, \YsQ_a + \left(\nB + \frac{1}{2}\right) \left[ \Kin{\MY} -\Krev{\MY}\right] \, \YsQ_d \right] , \\ 
\expval{(\YOp)_d} 
&= 
  - \Krev{\MY}\, \YsQ_d\,.
\end{split}
\end{equation}
Remembering that for the non-Markovian modes we parameterize the WIF as a functional of the hydrodynamic moduli, or equivalently the expectation value of the dual operator, \eqref{eq:nMVop}, we solve for the boundary source instead and thus end up with the analog of \eqref{eq:JOnMrel} 
\begin{equation}\label{eq:X1pt}
\begin{split}
 \XsJ_a &= 
  \Kin{\MX} \, \XP_a + \left(\nB+ \frac{1}{2}\right) \left[ \Kin{\MX} - \Krev{\MX}\right] \XP_d \,, \\
 \XsJ_d &= 
  \Krev{\MX}\,\XP_d \,.
\end{split}
\end{equation}

From these expressions we can read of the Green's functions for the operators $\XOp^\ai$ and 
$\YOp^\ai$, dual to the bulk fields $\MX_\ai$ and $\MY_\ai$, respectively. Using the results  \eqref{eq:M2ptfns} and \eqref{eq:nM2ptfns} reviewed in \cref{sec:wifc}  we can write down the retarded Green's function for these operators as 
\begin{equation}\label{eq:XYops2ptBare}
\begin{split}
\expval{\XOp^\ai(-\omega,-\bk) \, \XOp^{\ai'}(\omega,\bk)}^\text{Ret} 
&\propto 
  - i 
  \,  \frac{1}{\Kin{\MX}(\omega,\bk)} \, \delta_{\ai\,\ai'} \,, \\
\expval{\YOp^\ai(-\omega,-\bk) \, \YOp^{\ai'}(\omega,\bk)}^\text{Ret} 
& \propto 
  i\,  \Kin{\MY}(\omega,\bk)\,  \delta_{\ai\,\ai'} \,.
\end{split}
\end{equation}
The Keldysh Green's function  can be determined by the fluctuation-dissipation relation as in \eqref{eq:M2ptfns} and \eqref{eq:nM2ptfns}.

\subsubsection{The Einstein-Maxwell  boundary Wilsonian influence functional}
\label{sec:EMwif}

Starting with the Einstein-Maxwell action the dynamics of tensor and vector polarizations of gravitons and photons can be computed directly in terms of the fields $\tGR_\bi$, $\MX_\ai$ and $\MY_\ai$. We explain some of the elements behind this analysis in \cref{sec:bulkaction}. Using the solution on the grSK geometry, one can obtain the  boundary Wilsonian influence functional as a functional of the boundary sources $\JMarP^\bi_{a,c}$ and $\YsQ^\ai_{a,d}$ and the hydrodynamic field $\XsJ^\ai_{a,d}$. 

One finds after a bit of asymptotic analysis the result for the boundary WIF parameterized in terms of these variables takes the standard Schwinger-Keldysh form obtained in \cite{Ghosh:2020lel} at the Gaussian order: 
\begin{equation}\label{eq:WIFEin}
\begin{split}
 \frac{1}{\ceff} \, & \mathcal{S}_\text{WIF}[\JMarP^\bi_{a, d}, \YsQ_{a,d}^\ai, \XP_{a,d}^\ai] \\
&= 
  \mathcal{S}_\text{ideal} -    \int_k  \, \sum_{\bi=1}^{N_T} \, (\JMarP_d^\bi)^\dagger \,\Kin{d-1}
   \left[\JMarP_{a}^\bi+\left(\nB+\frac{1}{2}\right)\: \JMarP_d^\bi     \right]  \\
& \qquad  
 -\, \int_k \; k^2 \,  \sum_{\ai=1}^{N_V}  \left\{ 
  	\NX(\BQT) \, (\XP^\ai_d)^\dag \,
    \Kin{\MX} \left[\XP^\ai_{a}+\left(\nB+\frac{1}{2}\right)\: \XP^\ai_d \right]  \right.\\
&\left. \qquad \qquad   
	+ \;  \NY(\BQT) \, (\YsQ_d^\ai)^\dag \,
	\Kint{\MY}  \left[\YsQ_a^\ai+\left(\nB+\frac{1}{2}\right)\: \YsQ_d^\ai \right] 
    \right\} .
\end{split}
\end{equation}  
The renormalized retarded Green's function for the field $\MY_\ai$ appears with an additional contact term. Hence, we define a modified Green's function $\Kint{\MY}$ for the Markovian operator $\YOp^\ai$ absorbing the contact term and introduce
\begin{equation}\label{eq:KintY}
\Kint{\MY}(\omega,\bk) = (d-2)\, \RQ^{d-2}+ \Kin{\MY}(\omega,\bk)  \,.
\end{equation}  

The normalization factors appearing in the WIF for $\MX$ and $\MY$ are  relatively simple when written in terms of our deformed momentum parameter $\BQT$ defined in \eqref{eq:BQTdef}:
\begin{equation}\label{eq:knormXY}
\begin{split}
\NX(\BQT) &=  (\BQT^2+1)\, (\BQT^2+2)\,,\\ 
\NY(\BQT) &=   \frac{1}{\RQ^{2(d-2)}} \,\BQT^2\,(\BQT^2+1)\,.
\end{split}
\end{equation}  
The retarded correlation functions of $\XOp^\ai$ and  $\YOp^\ai$ in \eqref{eq:XYops2ptBare} are then normalized in the Einstein-Maxwell theory to be 
\begin{equation}\label{eq:XYops2pt}
\begin{split}
\expval{\XOp^\ai(-\omega,-\bk) \, \XOp^{\ai'}(\omega,\bk)}^\text{Ret} 
&= 
  - i 
  \,  \frac{1}{c_\text{eff}\, k^2\, \NX(\BQT)}\, \frac{1}{\Kin{\MX}(\omega,\bk)} \, \delta_{\ai\,\ai'} \,, \\
\expval{\YOp^\ai(-\omega,-\bk) \, \YOp^{\ai'}(\omega,\bk)}^\text{Ret} 
& = 
  i\, \frac{1}{c_\text{eff}\, k^2\, \NY(\BQT)}\,  \Kin{\MY}(\omega,\bk)\,  \delta_{\ai\,\ai'} \,.
\end{split}
\end{equation}

These factors account for the fact that in the Debye-like gauge where we parameterize the vector polarizations in terms of the fields $\MX_\ai$ and $\MY_\ai$ there are various momentum dependent factors (some of which arise in the process of decoupling the modes). We will account for these factors when we compute the current correlators, but note here that when we refer to the $\MX_\ai \,\MY_\ai$ system we will use the simpler probe action \eqref{eq:SXYprobe} and account for the normalization factors separately. We note that the relative dimensional factor between $\NX$ and $\NY$ involving powers of $\RQ$ is necesary to account for the different scaling of non-Markovian and Markovian modes ($\Kin{\MX} \sim r_+^{2-d}$ and $\Kin{\MY} \sim r_+^{d-2}$, respectively).

We now describe the each of these contributions in turn, explaining how they are obtained, and then turn to repackaging the information more directly in terms of the CFT currents. Details of the derivation of \eqref{eq:WIFEin} can be found in \cref{sec:bdyobs}.

\paragraph{1. The Class L fluid contribution:}
The first contribution,  $\mathcal{S}_\text{ideal}$, is the background thermal contribution to the Wilsonian influence functional. This arises because the background planar \RNAdS{d+1} geometry has a non-vanishing free energy. This leads to a local expression in terms of the induced boundary metric, which is best written in terms of a vector $\mathbf{b}^\mu$ and the ratio $\mu/r_+$.\footnote{These quanities are related to the thermal vector $\bm{\beta}^\mu$ and thermal twist $\Lambda_{\bm{\beta}} $ used to write hydrodynamic data in \cite{Haehl:2015pja}. We will however stick to the simpler parameters above as the functional form of the pressure in terms of the thermal vector and twist for a charged fluid is somewhat complex (owing to \eqref{eq:TRN}).} In the equilibrium background geometry these are 
\begin{equation}\label{eq:thermalvT}
\mathbf{b}^\mu\, \partial_\mu = \frac{1}{r_+}\, \partial_v \,, \qquad 
\Lambda_{\mathbf{b}} = \frac{\mu}{r_+} = \sqrt{\frac{d-1}{d-2}}\, Q \,.
\end{equation}  
The thermal free energy of the \RNAdS{d+1} black hole is $r_+^d (1+Q^2)$ and thus accounting for  contribution to the action from the free energy on both the left and right boundaries of the grSK geometry one finds:
\begin{equation}\label{eq:Sideal}
\begin{split}
\mathcal{S}_\text{ideal} 
&=    
   \, \int\, d^d x\, \sqrt{-\gamma_\skR}\, 
    \left[\sqrt{-(\gamma_\skR)_{\mu\nu} \vb{b}_\skR^\mu \, \vb{b}_\skR^\nu}\ \right]^{-d}\, \left(1+\frac{d-2}{d-1}\, \Lambda_{\mathbf{b},\skR}^2\right) \\
&\qquad  
-   \int\, d^d x\, \sqrt{-\gamma_\skL}\, \left[\sqrt{-(\gamma_\skL)_{\mu\nu} \vb{b}_\skL^\mu \, \vb{b}_\skL^\nu}\ \right]^{-d} \,\left(1+\frac{d-2}{d-1}\, \Lambda_{\mathbf{b},\skL}^2\right) .
\end{split}
\end{equation}  

\paragraph{2. The Markovian tensor modes:} Next is the contribution from the tensor polarizations  coupling to the source $\JMarP^\bi$. These are straightforward since the transverse tensor graviton fluctuations are captured by a Markovian field $\tGR_\bi$ with index $\ann = d-1$. The only data we need then is the boundary Green's function $\Kin{d-1}(\omega,\bk)$ which can be obtained from the general probe Markovian field analysis reported in \cref{sec:probeM}.

For the tensor polarization we specialize  \eqref{eq:KinMark} to the desired Markovianity index  $\ann =d-1$ and obtain
\begin{equation}\label{eq:KinMarkT}
\Kin{d-1}(\omega, \bk) 
=
 r_+^d \left[-i\,\bwt - \frac{\bqt^2}{d-2} + \Dfn{d-1}{2,0}(r_+)\, \bwt^2 + 2i\,\Mser{d-1}{0,2}(r_+)\, \bwt\,\bqt^2 -  2i\,\Mser{d-1}{2,0}(r_+)\, \bwt^3    + \cdots\right] .
\end{equation}  
The coefficients in the gradient expansion are horizon values of the functions defined in \eqref{eq:DeltaFns} and  tabulated in \cref{tab:Mgradsol}. These can be evaluated in terms of  digamma functions in a small charge expansion but are not amenable to closed form evaluation in general. We can nevertheless evaluate them numerically and the results are displayed in \cref{fig:KinMarT}. As noted above the same functions enter $\Kin{\MX}$.

\paragraph{3. The vector modes:} The interesting part of the story is in the final two terms, which are the contribution from the vector polarizations of the gravitons and photons. In \cref{sec:vectorsAct} we demonstrate that the Einstein-Maxwell action together with its boundary term can be simplified after a series of steps to the decoupled action given in \eqref{eq:SXYdecoupleA} and subsequent equations. If we evaluate the on-shell action (described in \cref{sec:osaction}) on the bulk solution, one ends up with a remarkably simple expression considering the complicated intermediate steps, which is the form presented in \eqref{eq:WIFEin}. We have already described the dynamical content of this sector in \cref{sec:XYwif} and can continue to use the results from there modulo keeping track of the normalization factors $\NX(\BQT)$ and $\NY(\BQT)$ (and the central charge). 

This completes the summary of the contributions to the WIF from the bulk analysis. While the Green's functions for the Markovian modes and dispersion functions for the non-Markovian modes are complicated, much of this owes to the nature of the gradient expansion. Per se, the coefficients at any given order of $\bwt^m\, \bqt^n$ are some functions of $\sdc$ once we scale out the overall dimensions in terms of $r_+$ as we have done.

\subsection{The boundary currents}
\label{sec:bdycurrents}

We have expressed the boundary observables in terms of the asymptotic sources and field configurations for the Markovian and non-Markovian modes, respectively. Our final task is to convert this data into the conserved currents $\Tcft_{\mu\nu} $ and $\Jcft_\mu$ which are obtained from the asymptotic behaviour of the solution by the standard AdS/CFT dictionary applied on the grSK geometry (see \cite{Ghosh:2020lel}). We find it convenient to remove the background ideal fluid contribution and therefore write:
\begin{equation}\label{eq:TJdecomp}
\begin{split}
\Jcft_\mu 
&= 
   J^\text{Ideal}_\mu + c_\text{eff} \, \widehat{J}_{\mu} \,, \qquad  \Tcft_{\mu\nu} 
 =  T^\text{Ideal}_{\mu\nu} + c_\text{eff} \, \widehat{T}_{\mu\nu} \,.
\end{split}
\end{equation}
It will be simplest to write the currents in the Fourier domain, so we will refrain from writing integrals over the momenta. 

For the stress tensor we will split the contribution into transverse traceless tensor and vector polarizations, respectively, as these are decoupled sectors. One finds the tensor part taking the form determined in \cite{Ghosh:2020lel}  
\begin{equation}\label{eq:TcftTT}
\begin{split}
\expval{\widehat{T}_{ij,\skR} } 
 &=
 	-\sum_{\bi=1}^{N_T}  \, \TT^\bi_{ij} \left( K_{_{d-1}}^\In \left[(\nB+1)\: \JMarP_\skR^\bi - \nB \: \JMarP_\skL^\bi \right] - \nB\, K_{_{d-1}}^\Rev \left[\JMarP_\skR^\bi-\JMarP_\skL^\bi\right]\right) , \\
\expval{\widehat{T}_{ij,\skL} } &=
  -\sum_{\bi=1}^{N_T}  \, \TT^\bi_{ij}\left( K_{_{d-1}}^\In \left[(\nB+1)\: \JMarP_\skR^\bi - \nB \: \JMarP_\skL^\bi \right] - (\nB+1) K_{_{d-1}}^\Rev  \left[\JMarP_\skR^\bi-\JMarP_\skL^\bi \right]\right)\,.
\end{split}
\end{equation}
Since the tensor modes are captured by a Markovian probe field in the \RNAdS{d+1} geometry one can simply  use the general relation for the one-point  function \eqref{eq:MarkJO}, which is now written in the L/R basis.

The vector polarizations of the stress tensor and the charge current are admixtures of the non-Markovian and Markovian operators. Instead of reporting the components we find it convenient to package the data into the following two linear combinations of currents\footnote{ Since $k^j\, \TT_{ij} =0$ contraction with $k^j$ projects onto the vector polarizations of the energy-momentum tensor, see \eqref{eq:TcftPA}.}
\begin{equation}\label{eq:CTJdef}
\begin{split}
\XT_i 
&= 
   \frac{\BQT^2}{\sdc}\, \omega \, \mu\, \widehat{J}_i   +  k^j \, \widehat{T}_{ij} \,, \\ 
\YJ_i 
&= 
  - \frac{\BQT^2+2}{\sdc}\, \omega \, \mu\, \widehat{J}_i  +   k^j \, \widehat{T}_{ij}  \,.
\end{split}
\end{equation}
These combinations are engineered to isolate the Markovian and the non-Markovian degrees of freedom. As the notation suggests $\XT_i$ picks out the momentum diffusion piece while $\YJ_i$ is the charge contribution. One can identify them with the boundary Schwinger-Keldysh operators $\XOp$ and $\YOp$  as 
\begin{equation}\label{eq:CTJXYops}
\begin{split}
(\XT_i)_{\skL/\skR} 
&= 
 - 2\,\omega\,k^2\,  (1+\BQT^2) \, \sum_{\ai=1}^{N_V}\, \VV_i^\ai \,( \XOp^\ai)_{\skL/\skR} \,,\\
(d-2)\, \RQ^{d-2}\, (\YJ_i)_{\skL/\skR} 
&= 
    2\,\omega\, k^2\, (1+\BQT^2) \, \sum_{\ai=1}^{N_V}\, \VV_i^\ai \, \left[ (\YOp ^\ai)_{\skL/\skR} -
    (d-2)\, \RQ^{d-2}\, \YsQ^\ai_{\skL/\skR} \right] .   
\end{split}
\end{equation}  
The polarization index sum converts back to the usual vectorial components in our coordinate basis, and we see that modulo factors of frequency and momenta (which are also present in $\BQT$) the current combinations in \eqref{eq:CTJdef} are precisely the boundary operators associated with the decoupled vector modes. The $vi$ component of the stress tensor by virtue of the above is not independent at this order, but rather is determined in terms of the currents $\XT_i$ and $\YJ_i$. This is to be expected since we only have two physical modes in the problem, which between them ought to parameterize all the current components. A derivation of these results can be found in \cref{sec:TJ} along with the expressions for the individual current components \eqref{eq:currentPA} and \eqref{eq:TcftPA}. 

Armed with this expression it is straightforward to write down the Schwinger-Keldysh expressions for the boundary one-point functions of these current combinations using \eqref{eq:MarkJO}, which are in turn,  
\begin{equation}\label{eq:lcTJcft}
\begin{split}
\expval{\XT_{i,\skR}} 
&= 
  -2 \, \omega\, k^2  \left(1+\BQT^2\right)  \sum_{\ai=1}^{N_V} \,   \VV_i^\ai \,  \XP^\ai_\skR \,, \\
\expval{\XT_{i,\skL} }
&= 
  -2 \, \omega\, k^2  \left(1+\BQT^2\right)  \sum_{\ai=1}^{N_V} \,   \VV_i^\ai \,  \XP^\ai_\skL \,, \\  
\expval{\YJ_{i,\skR} }
&= 
  -  \frac{2 \, \omega\, k^2   \left(1+\BQT^2\right) }{(d-2)\, \RQ^{d-2}} \sum_{\ai=1}^{N_V} \,   \VV_i^\ai\,
  \left( \Kint{\MY} \left[(\nB+1)\,\YsQ_{\skR}^\ai -\nB\, \YsQ_{\skL}^\ai \right]
- \nB\, \Krevt{\MY}\left[\YsQ_{\skR}^\ai - \YsQ_{\skL}^\ai\right]   \right)\,, \\ 
\expval{\YJ_{i,\skL} }
&= 
  -   \frac{2 \, \omega\, k^2   \left(1+\BQT^2\right)}{(d-2)\, \RQ^{d-2}  } 
    \sum_{\ai=1}^{N_V} \,   \VV_i^\ai\,
  \left( \Kint{\MY} \left[(\nB+1)\,\YsQ_{\skR}^\ai -\nB\, \YsQ_{\skL}^\ai \right]
 -(\nB+1)\, \Krevt{\MY}\left[\YsQ_{\skR}^\ai - \YsQ_{\skL}^\ai\right] \right) .
\end{split}
\end{equation}  

In our analysis thus far have directly worked with the operator relations for the currents in terms of the auxiliary operators $\XOp^\ai$ and $\YOp^\ai$. It is useful to also record the relation between the sources for these operators and those for the gravitons and photons. From the asymptotics of the solutions for the equations \eqref{eq:XYfinal} one can show that for the vector polarizations 
\begin{equation}\label{eq:sources}
\begin{split}
\left( \vMax^\ai \right)_{\infty, \skL/\skR} 
&= \lim_{r\to \infty \pm i  0} \, \vMax^\ai
= -\frac{\mu}{2\,\sdc}\, \BQT^2\, (\BQT^2+2)\, \left[ \XsJ^\ai_{\skL/\skR} + \frac{d-2}{\RQ^{d-2}}\, \YsQ^\ai_{\skL/\skR}\right] ,  \\
\left( \vGR_v^\ai \right)_{\infty, \skL/\skR} 
&= 
\lim_{r\to \infty \pm i  0} \, \vGR_v^\ai   = 
	-(\BQT^2 + 2)\, \XsJ^\ai_{\skL/\skR}  + \BQT^2\, \frac{d-2}{\RQ^{d-2}}\, \YsQ^\ai_{\skL/\skR}  . 
\end{split}
\end{equation}	
The sources for the spatial components of the metric are related to the spatio-temporal piece. These are the physical sources for the transverse vector polarizations of the  charge current and boundary energy-momentum tensor, respectively.  In this parameterization it is clear that in the $\mu \to 0$ limit the momentum diffusion mode decouples from the charge current mode.

\subsection{Current correlators, dispersion relations, and transport}
\label{sec:correlators}

The physical data in the WIF can be used to extract the correlation function of the conserved currents. To write these expressions in a compact form, we will pick the momentum vector to point in a particular spatial direction, say $\bk = k \, \hat{e}_z$, and consider the modes that are polarized in the $xy$-plane.

\paragraph{Current correlators:} First of all, it is straightforward to write down the  2-point correlation functions of the combinations of currents introduced in \eqref{eq:CTJXYops}. They can be  determined from the correlation functions of the operators $\XOp^\ai$ and $\YOp^\ai$ introduced earlier in \eqref{eq:XYops2pt}.  We have 
\begin{equation}\label{eq:TJops2pt}
\begin{split}
\expval{\XT_x(-\omega,-\bk) \, \XT_x(\omega,\bk)}^\text{Ret} 
&= 
	\left(2\,\omega \,k^2\,   (1+\BQT^2)\right)^2\, \expval{\XOp(-\omega,-\bk) \, \XOp(\omega,\bk)} \\
&=
	  -i\, \frac{4\,\omega^2\,k^2\,   (1+\BQT^2)^2}{c_\text{eff}\, \NX(\BQT)} \, 
 \frac{1}{\Kin{\MX} (\omega,\bk)} \,,  \\
\expval{\YJ_x(-\omega,-\bk) \, \YJ_x(\omega,\bk)}^\text{Ret} 
&=
	\left(2\, \omega\,k^2\,   (1+\BQT^2)\right)^2\, \expval{\YOp(-\omega,-\bk) \, \YOp(\omega,\bk)} \\
&= 
   i\,  \frac{4\,\omega^2\, k^2\, (1+\BQT^2)^2}{c_\text{eff}\, \NY(\BQT)} \,  \frac{\Kint{\MY}(\omega,\bk)}{(d-2)^2\, \RQ^{2(d-2)}} \,.
\end{split}
\end{equation}
Note that the physical currents have two-point functions that scale as $c_\text{eff}$ -- we defined $\XT_i$ and $\YJ_i$ by stripping of this factor which results in an answer that scales as $c_\text{eff}^{-1}$.

From the transverse tensor polarization of the gravitons which we take to be polarized in the $xy$-plane we can extract the correlator of shear-strain component of the stress tensor. Its retarded Green's function is given by $\Kin{d-1}(\omega, \bk)$, \eqref{eq:KinMarkT},  along with a contribution from $\mathcal{S}_\text{ideal}$, viz., 
\begin{equation}\label{eq:TxyT2ptRet}
\begin{split}
\expval{\Tcft_{xy}(-\omega,-\bk)\, \Tcft_{xy}(\omega,\bk)}^{\text{Ret}} 
&=  
 i\, \ceff\,  r_+^d\bigg[1+Q^2 
   -i\,\bwt - \frac{\bqt^2}{d-2} + \Dfn{d-1}{2,0}(r_+)\, \bwt^2  \\
&
\qquad\qquad \;
   + \;2i\,\Mser{d-1}{0,2}(r_+)\, \bwt\,\bqt^2 -  2i\,\Mser{d-1}{2,0}(r_+)\, \bwt^3    + \cdots \bigg]\,.
\end{split}
\end{equation}  
The constant piece arises from the pressure term, in fact from the ideal contribution to the WIF. One can verify that this reduces to the expression derived in the neutral black hole geometry in \cite{Ghosh:2020lel}. As explained there, one also finds from \eqref{eq:WIFEin} the Keldysh correlator satisfying the KMS relation 
\begin{equation}\label{eq:TxyT2ptKel}
\expval{\Tcft_{xy}(-\omega,-\bk)\, \Tcft_{xy}(\omega,\bk)}^{\text{Kel}} 
= \frac{1}{2} \coth\left(\frac{\beta\omega}{2}\right) \Re\left[\expval{\Tcft_{xy}(-\omega,-\bk)\, \Tcft_{xy}(\omega,\bk)}^{\text{Ret}} \right] ,
\end{equation}  
order by order in the gradient expansion. Furthermore, this expression reproduces the familiar expression 
$\eta/s = \frac{1}{4\pi}$ using the Kubo formula for shear viscosity.

The transverse vector polarization of gravitons and photons, with momentum  $\bk = k \, \hat{e}_z$, are captured by  the momentum density $\Tcft_{vx}$, momentum current $\Tcft_{zx}$, and  current density 
$\Jcft_x$. The current operators are expressed in terms of the boundary operators $\XOp^\ai$, $\YOp^\ai$ (and their sources) in \eqref{eq:currentPA} and \eqref{eq:TcftPA}. We can therefore write down the correlation functions in terms of those for the auxiliary operators $\XOp^\ai$ and $\YOp^\ai$ obtained in 
\eqref{eq:XYops2pt}. We will separate out the ideal contributions from the background sources and  indicate the physical transport (the non-ideal part) with subscript `non-ideal'.  The contribution to the ideal part comes from both the ideal term in the action $\mathcal{S}_\text{ideal}$ which gives a constant, momentum and frequency independent, contribution dependent on the background charge density and pressure.

We begin with the non-ideal part of the correlator. For the  current-current correlation function we find the holographic computation gives:
\begin{equation}\label{eq:Jx2ptRet}
\begin{split}
\expval{\Jcft_{x}(-\omega,-\bk)\, \Jcft_{x}(\omega,\bk)}_{\text{non-ideal}}^\text{Ret} 
= i\, c_\text{eff}\,\frac{\sdc^2}{\mu^2}  \frac{k^2}{\BQT^2+1}\left[-\frac{1}{\BQT^2+2}\, \frac{1}{\Kin{\MX}} 
  + \frac{1}{(d-2)^2\, \BQT^2} \, \Kint{\MY}\right] .
\end{split}
\end{equation}
Similarly, the correlation function for the current with the energy-momentum tensor can be simplified to a universal function up to an overall factor depending on the component of the latter, viz., 
\begin{equation}\label{eq:JxT2ptRet}
\begin{split}
\expval{\Jcft_{x}(-\omega,-\bk)\, \Tcft_{vx}(\omega,\bk)}^\text{Ret}_{\text{non-ideal}} 
&=  
  i\,c_\text{eff}\, \frac{\sdc}{\mu}  \, k^2 \, \mathcal{K}_1(\omega,\bk) \,, \\ 
\expval{\Jcft_{x}(-\omega,-\bk)\, \Tcft_{zx}(\omega,\bk)}^\text{Ret}_{\text{non-ideal}} 
&= 
   i\,c_\text{eff}\, \frac{\sdc}{\mu}  \, \omega\, k\,\mathcal{K}_1(\omega,\bk)  \,, 
\end{split}
\end{equation}
with 
\begin{equation}\label{eq:JTspectral}
 \mathcal{K}_1(\omega,\bk) 
 = 
 \frac{1}{\BQT^2+1}\, \left[ \frac{1}{\Kin{\MX}(\omega,\bk)} 
  + \frac{1}{(d-2)^2}\,  \Kint{\MY}(\omega,\bk) \right] . 
\end{equation}  
The two correlators in \eqref{eq:JxT2ptRet} are related by a Ward identity.  We have excluded in the above a potential constant contact term that we believe cancels against a contribution from the ideal part.

Finally, the correlation function of the energy-momentum tensors themselves take the form
\begin{equation}\label{eq:TT2ptRet}
\begin{split}
\expval{\Tcft_{vx}(-\omega,-\bk)\, \Tcft_{vx}(\omega,\bk)}^\text{Ret}_{\text{non-ideal}} 
&=
  i\, c_\text{eff}\, k^2\, \mathcal{K}_2(\omega,\bk) \,,\\ 
\expval{\Tcft_{vx}(-\omega,-\bk)\, \Tcft_{zx}(\omega,\bk)}^\text{Ret}_{\text{non-ideal}} 
&=
   - i\, c_\text{eff}\, \omega \,k\, \mathcal{K}_2(\omega,\bk) \,,\\ 
 \expval{\Tcft_{zx}(-\omega,-\bk)\, \Tcft_{zx}(\omega,\bk)}^\text{Ret}_{\text{non-ideal}} 
&=
   i\, c_\text{eff}\, \omega^2 \, \mathcal{K}_2(\omega,\bk) \,,
\end{split}
\end{equation}
with
\begin{equation}\label{eq:TTspectral}
 \mathcal{K}_2(\omega,\bk) 
 = 
   \frac{1}{\BQT^2+1}\, \left[- \frac{\BQT^2+2}{\Kin{\MX}(\omega,\bk)} 
  +  \frac{\BQT^2}{(d-2)^2}\, \Kint{\MY} (\omega,\bk) \right].
\end{equation}  
As above, the three correlators in \eqref{eq:TT2ptRet} are related by Ward identities. 

The two-point functions given above can also be obtained by varying the one-point functions \eqref{eq:JcftSKad}, \eqref{eq:TijCFTSKad} and \eqref{eq:TviCFTSKad}  with respect to the sources
$\left( \vMax^\ai \right)_{\infty, \skL/\skR} $ and $\left( \vGR_v^\ai \right)_{\infty, \skL/\skR} $ for the charge current and energy-momentum tensor, respectively,  given in \eqref{eq:sources}. For example, 
\begin{equation}\label{eq:TJvariations}
\begin{split}
\expval{\Jcft_x (-\omega,-\bk)\, \Jcft_x(\omega,\bk)} _\text{non-ideal} 
&= 
\frac{1}{i} \, \fdv{ \left(\vMax_x \right)_{\infty}} \, \expval{\widehat{J}_x(\omega,\bk)} \,, \\
\expval{\Tcft_{vx}(-\omega,-\bk)\, \Jcft_x (\omega,\bk)} _\text{non-ideal}
&= 
\frac{1}{i} \, \fdv{ \left(\vGR_{vx} \right)_{\infty}} \, \expval{\widehat{J}_x(\omega,\bk)} \,, \\
\expval{ \Jcft_x (-\omega,-\bk) \,\Tcft_{vx}(\omega,\bk)\,} _\text{non-ideal}
&=
	\frac{1}{i} \, \fdv{ \left(\vMax_{x} \right)_{\infty}} \, \expval{\widehat{T}_{vx}(\omega,\bk)} 
\,.
\end{split}
\end{equation}
The variations above are to be taken in a suitable direction in the space of sources; while varying with respect to the source for the current one should hold the source for the energy-momentum tensor  fixed.  We also record here that the second and third equations in \eqref{eq:TJvariations} give the same result up to contact terms (and complex conjugation).

\paragraph{Momentum diffusion:}  One of the key physical features of the low lying quasinormal modes in the vector sector of Einstein-Maxwell dynamics is momentum diffusion. For a neutral black hole the momentum diffusion mode is purely a gravitational perturbation, but in the charged black hole there is an admixture of the charge transport involved. Nevertheless, as we have argued extensively, it is possible to decouple this mode, and operationally, the physical operator that isolates the momentum diffusion is the combination $\XT_i$ defined in \eqref{eq:CTJdef}. The location of the diffusion pole is given by the vanishing locus of the function $\Kin{\MX}(\omega,\bk)$, which to quadratic order is given by 
\begin{equation}\label{eq:zerolocus}
\begin{split}
0&= 
  -i\, \bwt 
  + \left[\frac{1}{d} - \frac{\sdc}{2(d-1)} \right] \bqt^2  - \Dfn{d-1}{2,0}(r_+) \, \bwt^2 + \cdots \,,
\end{split}
\end{equation}  
where we have dropped the cubic order terms in \eqref{eq:KinX} for simplicity.

Solving for the location of the  diffusion pole we find the dispersion relation to quartic order\footnote{Note that to obtain the correct coefficient at $\order{k^4}$ we also need to include the contribution to $\Kin{\MX}$ from the quartic order solution in momenta, which can be found in \eqref{eq:KinXA}. }
\begin{equation}\label{eq:Xlocus}
\omega = - i\, \mathcal{D}(r_+,\RQ)\, k^2 + i\, \mathcal{D}_4(r_+,\RQ)\, k^4 + \cdots\,,
\end{equation}  
with 
\begin{equation}\label{eq:shearpole4}
\mathcal{D}(r_+,\RQ) = \frac{1}{r_+}\, \left[ \frac{1}{d} - \frac{\sdc}{2\, (d-1)}\right] .
\end{equation}  
Using the relation between the diffusion constant and the shear viscosity $\mathcal{D}= \frac{\eta}{\epsilon+p}$ and noting that the thermodynamic equation of state implies $\epsilon+ p = c_\text{eff}\, d\, r_+^d (1+Q^2)$ we find using \eqref{eq:RQdef} the familiar relation \cite{Kovtun:2004de}
\begin{equation}\label{eq:etas}
\frac{\eta}{s} = 4\, G_N\, c_\text{eff}\left[1 - \frac{d}{2\, (d-1)}\, \sdc\right] (1+Q^2)  = \frac{1}{4\pi} \,.
\end{equation}  

 \begin{figure}[hbt!]
 \centering
 \begin{minipage}[t]{\textwidth}
\vspace{0.4cm}
\hspace{0pt}
\begin{tikzpicture}
  \node (img)  {\includegraphics[scale=0.5]{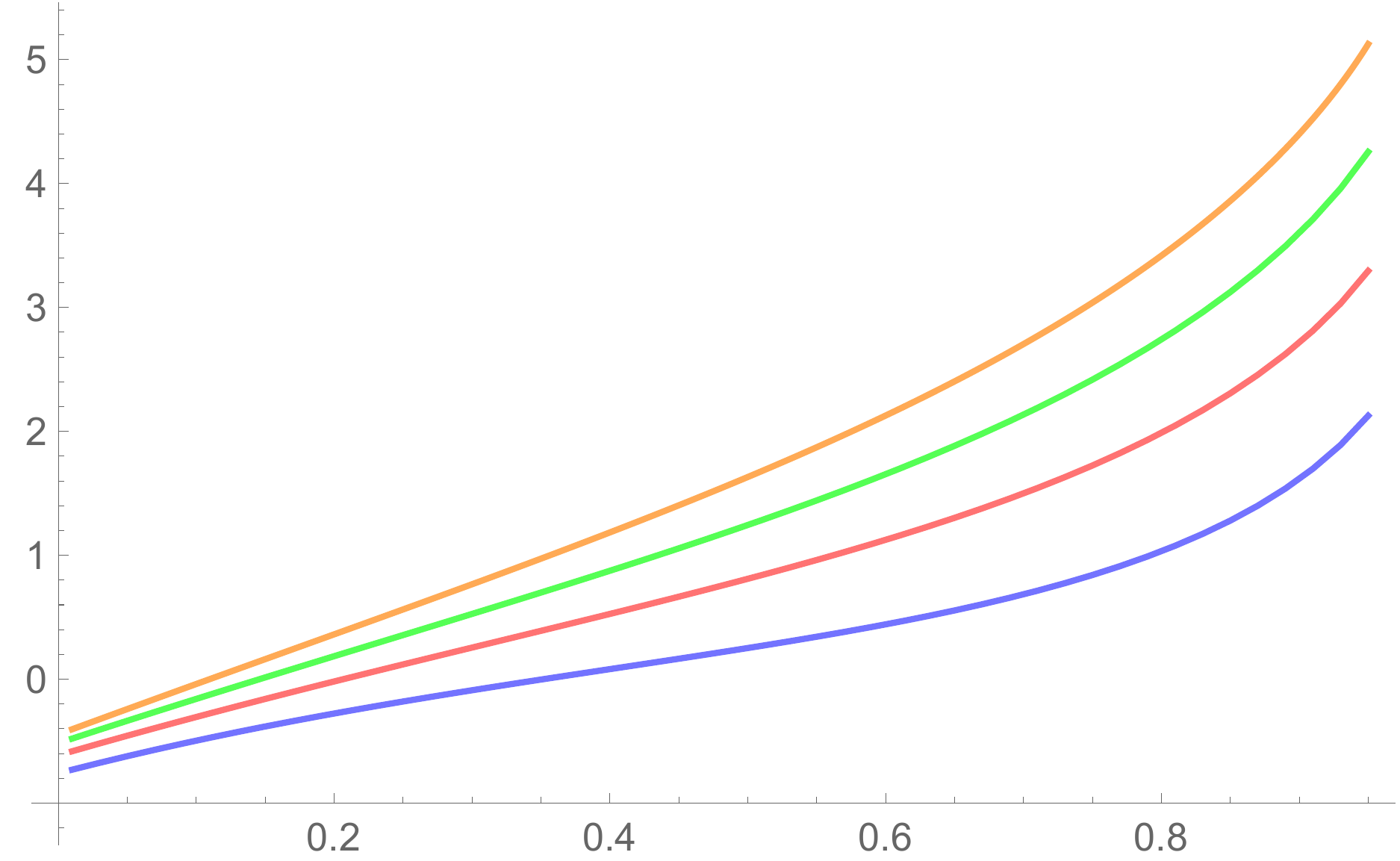}};
  \node[below=of img, node distance=0cm,font=\small,yshift=1.0cm] {$\sdc$};
  \node[left=of img, node distance=0cm, anchor=center,font=\footnotesize,xshift=0cm,yshift=1.2cm] {$(dr_{+})^3\mathcal{D}_4(r_+,\RQ)$};
  \node[right=of img, node distance=0cm,font=\normalsize,xshift=-1.25cm,yshift=2.6cm] {$d=6$};
  \node[right=of img, node distance=0cm,font=\normalsize,xshift=-1.25cm,yshift=1.85cm] {$d=5$};
  \node[right=of img, node distance=0cm,font=\normalsize,xshift=-1.25cm,yshift=1cm] {$d=4$};
   \node[right=of img, node distance=0cm,font=\normalsize,xshift=-1.25cm,yshift=0.1cm] {$d=3$};
 \end{tikzpicture}
 \end{minipage}
 \caption{The coefficient of the quartic term in the momentum diffusion dispersion relation given in (\ref{eq:D4}). We have scaled out dimensions using the horizon radius $r_+$ and have  plotted the coefficient $\mathcal{D}_4$ rescaled by a factor of $(dr_+)^3$. The additional dimension dependent factor makes the comparison to the neutral case in (\ref{eq:Dneutral}) straightforward since the is a relative factor of $d$ between the horizon size and the boundary temperature. }
 \label{fig:diffusion4}
 \end{figure}
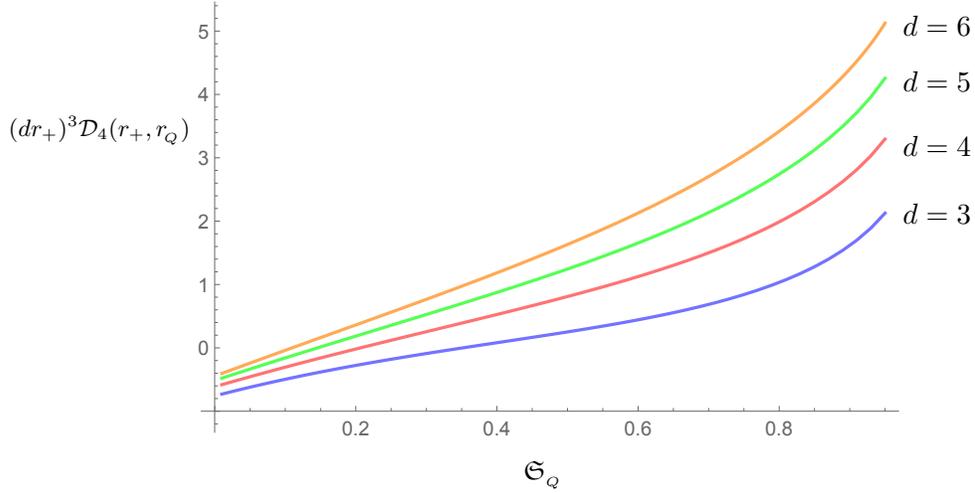

The quartic coefficient $\mathcal{D}_4(r_+,\RQ)$  is given by the horizon values of the functions introduced in \cref{sec:probenM} and is given by 
 \begin{equation}\label{eq:D4}
 \begin{split}
 (dr_{+})^3\,\mathcal{D}_4\left(r_{+},\RQ\right) 
 &= d^2 \, \Mser{d+1}{0,2}(r_+)-d\,(d-2)\Mser{d-1}{0,2}(r_+)\\
 & \quad -\frac{d^3\,\sdc}{4(d-1)}\left[\frac{1}{\bRQ^2}-\frac{2(d+2)(d-1)}{d^2(d-2)}+\frac{4}{d}\left(\Dfn{d-1}{2,0}(r_+)-(d-2)\Mser{d-1}{0,2}(r_+)\right)\right]\\
 &\qquad +d^3\, \sdc^2\left[\frac{\Dfn{d-1}{2,0}(r_+)}{4(d-1)^2}-\frac{2}{d(3d-2)(d-2))}\right] .
 \end{split}
 \end{equation}
While there is a  relatively simple expression in the case of the neutral \SAdS{d+1} black hole in terms of digamma functions,  see \eqref{eq:Dneutral},\footnote{ This computation for the neutral plasma fills a gap in the earlier analysis of \cite{Ghosh:2020lel} who only obtained the Green's functions to cubic order in gradients.}  there does not appear to be a simple analytic expression in the case of the charged plasma. It is nevertheless possible to numerically evaluate the coefficient; scaling out an overall dimensionful scale using $r_+$ we plot $\mathcal{D}_4(r_+, \RQ)$ suitably (non-dimensionalized) in  \cref{fig:diffusion4}.

\paragraph{Charge conductivity:}  The second physical phenomena we should analyze in the charged plasma is that of charge conductivity. While the charge diffusion mode is not present in the sectors we have analyzed (see \cref{sec:discuss}) one can recover the conductivity from the Kubo formula given the current-current correlator 
\begin{equation}\label{eq:kubosigma}
\begin{split}
\sigma_\text{dc}
&= 
  \lim_{\omega\to 0} \frac{1}{\omega}\, \Re \left[ \expval{\Jcft_{x}(-\omega,0)\, \Jcft_{x}(\omega,0)}_{\text{non-ideal}}^\text{Ret}\right] .
\end{split}
\end{equation}  
Using \eqref{eq:KinY} we find the promised expression \eqref{eq:sigmadc} obtained in \cite{Hartnoll:2007ip}
\begin{equation}\label{eq:sigma2}
\sigma_\text{dc} = r_+^{d-3}\, (1-\sdc)^2 \,.
\end{equation}  
As noted in \cref{sec:setup} the Ohmic parameter $\sdc$ effectively is a proxy for the dimensionless conductivity.

\section{Discussion}
\label{sec:discuss}

We have extended the analysis of open quantum systems with long-lived modes using holography initiated in \cite{Ghosh:2020lel} to systems with multiple degrees of freedom, focusing on the dynamics of momentum diffusion in a charged plasma. The key novel feature here is the fact that the momentum diffusion mode  mixes with the charge current, leading to an imprint of diffusive dynamics in the current itself. One of the advantages of the holographic analysis is that the bulk fields dual to the charge and energy-momentum tensor  currents, the photons and gravitons, naturally suggest a manner of decoupling the long-lived diffusive mode from the short-lived charge transport mode. 

In practical terms, the strategy we use here to analyze the coupled Einstein-Maxwell dynamics is well known from the study of  black hole perturbations \cite{Kodama:2003kk}. For our analysis we need a bit more than the decoupling of the dynamical equations of motion. As explained in \cite{Ghosh:2020lel}, one needs to understand the variational principle associated with the gauge invariant modes. The general rule of thumb is that the short-lived Markovian modes are quantized in the bulk with Dirichlet (standard) boundary conditions, while the long-lived ones are quantized with Neumann (alternate) boundary conditions. This choice is not ad-hoc, but rather a consequence of the parent gravitational dynamics. Standard Dirichlet boundary conditions for gravitons and photons in the Einstein-Maxwell theory descend naturally to the alternate quantization choice for the non-Markovian modes. 

Our analysis broadly supports the general thesis of \cite{Ghosh:2020lel} that the dynamics of Markovian and non-Markovian degrees of freedom are well modeled in holography by designer scalar fields, whose gravitational coupling with the background geometry is modulated by an auxiliary dilaton, as illustrated in \cref{fig:cartoon}. For the diffusive modes analyzed herein, we found two sets of Markovian modes from tensor polarization and the effective charge current, along with a single set of non-Markovian momentum diffusion modes. One difference from the earlier analysis is that the dilaton profiles are not necessarily simple monomials in the radial coordinate. The distinction between Markovianity and non-Markovianity is however nicely encoded in the asymptotic fall-off of the dilaton.

 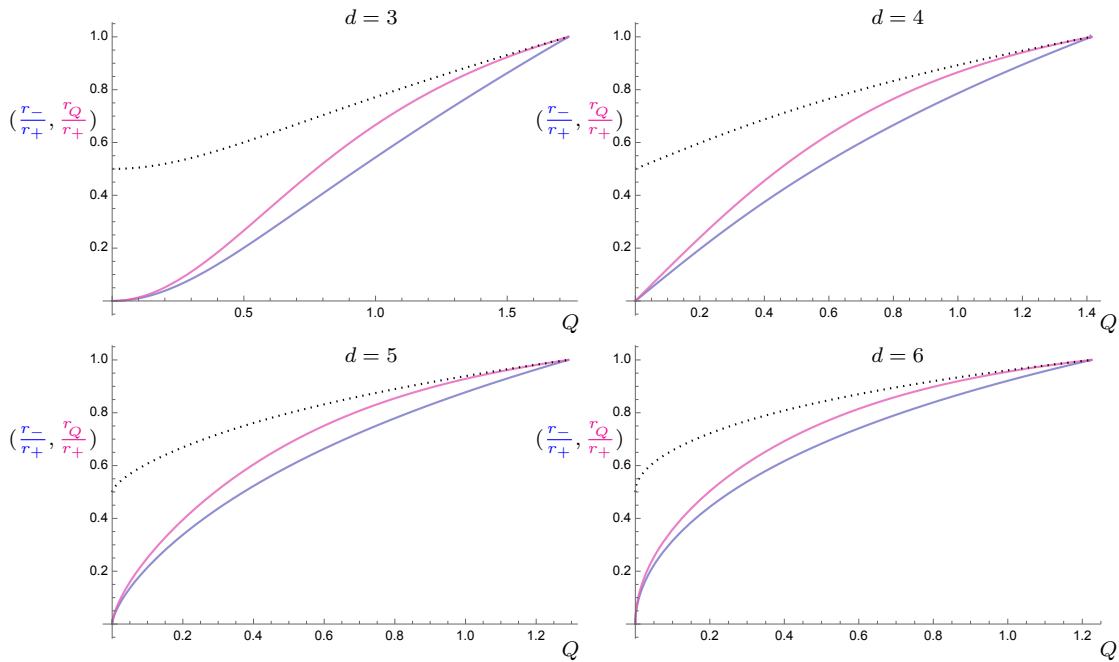
\begin{figure}[th]
 \centering
 \begin{minipage}[t]{\textwidth}
\vspace{0.4cm}
\hspace{0pt}
\begin{tikzpicture}
 \node (img)  {\includegraphics[scale=0.5]{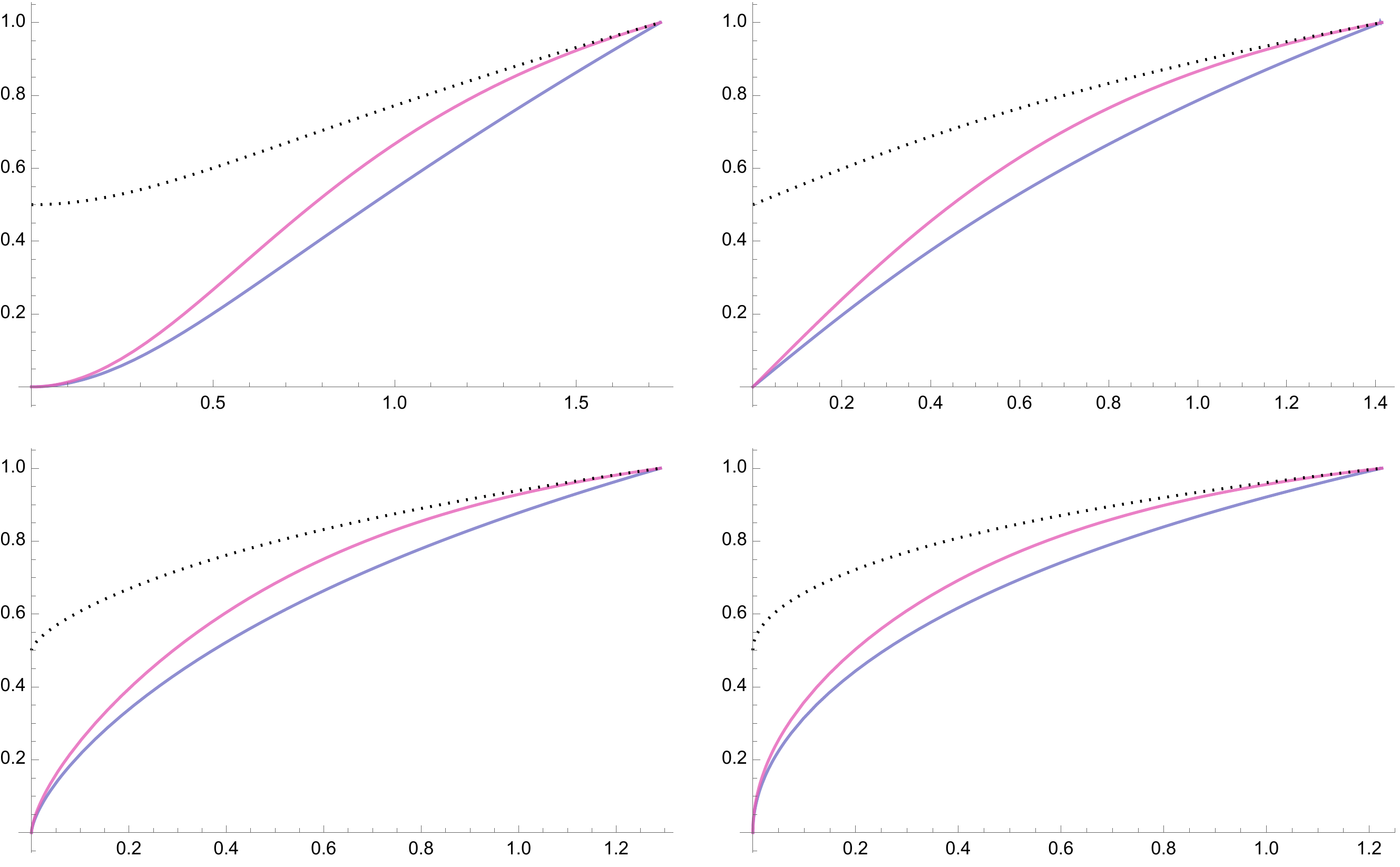}};
 \foreach \x in {-0.6,6.55}
 {\foreach \y in {0.15,-4.2}
  {\node[right,font=\scriptsize] at (\x,\y) {$Q$};
}}
 \foreach \xa in {-6.5,0.5}
 {\foreach \yb in {2.85,-1.4}
  {\node[left,font=\scriptsize]  at (\xa,\yb) {$({\color{blue}{\frac{r_-}{r_+}}},{\color{magenta}{\frac{\RQ}{r_+}}})$};
}}
\node[above,font=\scriptsize] at (-3,4) {$d=3$};
\node[above,font=\scriptsize]  at (4,4) {$d=4$};
\node[above,font=\scriptsize]  at (-3,-0.5) {$d=5$};
\node[above,font=\scriptsize]  at (4,-0.5) {$d=6$};
 \end{tikzpicture}
 \end{minipage}
 \caption{A plot of the dimensionless inner horizon radius $r_-$ (blue) and $\RQ$ (magenta), normalized by $r_+$,  as a function of the charge parameter $Q$. We also plot the arithmetic mean $\frac{r_++r_-}{2}$ (dotted curves) to illustrate that  $\RQ$ approaches this value from below in the near-extremal limit. Note that the range of $Q$ varies across dimensions owing to \eqref{eq:Qbound}.}
 \label{fig:RQrm} 
 \end{figure}

One curious feature of the dilatonic coupling for the Markovian charge mode is that the gravitational description naturally picks out the Ohmic radius $\RQ$, defined in \eqref{eq:RQdef}, which is related to the conductivity of the plasma \eqref{eq:sigmadc}. At the locus $r=\RQ$ the dilaton for the charge current mode vanishes. While this locus is not part of the minimal grSK geometry -- our contour in the complexified radial coordinate only encircles the outer horizon at $r=r_+$ as depicted in \cref{fig:mockt}  -- it is nevertheless curious that  in this region the Markovian mode is free to fluctuate without penalty. We note that the length scale $\RQ$ is sandwiched  between the inner horizons at  $r_-$ and $r_+$ and approaches their arithmetic mean from below in the near-extremal limit, \cref{fig:RQrm}. While its physical significance is not entirely clear to us at the moment, it would interesting to examine how this scale appears in the field theory.

A related issue has to do with the extremal limit. As can be seen from various plots \cref{fig:KinMarY,fig:KinMarT,fig:KinMarT4th,fig:KinMarY4th}, as $\sdc \to 1$, the physical quantities diverge. This is expected from the fact that the geometry develops an infinite \AdS{2} throat in the IR and relatedly the spectrum of quasinormal modes develops a branch cut in the complex frequency plane emanating from the origin (in the negative imaginary direction). One understands this  as resulting from a local critical behaviour at extremality \cite{Faulkner:2009wj}; the hydrodynamic modes merge into a continuum supported in the near horizon \AdS{2} fixed point. It is also directly visible in the breakdown of the hydrodynamic gradient expansion as the functions parameterizing the gradient expansion are no longer regular owing to the double zero of $f(r)$. Once again in this limit the analytic structure of the charge current mode is different owing to the dilatonic modulation (the solution for the  fields $\MY_\ai$ involves inverting a differential operator that has a fourth order zero from $f h^2$). If we work in the near-extremal limit, then $\RQ \approx \frac{r_++r_-}{2}$, suggesting that there is still a sensible hydrodynamic gradient expansion one can perform. It would be useful to verify this directly and understand the approach to extremality in terms of the grSK contour  pinched between the branch point at $r=r_+$ and the locus $r=\RQ$.

We restricted our analysis herein to Einstein-Maxwell dynamics. \RNAdS{d+1} black holes in even $d$ are also solutions to Einstein-Maxwell-Chern-Simons theories, arising from Freund-Rubin compactifications of 10 and 11 dimensional supergravity. The Chern-Simons terms in the bulk capture the  't Hooft anomalies of the field theory $R$-symmetry, and lead to a non-trivial imprint in transport \cite{Banerjee:2008th,Erdmenger:2008rm}, in the form of parity-odd contributions to the current. It would be desirable to analyze the dynamics including these terms. One interesting question is whether the general structure of the anomalous contributions to the hydrodynamic effective action which were argued to take a Schwinger-Keldysh form in  \cite{Haehl:2013hoa}, can be justified holographically. 

The ideal part of the Einstein-Maxwell action computed herein when expanded out in gradients should compute the Class L hydrodynamic Lagrangian of a charged fluid. The authors of \cite{Haehl:2015pja} (see also \cite{Bhattacharyya:2014bha}) classified the seven hydrostatic coefficients that are admissible at quadratic order (in addition to the leading order pressure term), but they did not construct the adiabatic Lagrangian. We however have a clear prediction from the holographic analysis; given that the neutral fluid Class L Lagrangian predicted in \cite{Haehl:2015pja} was cleanly reproduced from gravity in \cite{Ghosh:2020lel} it would be interesting to generalize the same to the charged plasma.\footnote{We thank Akhil Sivakumar for raising this point.}   In fact, given that we have complete analysis to quartic order in gradients, one could potentially extract higher order hydrostatic transport data.\footnote{As we explain at the end of \cref{sec:probenM} existing claims in the literature regarding third order transport data for the neutral fluid are incorrect as they miss potential mixing between third and fourth order coefficients.}

Our analysis did not explore the scalar polarizations which physically capture the sound mode and the charge diffusion mode. This sector of Einstein-Maxwell dynamics is quite complicated as there are a-priori 10 functions characterizing the perturbations. There are however only two physical modes in the problem, the sound mode and the charge diffusion mode, both of which are non-Markovian. The general lessons learnt from our  analysis here and from that of sound propagation in neutral holographic plasmas \cite{Loganayagam:2021ab} is that it should be possible to characterize the dynamics in terms of a suitable designer scalar field. We hope to report on this analysis, which offers the novelty of untangling the mode mixing between two non-Markovian degrees of freedom in the near future \cite{He:2021ab}.

\section*{Acknowledgements}
It is a pleasure to thank Veronika Hubeny for helpful discussions.    RL would also like to acknowledge his debt to the people of India for their steady and generous support to research in the basic sciences.  
TH, MR, and JV are supported  by U.S.\ Department of Energy grant DE-SC0009999 and by funds from the University of California. 

\appendix

\section{The variational problem in the bulk}
\label{sec:bulkaction}

The bulk Einstein-Hilbert dynamics is described by the action \eqref{eq:SEMax} with appropriate boundary counterterms. The counterterm action accurate to quartic order in boundary gradients is given by\footnote{Often we will write boundary terms and counterterms in terms of Lagrangian densities $L$ without the measure factor.  Furthermore, to avoid cluttering up the equations in the appendices we will drop the dependence on the Newton's constant, setting $c_\text{eff}=1$. We reinstate this factor in the main text. \label{fn:ceffGN}} \cite{Emparan:1999pm,deHaro:2000vlm} 
\begin{equation}\label{eq:SEMct}
\begin{split}
S_\text{ct} 
&= 
	 \int\, d^dx\, \sqrt{-\gamma} \left[ L_\text{EH,ct} + L_\text{Max,ct} \right] ,\\
L_\text{EH,ct} 
&= 	-2 (d-1) - \frac{1}{d-2} \, \tensor[^\gamma]{R}{} 
	- \frac{1}{(d-4)\,(d-2)^2} \left(
		\tensor[^\gamma]{R}{_{\mu\nu}} \, \tensor[^\gamma]{R}{^{\mu\nu}}  - \frac{d}{4(d-1)}\, \tensor[^\gamma]{R}{}^2
	\right) ,\\
L_\text{Max,ct}
&=
	\frac{1}{4(d-4)} \left(\tensor[^\gamma]{F}{_{\mu\nu}} \, \tensor[^\gamma]{F}{^{\mu\nu}} 
	+ \frac{1}{(d-4)(d-6)}\, \tensor[^\gamma]{F}{_{\mu\nu}} \Box_\gamma\, \tensor[^\gamma]{F}{^{\mu\nu}}\right) 	.
\end{split}
\end{equation}
We show below that the gravitational action can be written in terms of three sets of decoupled modes (indexed by a polarization label)
\begin{equation}\label{eq:SEMresult}
\begin{split}
S_\text{EM} =  \sum_{\bi=1}^{N_T}\, S[\tGR_\bi] + \sum_{\ai=1}^{N_V}\, \left(S[\MX_\ai] + S[\MY_\ai]\right)
	+ 
	 \int\, d^d x\, \sqrt{-\gamma}\, \left[\sqrt{-\gamma_{\mu\nu} \vb{b}^\mu \, \vb{b}^\nu}\ \right]^{-d} \, \left(1+\frac{d-2}{d-1}\, \Lambda_{\mathbf{b}}^2\right) .
\end{split}
\end{equation}
The vector $\vb{b}^\mu$ and the parameter  $\Lambda_{\mathbf{b}}$  are introduced in \eqref{eq:thermalvT} -- they are related to the thermal vector and thermal twist introduced in \cite{Haehl:2015pja}. The latter are defined as $\bm{\beta}^\mu= \frac{u^\mu}{T} $ and the dimensionless ratio $\Lambda_{\bm{\beta}} = \frac{\mu}{T}$. We have chosen to use $r_+$ to non-dimensionalize the physical quantities in lieu of the temperature as the resulting expressions are more compact. 

The last term in \eqref{eq:SEMresult} arises because of the non-zero background free energy of the solution. The standard Gibbons-Hawking computation on the Euclidean \RNAdS{d+1} solution would give a free energy $r_+^d(1+Q^2)$ which we have re-expressed in terms of the thermal field theory parameters using the induced boundary metric data.  This contribution is the Class L fluid  Lagrangian in the nomenclature of \cite{Haehl:2015pja}. and receives contributions from the tensor and vector perturbations. As written it not only includes the background free energy but also captures additional  contributions quadratic in the fluctuations. We will explain below the individual contributions, but not illustrate how they combine nicely into the form quoted above (it follows along similar lines to the discussion in \cite{Ghosh:2020lel}).

\subsection{Tensor perturbations}
\label{sec:tensorsAct}

The dynamics of the tensor perturbations parameterized in \eqref{eq:hAtensor} can be straightforwardly obtained by plugging the ansatz into the Einstein-Maxwell action \eqref{eq:SEMax} and evaluating the appropriate counterterms in \eqref{eq:SEMct}. We find the following:\footnote{The superscript $T$ is to indicate that we are projecting onto the tensor perturbations in this section.}
\begin{equation}\label{eq:Tactionterms}
\begin{split}
S^\text{T}_\text{EM,bulk}
&= 
	 - \sum_{\bi=1}^{N_T} \, \int d^{d+1}x\sqrt{-g}\, \frac{1}{2}\, \nabla_A \tGR_\bi\nabla^A \tGR_\bi
	 -	 \int dr\,d^dx \dv{L^\text{T}_\text{bdy}}{r}\,,\\
S^\text{T}_\text{bdy}
&= 
	\int d^d x\left(L^\text{T}_\text{bdy}+ L^\text{T}_\text{ideal}\right) ,\\
S^\text{T}_\text{ct}
&=	\int d^dx\, \sqrt{-\gamma} \left[ L^\text{T}_\text{ideal,ct} + L^\text{T}_\text{scalar,ct}\right] .
\end{split}
\end{equation}
 The ideal fluid contribution $L^\text{T}_\text{ideal}$ is\footnote{We will quote the ideal fluid contribution in its entirely in both the tensor and vector sector for completeness. Note however that the contribution from the background (the terms independent of fluctuations) should only be included once. \label{fn:idealcaveat} }
\begin{equation}\label{eq:Tideal}
L^\text{T}_\text{ideal} = r^d\left[d + (d-2)\, f - \frac{(d-2)^2}{d-1}\, \frac{a^2}{r^2}\right] \left(1-\frac{1}{2}\,\sum_{\bi=1}^{N_T}\, \tGR_\bi^2\right) ,
\end{equation}	
where the counterterm contribution is given by 
\begin{equation}\label{eq:Tcts}
\begin{split}
L^\text{T}_\text{ideal, ct}  &= 
			2(d-1) \left[-1+ \frac{1}{2}\, \sum_{\bi=1}^{N_T}\,  \tGR_\bi^2\right] ,\\
L^\text{T}_\text{scalar, ct} 
&=	\frac{1}{2} \sum_{\bi=1}^{N_T} 
	\left[-\frac{1}{d-2} \tGR_\bi\Box_{\gamma}\tGR_\bi-\frac{1}{(d-2)^2(d-4)}\tGR_\bi\Box_{\gamma}^2\tGR_\bi\right] .
\end{split}
\end{equation}	
While the boundary contribution $L^\text{T}_\text{bdy}$ is irrelevant as it cancels between the Einstein-Hilbert and the Gibbons-Hawking term,  we record it here for completeness: 
\begin{equation}\label{eq:Tbdyterm}
L^\text{T}_\text{bdy} = 2\, r^d f(r)\left(1 - \frac{1}{2}\, \sum_{\bi=1}^{N_T} \dv{r} (r\,\tGR_\bi^2)\right) \,.
\end{equation}

\subsection{Vector perturbations: Equations of motion}
\label{sec:vectorEom}

We will now demonstrate that the dynamics of the vector perturbations can be repackaged into a paired of coupled scalar fields, which we will subsequently diagonalize into a set of decoupled non-Markovian and Markovian field, respectively.

\subsubsection{The coupled Markovian and non-Markovian system}
\label{sec:Vcoupled}

To set the stage for our discussion let us start by recording the gravitational perturbation packaged into the diffusive gauge system \eqref{eq:defAuxSA} in terms of  gauge invariant combinations (that can be viewed as conjugate momenta, or the field strengths):
\begin{equation}\label{eq:VginvarsA}
\begin{split}
\psGR_v^\ai &\equiv \dv{\vGR_v^\ai}{r} + i\omega  \, \vGR_r^\ai \,, \qquad
\pJGR_r^\ai \equiv  \dv{\vGR_x^\ai}{r}+ik \,\vGR_r^\ai \,, \\ 
\psGR_x &\equiv  \Dz_+\vGR_x^\ai+ik\,\vGR_v^\ai+ik\, r^2 f \,\vGR_r^\ai  \equiv r^2 f\, \pJGR_r^\ai + \pJGR_v^\ai\,,\\
\pJGR_v^\ai &\equiv i\, k\, \vGR_v^\ai  -i \omega\, \vGR_x^\ai \,,
\end{split}
\end{equation}	
These can be compared with the parameterization introduced in \cite[Eq.~(8.4)]{Ghosh:2020lel}. 
 
Since we are discussing a charged plasma we have a discrete charge conjugation symmetry $\text{C}$ in addition to time reversal $\text{T}$. Let us record $\text{CT}$-transformations; only the photons carry $\text{C}$ quantum number, so for the gravitons we can proceed as before:
\begin{itemize}[wide,left=0pt]
\item $\{\vGR_x^\ai, r^2 f\, \vGR_r^\ai + \vGR_v^\ai\}$ have even  time-reversal and hence even $\text{CT}$ parity.
\item $\vGR_v^\ai$ has odd time-reversal parity and hence odd $\text{CT}$ parity.
\item  $\psGR_v^\ai$ and $\pJGR_v^\ai$  have odd time reversal parity and hence odd $\text{CT}$ parity.
\item $\psGR_x^\ai$ has even time reversal and $\text{CT}$ parity. 
\item $\vMax_\ai$ is $\text{C}$ odd and $\text{T}$ even, so altogether $\text{CT}$-odd.
\end{itemize}
This should constrain the perturbation equations and serve as a useful check on the results. 

The perturbative Maxwell equation leads to:\footnote{The equations $\EMax$ and $\EEin$ are rescaled by a factor of $\sqrt{(d-1)(d-2)}$ for convenience.}
\begin{equation}\label{eq:MaxVpert}
\frac{r^{d-2}}{d-3}\EMax_i 
:\ 
\dv{r}(r^{d-1} f\, \dv{\vMax_\ai}{r} ) - i\omega\, \left[\dv{r}(r^{d-3} \vMax_\ai) + r^{d-3}\, \dv{\vMax_\ai}{r} \right] -k^2\, r^{d-5}\, \vMax_\ai = r^{d-1} a'(r) \, \psGR_v^\ai \,,
\end{equation}	
where we recognize that the l.h.s is the wave operator acting on a designer scalar $\vMax_\ai$ with Markovianity index $\ann =d-3$. This is indeed what we should expect; a probe Maxwell system would exhibit this equation as discussed in \cite{Ghosh:2020lel}.\footnote{To be clear, here we indicate a probe Maxwell field which is different from the one that generates the \RNAdS{d+1} background.}   Furthermore, the $\text{CT}$ transformation properties indicate that only  $\vGR_v^\ai$ can source the odd-parity $\vMax_\ai$.

From Einstein's equations we find: 
\begin{equation}\label{eq:EinVpert}
\begin{split}
\frac{2\,r^{2}}{d-2} \, \EEin_{ri} 
&:\  
	\dv{r}\left(r^{d+1} \psGR_v^\ai\right) -ik\, r^{d-1} \pJGR_r^\ai  - \frac{d-1}{d-2}\, \kappa_q\, a'\, r^{d-1}\,\vMax_\ai' =0 \,,  \\
\frac{\sqrt{d-1}}{(d-2)\,r} \, \EEin_{ii} 
&:\
	\dv{r}(r^{d-1}\, \psGR_x^\ai) - i \omega\, r^{d-1}\, \pJGR_r^\ai =0 \,,  \\
-\frac{2r^{d-1}}{d-2}(r^2f\, \EEin_{ri} + \EEin_{vi} )&:\
 -i\omega\, r^{d+1}\, \psGR_v^\ai + i k\,	 r^{d-1}\, \psGR_x^\ai -  i\omega (d-1)\, q\,\kappa_q\,\vMax_\ai = 0 \,.
\end{split}
\end{equation}	
The last equation above is the momentum conservation equation -- however, now $\vMax_\ai$ acts a source of momentum flux. Again only the first and third equations are for $\text{CT}$-odd gravitons, so these can see a contribution from the photon mode; the second equation is $\text{CT}$-even and hence unmodified.  
 
To simplify the system we adopt the Debye gauge parameterization from \cite[\S8.2]{Ghosh:2020lel}. Let us introduce two scalar fields $X_\ai$ and $Y_\ai$ as follows: 
\begin{equation}\label{eq:Debyeinspire}
\begin{split}
r^{d+1}\, \psGR_v^\ai &= k^2 (X_\ai +2\,Y_\ai)  \,, \\
r^{d-1}\, \psGR_x^\ai &= \omega\, k\, X_\ai \,, \\
\vMax_\ai &= -\frac{k^2}{(d-2)\, \mu\,r_+^{d-2} } \, Y_\ai \,,\\
r^{d-1}\,\pJGR_r^\ai &= -ik \, \dv{X_\ai}{r}\,.
\end{split}
\end{equation}	
This parameterization immediately solves all  three Einstein's equation. Maxwell's equation gives one constraint relating $X_\ai$ and $Y_\ai$. The other  dynamical equation comes from demanding consistency of \eqref{eq:Debyeinspire}, with the definitions in \eqref{eq:VginvarsA} as in the analysis of \cite[\S8.2]{Ghosh:2020lel} .
We first solve
\begin{equation}
	\pJGR_v^\ai = \psGR_x^\ai - r^2 f\, \pJGR_r^\ai = \frac{ik}{r^{d-1}}\, \Dz_+ X_\ai 
	\;\; \Longrightarrow \;\;
	\vGR_v^\ai = \frac{1}{r^{d-1}}\, \Dz_+ X_\ai+ \frac{\omega}{k}\, \vGR_x^\ai\,.
\end{equation}	
Plugging this back again into the definition for $\psGR_v^\ai$ we obtain the equation for $X_\ai$. The Maxwell equation \eqref{eq:MaxVpert} upon using  \eqref{eq:Debyeinspire} also simplifies. The system of Einstein-Maxwell equations reduces to a clean time-reversal symmetric form:
\begin{equation}\label{eq:XYcoupled}
\begin{split}
&
	r^{d-1}\,\Dz_+\left(\frac{1}{r^{d-1}}\, \Dz_+ X_\ai\right) +\omega^2 \, X_\ai-k^2 f (X_\ai + 2\, Y_\ai)
=
	0\,,\\
&
	\frac{1}{r^{d-3}}\,\Dz_+\left(r^{d-3}\, \Dz_+ Y_\ai \right) + (\omega^2 -k^2\,f)\, Y_\ai
= 
	(d-2)^2\, a^2\,f\, (X_\ai + 2 \,Y_\ai )\,.
\end{split}
\end{equation}	
In other words the dynamics of the Einstein-Maxwell system can be encapsulated in a set of coupled non-Markovian and Markovian degrees of freedom, for using the operator \eqref{eq:DesOp}  we can rewrite \eqref{eq:XYcoupled} as
\begin{equation}\label{eq:XYdiffuseM}
\begin{split}
\mathfrak{D}_{1-d} X_\ai
 	&=2\, k^2\, f\, Y_\ai  \,,\\
\mathfrak{D}_{d-3} Y_\ai
	&= 
	(d-2)^2\, a^2\,f\, \left(X_\ai + 2\, Y_\ai\right) . 
\end{split}
\end{equation}
%

\subsubsection{Decoupling the vector modes}
\label{sec:Vdecoupled}

The coupled equations \eqref{eq:XYcoupled} can be decoupled into a pair of Markovian and non-Markovian fields. The strategy is straightforward: we homogenize the kinetic operator $\Dann$ by a suitable field redefinition (say by rescaling $X_\ai$ so that the kinetic operator acting on it is $\mathfrak{D}_{d-3}$) and rotate basis. After a bit of algebra we see that the following field redefinition suffices to decouple the system:
\begin{equation}\label{eq:XYredef}
\begin{split}
X_\ai 
&=
	 -(\BQT^2+2) \, \MX_\ai + \BQT^2\, \frac{h}{1-h} \, \MY_\ai\,,\\
Y_\ai 
&=
	 (1-h) \,\MX_\ai + h\, \MY_\ai \,.
\end{split}
\end{equation}
The parameter $\BQT$ was introduced earlier in \eqref{eq:BQTdef} and arises from the diagonalization of the modes as presaged. Plugging the field redefinition into \eqref{eq:XYcoupled} we recover the equations \eqref{eq:XYfinal} given in the main text.

We see that $\MX$ is a designer scalar with dilaton proportional to  $r^{d-3}\, (1-h)^2 $ which effectively makes it non-Markovian since the index shifts from $\ann_+ = d-3$ to $\ann_-=1-d$. 
 On the other hand $\MY$ looks Markovian near the boundary, but switches character as one hits nears the Ohmic radius owing to dilaton profile $r^{d-3}\, h^2$. 

We note here that the field redefinition \eqref{eq:XYredef} has a simple action on the equations of motion. Letting $\mathcal{E}_X$, $\mathcal{E}_Y$, $\mathcal{E}_\MX$ and $\mathcal{E}_\MY$ be the equations of motion of the four fields, one can check that:
\begin{equation}\label{eq:EXYmap}
\begin{split}
\mathcal{E}_X 
&=
	- (\BQT^2+2) \, \mathcal{E}_{\MX} + \BQT^2 \,\frac{h}{1-h} \, \mathcal{E}_{\MY}  \,, \\
\mathcal{E}_Y
&=
	 (1-h) \,\mathcal{E}_{\MX}  + h\,  \mathcal{E}_{\MY}  \,. 
\end{split}
\end{equation}	
%

\subsection{Vector perturbations: Variational principle}
\label{sec:vectorsAct}

Now we turn to vector perturbations which we will tackle in several stages. First in this section we will demonstrate that the perturbations can be repackaged into an Markovian scalar of index $\ann = d-3$ and an auxiliary diffusive gauge field. The latter in turn can be rewritten in terms of a non-Markovian scalar as explained in \cite{Ghosh:2020lel}. 

\subsubsection{The diffusive vector and scalar parameterization}
\label{sec:vectorsActA}

First consider encoding graviton fluctuations from \eqref{eq:hAvector} into a diffusive gauge field. As in \cite[Eq (9.2)]{Ghosh:2020lel}  we define an auxiliary diffusive gauge field $\AGR$ with only scalar polarizations:
\begin{equation}\label{eq:defAuxSA}
\begin{split}
\AGR^\ai_B(v,r,\bx) \,dx^B 
&\equiv  \int_k \, 
	 \bigg( (\vGR_r^\ai(r,\omega,\bk) \, dr +\vGR_v^\ai (r,\omega,\bk) dv)\,
  	\ScS(\omega,\bk|v,\bx) \\
  & \qquad \qquad \quad 
  	- i\,\vGR_x^\ai(r,\omega,\bk)\ \ScS_i\,  dx^i \bigg),
 \end{split}
\end{equation}
In addition we have the transverse photon degree of freedom $\vMax_\ai$. We again compute directly
\begin{equation}\label{eq:Vactionterms}
\begin{split}
S^\text{V}_\text{EM,bulk}
&=	 
	 -\int d^{d+1}x\sqrt{-g} \sum_{\ai=1}^{N_V}
	 \bigg\{
	 r^{-2} \, \nabla_{A}\vMax_\ai\nabla^{A}\vMax_\ai 
	 +\frac{1}{4}\, r^2 \, \FGR_{AB}^\ai\, \FGR^{AB}_\ai 
	 +\vMax_\ai\, \FGR_{AB}^\ai\, F^{AB}\bigg\} \\
&\qquad \qquad 
		- \int dr\, d^dx \, \dv{r} L^\text{V}_\text{bdy} \,, \\
S^\text{V}_\text{bdy}
& = 
	 \int  d^dx \, \left(L^\text{V}_\text{bdy}+ L^\text{V}_\text{ideal}\right) ,\\
S^\text{V}_\text{ct}
&=
		 \int  d^dx \, \sqrt{-\gamma} \left[L^\text{V}_\text{ideal,ct}+ L^\text{V}_\text{Aux,ct} +L^\text{V}_{\ann=d-3}\right] .
\end{split}
\end{equation}	
The ideal fluid contribution from the vectors is given in terms of the boundary values of the diffusive gauge field, which we denote by $\JMarQ_{\mu}^\ai dx^\mu = \AGR_A^\ai\, dx^A\big|_\text{bdy}$, and the Markovian scalar (cf., \cref{fn:idealcaveat} for the caveat about this ideal part), 
\begin{equation}\label{eq:Videal}
\begin{split}
 L^\text{V}_\text{ideal}
 &= 
	r^d\left[\left(d+(d-2)f - \frac{(d-2)^2}{d-1}\frac{a^2}{r^2}\right)
		\left(1-\frac{1}{2} \, \sum_{\ai=1}^{N_V} \JMarQ_{i}^\ai \, \JMarQ_{i}^\ai\right) \right.\\
&\left.		\qquad \qquad 
	+ \sum_{\ai=1}^{N_V} \left( (d-1)\, \JMarQ_{v}^\ai\, \JMarQ_{v}^\ai -\frac{2(d-2)a}{r^2}\, \vMax_\ai \JMarQ_{v}^\ai\right)\right] . 
\end{split}
\end{equation}	

The counterterm action in turn is given by the contributions from the auxiliary gauge system and the Markovian scalar mode of index $\ann =d-3$, along with an ideal fluid contribution. To wit, 
\begin{equation}\label{eq:Vcts}
\begin{split}
L^\text{V}_\text{ideal, ct}
&=
	(d-1)\,\sum_{\ai=1}^{N_V}  \left[\JMarQ_{i}^\ai \, \JMarQ_{i}^\ai  - \frac{1}{f}\, \JMarQ_{v}^\ai\, \JMarQ_{v}^\ai\right] , \\
 L^\text{V}_\text{dv,ct}
 &= 
 	\frac{r^2}{4} \, 
 	\sum_{\ai=1}^{N_V}  
 		\frac{\gamma^{\mu_1\mu_2}\,\gamma^{\nu_1\nu_2}}{d-2} \left(\JMarQF^\ai_{\mu_1\nu_1}  \,\JMarQF^\ai_{\mu_2\nu_2}  + \frac{1}{(d-2)\,(d-4)} \
 		\JMarQF^\ai_{\mu_1\nu_1} \Box_{\gamma} \JMarQF^\ai_{\mu_2\nu_2} \right),\\
L^\text{V}_\text{ms,ct}
&=
	 -\frac{1}{2\,r^2} 
	 \sum_{\ai=1}^{N_V}  \left[\frac{1}{d-4} \, \vMax_\ai \Box_{\gamma}\, \vMax_\ai+\frac{1}{(d-4)^2\,(d-6)}\, \vMax_\ai\Box_{\gamma}^2\, \vMax_\ai\right].
\end{split}
\end{equation}
where $\JMarQF_{\mu\nu} =2\,\partial_{[\mu} \JMarQ_{\nu]}$ is the field strength of the diffusive gauge field.

Finally, for the vector perturbations the (irrelevant) boundary term itself evaluates to the following contribution:
\begin{equation}\label{eq:Vbdyterm}
\begin{split}
L^\text{V}_\text{bdy} 
&= 
	2 r^d f + r^d \sum_{\ai=1}^{N_V} \left[ 
		-f\,  \dv{r}(r^3\, \AGR^\ai_{\mu}\AGR_\ai^{\mu})
		+\frac{d}{2f}  	\left((1+f) \left(\AGR_v^\ai + r^2f\,\AGR_r^\ai\right)^2+(1-f) (\AGR_v^\ai)^2\right) \right.\\
&\left.\qquad\qquad \qquad \qquad
	-\frac{a^2}{4\,r^2f}\left(\left(\AGR_v^\ai + r^2f\, \AGR_r^\ai\right)^2+ (\AGR_v^\ai)^2\right)
		+2(d-2) \frac{a}{r^2}\, \AGR_v^\ai\, \vMax_\ai\right] .
\end{split}
\end{equation}
which we have left in terms of the bulk diffusive gauge field $\AGR_A^\ai$.  As should be clear from \eqref{eq:Vactionterms} this contribution cancels between the bulk Einstein-Maxwell action and the boundary Gibbons-Hawking term.

\subsubsection{The coupled designer scalars parameterization}
\label{sec:vectorsActB}

In \cref{sec:vectorsActA} we argued that the dynamics of the vector perturbations of gravitons and gauge fields can be re-expressed in terms of a diffusive gauge field and a Markovian scalar. We now rewrite this action in terms of the scalar system used to simplify the system in \cref{sec:vectorEom}. While we would jump directly to the final parameterization in terms of the decoupled fields $\MX_\ai$ and $\MY_\ai$ used in the main text, it will be edifying to analyze the intermediate steps in terms of the coupled scalars $X_\ai$ and $Y_\ai$ to understand the variational principle.

The passage from the parameterization in terms of the auxiliary diffusive gauge field $\AGR_\ai$ and the transverse mode of the photon $\vMax_\ai$ to the coupled scalars $X_\ai$ and $Y_\ai$ simply involves using the Debye gauge inspired parameterization introduced in \eqref{eq:Debyeinspire}. Substituting this we directly obtain the following decomposition of the bulk action and variational boundary terms:
\begin{equation}\label{eq:XYactioncoupled}
\begin{split}
& \hspace{2cm} 
	S^\text{V}_\text{EM,bulk} + S^\text{V}_\text{bdy} 
= 	
 	S_X+ S_Y + S_{XY} \,,  \\
S_X 
&=
	-\frac{1}{2}\int\, d^{d+1}x\, \frac{\sqrt{-g}}{r^{2(d-1)}}\,
	 \sum_{\ai=1}^{N_V} \,  \partial_i \nabla^A X_\ai	\partial^i\nabla_A X_\ai 
	+ \int d^d x \, r^{1-d}\,  \sum_{\ai=1}^{N_V} \,\partial_i X_\ai \partial^i \Dz_+ X_\ai \\
& \qquad \qquad 
	+  
	\frac{1}{2}\, \int_k dr \,\frac{1}{r^{d+1}}\, 	\left(\frac{\mathcal{E}_X}{f}\right)^2  ,\\
S_Y
&= 
	-\frac{C_Y}{2}\,  \int d^{d+1}x\, \frac{\sqrt{-g}}{r^2} \, 
	\sum_{\ai=1}^{N_V}  \left\{
	 	\partial^2\nabla_A Y_\ai\, \partial^2\,\nabla^A Y_\ai    +\frac{4}{C_Y}\, 
	 	\frac{1}{r^{2(d-1)}} \, \partial^2\, Y_\ai\, \partial^2 Y_\ai \right\} , \\
S_{XY}
&=
	- 2\, \int d^{d+1}x\, \frac{\sqrt{-g}}{r^{2d}} \, \sum_{\ai=1}^{N_V} \, \partial^i\, X_\ai\, \partial_i\partial^2 Y_\ai  \,.
\end{split}
\end{equation}	
The normalization factor $C_Y$ is 
\begin{equation}\label{eq:Ynormalize}
C_Y=   \frac{2}{\bRQ^2\, \RQ^{2(d-2)}\, r_+^2 } \,.
\end{equation}	
We have written the action in position space by converting momentum factors to spatial derivatives $\partial_i$ and are using $\partial^2$ to denote the Fourier transform of $k^2$.

As explained in \cite{Ghosh:2020lel} the variational problem for field $X_\ai$ is dictated by the first line in $S_X$, the presence of the boundary term $k^2\, X_\ai\, \Dz_+ X_\ai$, which dictates that it be quantized with Neumann boundary conditions. On the other hand we see that the field $Y_\ai$ behaves like a Markovian scalar of index $\ann=d-3$, so can be quantized with standard Dirichlet boundary conditions.  

There are two additional pieces of the action: the terms $L^\text{V}_\text{ideal} + L^\text{V}_\text{ideal,ct}$ which  assemble together into the ideal fluid contribution indicated in \eqref{eq:SEMresult}. We won't try to rewrite this contribution from the vector sector in terms of the scalars. On the other hand, the remaining part of the counterterm action from the diffusive gauge field and the Markovian scalar \eqref{eq:Vbdyterm} simplifies to
\begin{equation}\label{eq:XYcoupledCterms}
\begin{split}
L^\text{V}_\text{ct} 
&=
	L^\text{V}_\text{dv,ct} + L^\text{V}_\text{ms,ct}   \\
&	= 
	-\frac{1}{2\,(d-2)} \,\frac{1}{r^{2d}\, f} \sum_{\ai=1}^{N_V} \left[(\partial_i \Dz_+X_\ai)^2  
	+ \frac{1}{(d-2)\,(d-4)}\partial_i \Dz_+X_\ai\Box_{\gamma} \partial^i \Dz_+X_\ai\right] \\
&\qquad\qquad
-
	\frac{1}{2\,r^2}\, C_Y\,\sum_{\ai=1}^{N_V} 
	 \left[\frac{1}{d-4} \, Y_\ai \Box_\gamma Y_\ai + \frac{1}{(d-4)^2(d-6)} \, Y_\ai\,\Box_\gamma^2  Y_\ai\right] .
\end{split}
\end{equation}
%

\subsubsection{The designer decoupled scalars parameterization}
\label{sec:vectorsActC}

Our final step is to convert the action for the vector perturbations in terms of the decoupled gauge invariant scalars $\MX_\ai$ and $\MY_\ai$. To achieve this we start from the contributions given in \eqref{eq:XYactioncoupled} for the bulk action and plug in the field redefinition \eqref{eq:XYredef}. We find after a bit of algebra the following decomposition:
\begin{equation}\label{eq:SXYdecoupleA}
	S^\text{V}_\text{EM,bulk} + S^\text{V}_\text{bdy} 
	= S_{\MX} + S_{\MY} +  S_{\MX\MY,\text{bdy}} + \frac{1}{2}\, \int_k dr \,\frac{1}{r^{d+1}}\, 	\left(\frac{\mathcal{E}_X}{f}\right)^2 \,.
\end{equation}	
Note that we have chosen not to transform the term proportional to $\mathcal{E}_X^2$ in the action, which however can be repressed using \eqref{eq:EXYmap} in terms of the fields $\MX$ and $\MY$. It will play no role in our variational principle and vanishes on-shell. 

The bulk action is 
\begin{equation}\label{eq:SXandSY}
\begin{split}
S_{\MX}
& = 
	-  
	\int_k\,\NX(\BQT)\,  \int dr\, \sqrt{-g}\, e^{\chi_{_\MX}} 
	\,\sum_{\ai=1}^{N_V}  \,k^2\left[ \nabla^A \MX_\ai \, \nabla_A\MX_\ai
	-  \BQT^2\, \bRQ^2\,\frac{r_+^2}{r^2}\,  (1-h) \,  \MX_\ai \MX_\ai\right] , \\
S_{\MY}
& = 
	-  \int_k\,\NY(\BQT)\,  \int dr\, \sqrt{-g}\, e^{\chi_{_\MY}} \,\sum_{\ai=1}^{N_V} \,k^2
	\left[ \nabla^A \MY_\ai \, \, \nabla_A\MY_\ai
	+  \BQT^2\, \bRQ^2\,\frac{r_+^2}{r^2}\, (1-h) \, \MY_\ai\, \MY_\ai\right]	,
\end{split}
\end{equation}
with dilatons
\begin{equation}\label{eq:dilXYA}
e^{\dilX}= \frac{1}{r^{2(d-1)}} \,, \qquad e^{\dilY} = \frac{h^2}{r^2} \,,
\end{equation}	
corresponding to a non-Markovian fall-off with $\ann = (1-d)$ and a Markovian fall-off $\ann = d-3$, respectively. The normalization factors are given in \eqref{eq:knormXY} and reproduced here for convenience (though cf., \cref{fn:ceffGN}):
\begin{equation}\label{eq:knormXYa}
\begin{split}
\NX(\BQT) &=  (\BQT^2+1)\, (\BQT^2+2)\,,\\ 
\NY(\BQT) &=  \frac{1}{\RQ^{2(d-2)}} \,\BQT^2\,(\BQT^2+1)\,.
\end{split}
\end{equation}  
In deriving these expressions it is useful to use \eqref{eq:BQTdef} to write 
\begin{equation}\label{eq:kBQTrel}
\frac{k^2}{(d-2)^2} = \frac{\BQT^2 \, (\BQT^2+2)}{2}\, \frac{\mu^2}{\sdc^2} \,. 
\end{equation}	

The complication is hidden in the boundary term $k^2 X_\ai \Dz_+ X_\ai$  in the second line of  
\begin{equation}\label{eq:LbdyXY}
\begin{split}
\frac{1}{r_+^2}\, L_{\MX\MY,\text{bdy}} 
&= 
	-\left( \BQT^2+2 \right)^2 \, \bqt^2  \, \cpen{\MX}^\ai \, \MX_\ai - \left( \frac{\BQT^2}{\RQ^{d-2}} \right)^2 \,  \bqt^2  \, \cpen{\MY}^\ai \, \MY_\ai + \frac{(d-2)}{\bRQ^2}\, \bqt^4 \, \frac{f}{r^{d-2}}\,\MX_\ai^2 \\
& \qquad 
	+ \frac{(d-2)\, \BQT^2}{2\,\RQ^{d-2}} \, \bqt^2\, f \, h  \left(\frac{h}{1-h}\BQT^2 - 1 \right) \, \MY_\ai^2 - 
	2\, \frac{d-2}{\bRQ^2}\, \bqt^4\, \frac{f}{r^{d-2}} \, \MX_\ai\, \MY_\ai\\
& \qquad
	+ \frac{ \BQT^2 (\BQT^2+2) }{\RQ^{2(d-2)}} \,  \bqt^2\left(\frac{1-h}{h}\, \MX_\ai\, \cpen{\MY}^\ai + \frac{h}{1-h} \,\RQ^{2(d-2)}\,  \MY_\ai\, \cpen{\MX}^\ai\right) .
\end{split}
\end{equation}	
where we defined the conjugate momenta
\begin{equation}\label{eq:XYconjmom}
\cpen{\MX}^\ai = -r^{1-d}\, \Dz_+\MX_\ai \,, \qquad
\cpen{\MY}^\ai = - r^{d-3}\, h^2\, \Dz_+\MY_\ai \,.
\end{equation}	

The final piece of information we need is the counterterm action which can be obtained by substituting the field redefinition into \eqref{eq:XYcoupledCterms}. This is also a bit involved and reads: 
\begin{equation}\label{eq:LctermsXY}
\begin{split}
L^\text{V}_\text{ct} 
&=
	-\frac{\bqt^2 \, r_+^2}{2}\, \frac{r^{d-2}}{\RQ^{2(d-2)}} 
	\left[\frac{\RQ^{2(d-2)}}{d-2}\, 
		\cpen{\MX}^\ai \, \mathbb{O}_{\text{ct},1} \cpen{\MX}^\ai +  \MY_\ai \, \mathbb{O}_{\text{ct},2}  \MY_\ai -  2\,\RQ^{d-2} \, \cpen{\MX}^\ai \, \mathbb{O}_{\text{ct},3}  \,\MY_\ai \right] ,\\
\mathbb{O}_{\text{ct},1}
&= 
	  ( \BQT^2+2)^2  + \frac{1}{(d-2)(d-4)}\left( (\BQT^2+2)^2 +  \frac{\bqt^2}{\bRQ^2} \right)\Box_\gamma
	+ \frac{\bqt^2}{(d-2)\,(d-4)^2\,(d-6) \bRQ^2} \, \Box_\gamma^2 \,, \\
\mathbb{O}_{\text{ct},2}
&=
	 (d-2) \BQT^4 + \frac{1}{d-4} \left( \BQT^4 + \frac{\bqt^2}{\bRQ^2} \right) \Box_\gamma + \frac{\bqt^2}{(d-4)^2\, (d-6) \bRQ^2 }\, \Box_\gamma^2	 \,,\\
\mathbb{O}_{\text{ct},3}
&=
	\BQT (\BQT^2+2) + \frac{1}{(d-2)(d-4)} \left( \BQT^2 (\BQT^2 + 2) - \frac{\bqt^2}{\bRQ^2}\right) \Box_\gamma - \frac{\bqt^2}{(d-2)\,(d-4)^2\, (d-6) \bRQ^2 }\, \Box_\gamma^2\,.
\end{split}
\end{equation}

\paragraph{A variational principle  for the decoupled scalars:}

In \cref{sec:vectorsActC} we argued that the boundary conditions for the fields $X_\ai$ and $Y_\ai$ were Neumann and Dirichlet, respectively, as was clear from the boundary terms which dictate the variational principle. Let us understand how this translates to the decoupled scalars $\MX_\ai$ and $\MY_\ai$ and use it to derive the on-shell action.

To translate the boundary conditions from $(X_\ai,Y_\ai)$ to $(\MX_\ai, \MY_\ai)$ we record the relation between the time-reversal derivatives:  
\begin{equation}\label{eq:DzXYmap}
\begin{split}
\Dz_+ X_\ai 
&= 
	-(\BQT^2+2)\, \Dz_+ \MX_\ai + \BQT^2 \, \frac{h}{1-h}\, \Dz_+ \MY_\ai 
	+(d-2) \,\BQT^2\, \frac{r f }{1-h}\, \MY_\ai\,, \\
\Dz_+Y_\ai 
&=
	 (1-h) \,\Dz_+\MX_\ai + h\, \Dz_+\MY_\ai - (d-2)\,r f\, (1-h)\,  (\MX_\ai -\MY_\ai) 	 \,.	 
\end{split}
\end{equation}
To derive this expression we used the modified Leibniz rule for the operator $\Dz_+$; assuming that $\mathfrak{g}$ is a function only of $r$, but $\mathfrak{f} = \mathfrak{f}(\omega,r)$ we have in Fourier domain
\begin{equation}\label{eq:DzLieb}
\Dz_+( \mathfrak{f}\, \mathfrak{g})= r^2 f \left(\mathfrak{f} \, \mathfrak{g}' + \mathfrak{f}' \, \mathfrak{g} \right) - i\omega\, \mathfrak{f}\, \mathfrak{g} = \mathfrak{g} \, \Dz_+\, \mathfrak{f} + r^2 f\, \mathfrak{f} \, \mathfrak{g}'\,.
\end{equation}	

Given the definition of the conjugate momenta \eqref{eq:XYconjmom} the boundary conditions $\delta Y_\ai =0$ and $\delta \cpen{X}^\ai =0$ translate to 
\begin{equation}\label{eq:XYbcs}
(1-h)\, \delta \MX_\ai + h\, \delta\MY_\ai =0 \,, \qquad 
(\BQT^2+2)\,\RQ^{d-2} \, \delta \cpen{\MX}^\ai - \frac{\BQT^2}{r^{d-2}\, h} \, \delta \cpen{\MY}^\ai + (d-2)\, \BQT^2\, f\, \delta \MY_\ai =0\,.
\end{equation}	
If we work with a fixed radial cut-off then these boundary conditions would indeed be the correct ones to impose to ensure that we obtain the equations of motion \eqref{eq:XYfinal} by varying the action \eqref{eq:SXYdecoupleA}. 

As we are interested working at the asymptotic boundary we can however exploit the asymptotic behaviour to simplify our boundary conditions. We start by noting that the general behaviour of Markovian and non-Markovian fields in the $r\to \infty$ limit discussed in \cite{Ghosh:2020lel} and reviewed briefly in \cref{sec:probeM} and \cref{sec:probenM}, respectively, gives us the asymptotic relations: 
\begin{equation}\label{eq:XYpXYrel}
r\to \infty: \quad \MX_\ai = -\frac{r^{d-2}}{d-2} \,  \cpen{\MX}^\ai \,, \qquad
 \cpen{\MY}^\ai = \frac{\omega^2-k^2}{d-4}\, r^{d-4}\, \MY_\ai \,.
\end{equation}	
If we plug this into \eqref{eq:XYbcs} and retain the terms that contribute as we remove the radial cut-off we find that the fields $\MX_\ai$ and $\MY_\ai$ are constrained to satisfy:
\begin{equation}\label{eq:}
 \delta \cpen{\MX}^\ai- \frac{d-2}{\RQ^{d-2}} \, \delta\MY_\ai =0 \,, \qquad 
(\BQT^2+2)\, \delta \cpen{\MX}^\ai + \frac{d-2}{\RQ^{d-2}} \, \BQT^2\, \delta \MY_\ai  =0\,.
\end{equation}	
Since the coefficients are constants, we therefore learn that asymptotically one has to impose Neumann boundary conditions on $\MX_\ai$ and Dirichlet boundary condition on $\MY_\ai$, consistent with the intuition one would have from the Makovianity properties following from the dilatonic modulation \eqref{eq:dilXY}.

\section{Probe Markovian scalars}
\label{sec:probeM}

As explained in \cite{Ghosh:2020lel} the prototype problem we need to study to understand the real-time correlation functions is that of a designer scalar, coupled to gravity with an coupling that is modulated across energy scales holographically. Focusing first on the case of scalars which have no long-lived modes, the Markovian scalars, we characterize them by a Markovianity index $\ann > -1$. Consider the following probe action:
\begin{equation}\label{eq:probeMAct}
\begin{split}
S_\ann 
&=
	-\frac{1}{2}\, \int d^{d+1}x \, \sqrt{-g}\, e^{\dils}\, g^{AB}\, \nabla_A\sen{\ann} \,\nabla_B\sen{\ann} 
	+ S_{\ann,\text{ct}} \,, \\  
 S_{\ann,\text{ct}} 
 &= 
 	\frac{1}{2} \int d^dx\, \sqrt{-\gamma} \, e^{\dils} 
 	\left[ \frac{1}{1-\ann} \,\sen{\ann} \Box_\gamma \sen{\ann} +
 		\frac{1}{(1-\ann)^2(3-\ann)} \,\sen{\ann} \Box_\gamma^2 \sen{\ann} 
 	\right],
\end{split}
\end{equation}	
$\dils$ here denotes the designer dilaton which is characterized by its asymptotic behaviour. For Markovianity index $\ann$ we require
\begin{equation}\label{eq:Mchidef}
\lim_{r\to \infty} e^{\dils} \to r^{\ann + 1-d} \,, \qquad \ann > -1 \,. 
\end{equation}	
In our analysis we encounter two Markovian scalars:
\begin{itemize}[wide,left=0pt]
\item  the field $\tGR_\bi$ which has $e^{\dils} = 1$ (viz., $\ann =d -1$) from the tensor polarization of gravitons, and 
\item  the field $\MY_\ai$ which has a $e^{\dils} = \frac{h^2}{r^2} $ (hence $\ann = d-3$), the decoupled mode, from the vector perturbations. 
\end{itemize} 

For the rest of the discussion, we will take $e^{\dils}$ to be a simple monomial as dictated by the asymptotic behaviour in  \eqref{eq:Mchidef}. This can easily be generalized to more involved dilaton profiles with suitable modifications.  When $e^{\dils} = r^{\ann+1-d}$ the dynamics of the probe designer scalar $\sen{\ann}$ takes the form
\begin{equation}\label{eq:MprobeEq}
\Dann \sen{\ann} =0 \,,
\end{equation}	
where $\mathfrak{D}_\ann$ is defined to be the following operator designer scalar wave operator in Fourier domain:
\begin{equation}\label{eq:DesOp}
\Dann  = r^{-\ann} \Dz_+ \left(r^\ann\, \Dz_+ \right) + (\omega^2 - k^2 f) \,.
\end{equation}	
This defines a time-reversal invariant system on the grSK geometry.  For future reference we introduce here the  normal derivatives and conjugate momenta for radial evolution:
\begin{equation}\label{eq:normalD}
\partial_n\sen{\ann} = n^A\, \nabla_A \sen{\ann}  = \frac{1}{r\,\sqrt{f}} \, \Dz_+ \sen{\ann}\,, \qquad 
\cpen{\ann} = -\sqrt{-\gamma}\, e^{\chi_s}\, \partial_n\,\sen{\ann}  = - r^{\ann}\, \Dz_+ \sen{\ann}\,.
\end{equation}	

As described in \cite{Jana:2020vyx} such systems are simply solved by first evaluating the ingoing bulk-boundary Green's function and thence using the time-reversal isometry of the grSK geometry to obtain the outgoing Green's function.  We describe below the solutions for the ingoing Green's function in the boundary gradient expansion. 

\paragraph{Solutions in gradient expansion:} For the purposes of writing the solutions, it will be helpful to work with a dimensionless coordinate $\ri$\footnote{Readers comparing the expressions here with those in \cite{Ghosh:2020lel} should note that the latter works with the dimensionless coordinate $\xi = \frac{1}{\ri}$. } 
\begin{equation}\label{eq:inverserad}
\ri \equiv \frac{r_+}{r}
\end{equation}	
and rewrite  \eqref{eq:MprobeEq} using \eqref{eq:dimlesswk} as
\begin{equation}\label{eq:MeqA}
\begin{split}
\ri^{\ann+1} \, \dv{\ri}(\ri^{-\ann} \, f\, \dv{\sen{\ann}}{\ri}) + i\, \bwt\, \left( 2\ri\, \dv{\sen{\ann}}{\ri} - \ann\, \sen{\ann} \right) -\bqt^2\, \ri\, \sen{\ann} 
&=0 \,.
\end{split}
\end{equation}	
Since we will exclusively use $\ri$ in the appendices note that we therefore will write the background functions in terms of $\ri$, i.e., 
\begin{equation}\label{eq:bgfnsri}
f(\ri) = 1 - (1+Q^2)\,\ri^d + Q^2 \, \ri^{2(d-1)} \,, \qquad h(\ri) = 1 - \sdc\, \ri^{d-2}\,.
\end{equation}	

Our goal is to solve \eqref{eq:MeqA} order by order in gradients along the boundary directions, viz., in Taylor series in $\bwt$ and $\bqt$.  At zeroth order  we have the simple solution
\begin{equation}\label{eq:Masym}
\begin{split}
\sen{\ann}^{(0)} 
&= 
	c_a + c_m \int_{\ri_c + i0}^{\ri} \, \frac{d\rib}{f(\rib)}\, \rib^\ann  \,.
\end{split}
\end{equation}	
 Here $\ri_c = \frac{r_+}{r_c}$ is the location of a radial cutoff surface chosen to regulate the solution, cf., \cref{fig:mockt}. The ingoing boundary condition for the field requires $c_m = 0$. We will further normalize the source of the field $\sen{\ann}$ to unity at the boundary by setting $c_a =1$. 

 The ingoing Green's function with unit source can therefore be parameterized as\footnote{
We have chosen to write the gradient expansion ansatz  somewhat differently from the form employed in \cite{Ghosh:2020lel}.  The exponential ansatz allows us to isolate the contributions from lower order terms more effectively (this was already noted in \cite{Chakrabarty:2019aeu}).}
\begin{equation}\label{eq:Ginexp}
\begin{split}
\Gin{\ann}(\ri,\bwt, \bqt)  
&= 
	\exp\left(\sum_{n,m=1}^\infty \,  (-i)^m \,\bwt^m\, \bqt^{2n}\, \Mser{\ann}{m,2n}(\ri) \right) ,
\end{split}
\end{equation}	
where we have exploited the spatial parity symmetry which ensures that the odd powers of $\bqt$ vanish.
It is easy to see that the general solutions to the equation at each order in the gradient expansion will take the form 
\begin{equation}\label{eq:Fgradgen}
\mathfrak{F}(\ri) = \int_0^\ri \, \frac{d\rib}{f(\rib)} \, \rib^\ann \, \int_1^\rib \, \mathfrak{J}(\rib') d\rib' \,, 
\end{equation}	
for some source function $\mathfrak{J}(\ri)$ obtained from the lower order terms in the gradient expansion.  The boundary conditions for the ingoing Green's function require that 
\begin{equation}\label{eq:}
\lim_{\ri \to 0}\Mser{\ann}{m,n} (\ri) = 0\,, \qquad  \dv{\ri} \Mser{\ann}{m,n} (\ri=1) = \text{regular} \,.
\end{equation}	
In writing the expressions we will employ the notation 
\begin{equation}\label{eq:horsub}
\hat{\mathfrak{F}}(\ri) \equiv \mathfrak{F}(\ri) - \mathfrak{F}(1) \,,
\end{equation}	
to denote the function with its value at the horizon subtracted out to enforce the aforesaid boundary conditions.\footnote{ All functions in the appendices evaluated at $\ri =1$ correspond to horizon values, and as noted before are functions of $\sdc$ since we have extracted overall dimensions with appropriate powers of $r_+$.  In the main text we will write the these constants as function values at $r=r_+$ as in \eqref{eq:KinY}, \eqref{eq:KinX}, \eqref{eq:KinMarkT}, etc.}

The solutions for the functions $\Mser{\ann}{m,n}$ up to fourth order in the gradient expansion can be written down in terms of a single integral by introducing a set of auxiliary functions. These will be important for analytic continuation to the non-Markovian regime $\ann < -1$. In fact, one can arrange the computation so that all the functions we encounter are integral transforms with respect to the kernel $\frac{1}{f}$. 
We define: 
\begin{equation}\label{eq:MItransform}
\mathfrak{T}\big[\mathfrak{g}\big](\ri) \equiv \int_0^\ri\, \frac{d\rib}{f(\rib)}\, \mathfrak{g}(\rib) , \ \hspace{0.5cm} 
\hat{\mathfrak{T}}\big[\mathfrak{g}\big](\ri) \equiv \int_1^\ri\, \frac{d\rib}{f(\rib)}\, \mathfrak{g}(\rib) \,.
\end{equation}	
We then have:
\begin{equation}\label{eq:DeltaFns}
\begin{split}
\Dfnh{\ann}{2,0} (\ri) 
& = 
	\int_1^{\ri} \, \frac{d\rib}{f(\rib)}  \left[ \rib^\ann - \rib^{-\ann}\right]
	= \hat{\mathfrak{T}}\big[\ri^\ann - \ri^{-\ann}\big] \,,\\
\Dfnh{\ann}{1,2} (\ri) 
& = 
	- \int_1^{\ri} \, \frac{d\rib}{f(\rib)} \, \rib\, \Dfnh{\ann}{2,0}(\rib) 
	=  -\hat{\mathfrak{T}}\big[\ri\, \Dfnh{\ann}{2,0}(\ri) \big]  \,,\\
\Dfnh{\ann}{3,0} (\ri) 
& = 
	- \int_1^{\ri} \, \frac{d\rib}{f(\rib)} \, \rib^{\ann}\, \Dfnh{\ann}{2,0}(\rib)^2 	
	= - \hat{\mathfrak{T}}\big[\ri^{\ann}\, \Dfnh{\ann}{2,0}(\ri)^2 \big] \,.
\end{split}
\end{equation}
Near the boundary, these functions have the following asymptotic behavior
\begin{equation}\label{eq:DeltaDiv}
\begin{split}
\Dfn{\ann}{2,0} (\ri) 
& = 
	-\frac{1}{1-\ann}\ri^{1-\ann} + \frac{1}{1+\ann}\ri^{1+\ann} \,,
	\\
\Dfn{\ann}{1,2} (\ri) 
& =
	 \frac{1}{(3-\ann)(1-\ann)}\ri^{3-\ann}-\frac{1}{(3+\ann)(1+\ann)}\ri^{3+\ann}+\frac{\Dfn{\ann}{2,0}(1)}{2}\ri^2\,,
	 \\
\Dfn{\ann}{3,0} (\ri) 
& = 
	-\frac{1}{(3-\ann)(1-\ann)^2}\ri^{3-\ann} + \frac{2}{(3+\ann)(1-\ann^2)}\ri^{3+\ann} - \frac{\Dfn{\ann}{2,0}(1)}{1-\ann}\ri^2 \\ 
&\qquad 
	-\frac{\Dfn{\ann}{2,0}(1)^2}{1+\ann}\ri^{1+\ann} \,.
\end{split}
\end{equation}
Armed with these auxiliary functions one can give compact integral representations for the functions appearing in the gradient expansion. The solutions to all orders then can be written as $\Mser{\ann}{n,m}(\ri) = \mathfrak{T} \big[\mathfrak{g}\big]$. The explicit expressions up to the fourth order in gradient expansion are tabulated in \cref{tab:Mgradsol}.

\renewcommand{\arraystretch}{1.8}
\begin{table}[th!]
\centering
\begin{tabular}{|c|c|c|}
\hline
\shadeR{$\mathfrak{T}\big[\mathfrak{g}\big]$}	&	\shadeB{$\mathfrak{g}$ }		&		\shadeB{Asymptotics} \\		\hline

$\Mser{\ann}{1,0}$ 	& 
	$1 - \ri^{\ann}$ 	&  
		$\ri - \frac{\ri^{1+\ann}}{1+\ann} $ \\ \hline

$\Mser{\ann}{0,2}$ 	& 
	$\frac{\ri}{1-\ann}\left(1-\ri^{\ann-1}\right)$  	&  
			$\frac{1}{1-\ann}\left(\frac{\ri^2}{2} -\frac{\ri^{1+\ann}}{1+																																	\ann}\right)$ \\ \hline

$\Mser{\ann}{2,0}$ 	& $-\ri^{\ann}\Dfnh{\ann}{2,0}(\ri)$ 						  	& $\frac{1}{2(1-\ann)}\ri^{2} + \frac{\Dfn{\ann}{2,0}(1)}{1+\ann}																																	\ri^{1+\ann}$ \\ \hline

$\Mser{\ann}{3,0}$	& $2\ri^{\ann}\Mserh{\ann}{2,0}(\ri)$							& $-\frac{2\Mser{\ann}{2,0}(1)}{1+\ann}\ri^{1+\ann}+																																					\frac{\ri^{3+\ann}}{(3+\ann)(1-\ann)}$ \\ \hline

$\Mser{\ann}{1,2}$ 	& $2\ri^{\ann}\Mserh{\ann}{0,2}(\ri)$ 						&  $-\frac{2\Mser{\ann}{0,2}(1)}{1+\ann}\ri^{1+\ann}+																																					\frac{\ri^{3+\ann}}{(3+\ann)(1-\ann)}$ \\ \hline

$\Mser{\ann}{4,0}$	& 	$2\ri^{\ann}\left(\Mserh{\ann}{3,0}(\ri)+\frac{1}{2}\Dfnh{\ann}{3,0}(\ri)\right)$ 
	& 
		\scriptsize $-\frac{\ri^4}{4(3-\ann)(1-\ann)^2}-\frac{\Dfn{\ann}{3,0}(1)+
		2\Mser{\ann}{3,0}(1)}{1+\ann}\ri^{1+\ann}$\normalsize 
											\scriptsize $-\frac{\Dfn{\ann}{2,0}(1)}{(3+\ann)(1-\ann)}																																			\ri^{3+\ann}$\normalsize	\\ \hline

$\Mser{\ann}{2,2}$	& \scriptsize$2\ri^{\ann}\left(\Mserh{\ann}{1,2}(\ri)-\frac{1}{1-\ann}\left(\Dfnh{\ann}{1,2}(\ri)-\Mserh{\ann}{2,0}(\ri)\right)\right)$\normalsize																				&	\scriptsize$\frac{1}{2(\ann-3)(1-\ann)^2}\ri^{4}+\frac{1-(1-																																		\ann)\Dfn{\ann}{2,0}(1)}{(3+\ann)(1-\ann)^2}\ri^{3+\ann}$																																			\normalsize \\ 
										&																					& \scriptsize $-\frac{2\left(\Mser{\ann}{2,0}(1)-\Dfn{\ann}{1,2}																																		(1)+(1-\ann)\Mser{\ann}{1,2}(1)\right)}{(1+\ann)(1-\ann)}																																			\ri^{1+\ann}$ \normalsize\\ \hline

$\Mser{\ann}{0,4}$	& $\frac{1}{1-\ann}\left(\Mserh{\ann}{0,2}(\ri)-\Mserh{2-\ann}{0,2}(\ri)\right)$ 
	& 
		\scriptsize $-\frac{\ri^4}{4(3-\ann)(1-\ann)^2}-\frac{\Mser{\ann}{0,2}(1)
			-\Mser{2-\ann}{0,2}(1)}{1-\ann^2}			\ri^{1+\ann}+\frac{\ri^{3+\ann}}{(3+\ann)(1-\ann)^2}$																																			\normalsize \\ \hline
\end{tabular}
\caption{The functions appearing in the gradient expansion of the Markovian $\sen{\ann}$ up to the fourth order in gradients, given in the form of an integral transform defined in Eq.~(\ref{eq:MItransform}). We  also present the leading asymptotic behaviour of the functions which is used for computing boundary observables.}
\label{tab:Mgradsol}
\end{table}

\paragraph{Renormalized Green's function:}
The on-shell action on the solution may be evaluated directly given this data. We find 
\begin{equation}\label{eq:pMosact}
\begin{split}
S_\ann \bigg|_\text{finite}
&=
	 \lim_{\ri \to 0} \frac{1}{2} \int\, d^d x\, \sqrt{-\gamma}\, \ri^{d-1-\ann} \left[
	 	-\frac{\ri}{f(\ri)} \, \sen{\ann} \Dz_+ \sen{\ann} + \frac{1}{1-\ann}\, \sen{\ann} \Box_\gamma\sen{\ann} 
	 	\right.\\
&\left.\hspace{5cm}	
	+ \frac{1}{(1-\ann)^2(3-\ann)}\, \sen{\ann}\Box_\gamma^2 \sen{\ann}\right] .
\end{split}
\end{equation}	
Plugging in our solution in the gradient expansion we find the following expression for the renormalized ingoing Green's function to quartic order in gradients:
\begin{equation}\label{eq:KinMark}
\begin{split}
\Kin{\ann}(\omega, \bk) 
&=
	-\lim_{r\to \infty} \cpen{\ann}\bigg|_\text{ren} \\
\frac{1}{r_+^{\ann+1}}\, \Kin{\ann}(\omega, \bk) 
& =
	-i\bwt + \frac{\bqt^2}{1-\ann} + \Dfn{\ann}{2,0}(1)\, \bwt^2 + 2i\,\Mser{\ann}{0,2}(1)\,   \bwt\,\bqt^2 -  2i\,\Mser{\ann}{2,0}(1)\, \bwt^3 \\
& \qquad
	+ 
		\left[\Mser{\ann}{0,2}(1)-\sen{2-\ann}^{0,2}(1)\right]\, \frac{\bqt^4}{1-\ann} - \left[2\, \Mser{\ann}{3,0}(1) - \Dfn{\ann}{3,0}(1)\right]\, \bwt^4 \\
& \qquad
	+ 2\left[\Mser{\ann}{1,2}(1) +\frac{\Mser{\ann}{2,0} (1)+ \Dfn{\ann}{1,2}(1) }{\ann-1}  \right]\, \bwt^2\, \bqt^2 \,.
\end{split}
\end{equation}	
%

\begin{figure}[h]
\centering
\subfloat{
\begin{minipage}[t]{0.5\textwidth}
\vspace{0.5cm}
\hspace{0pt}
\begin{tikzpicture}
  \node (img)  {\includegraphics[scale=0.275]{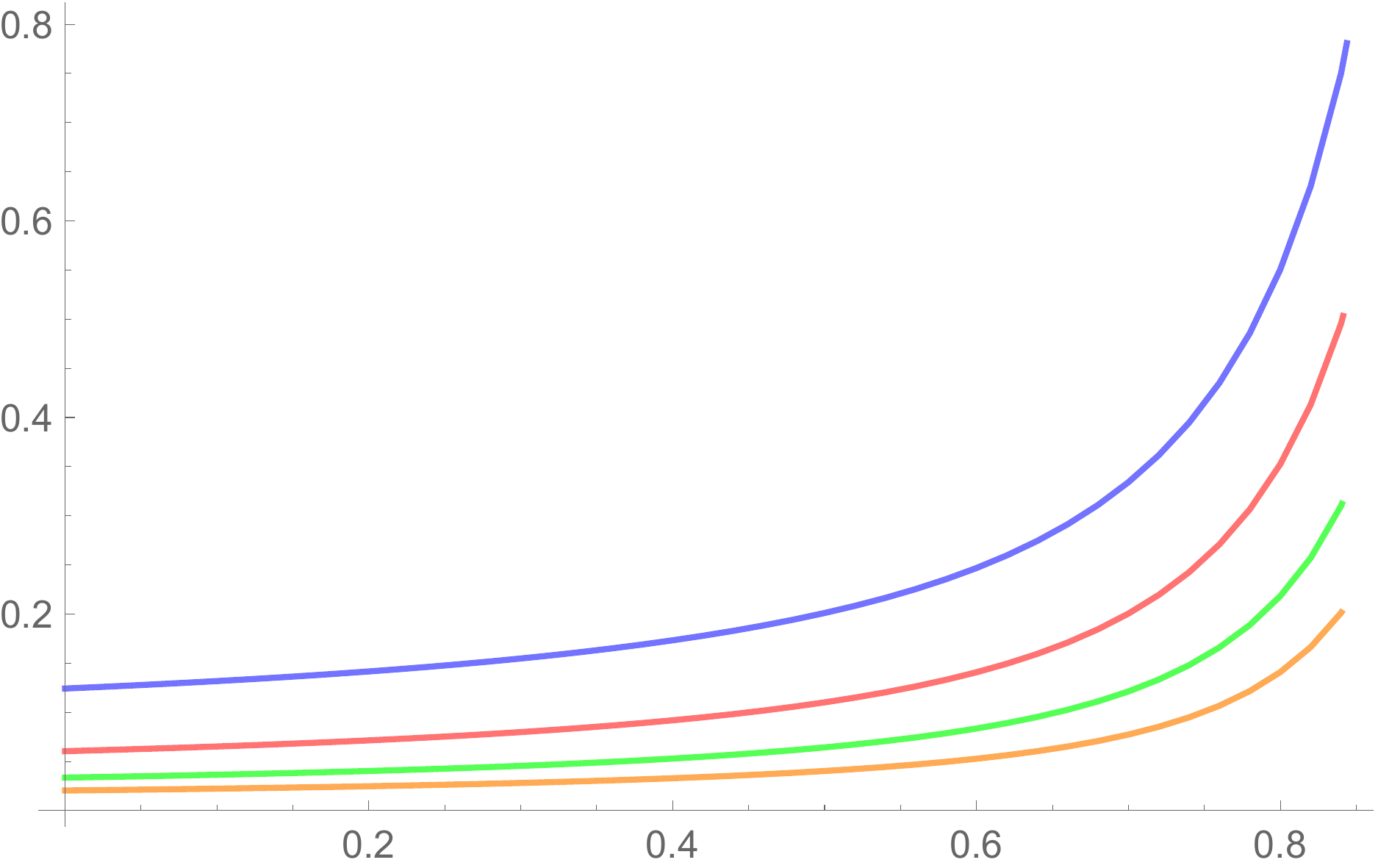}};
  \node[below=of img, node distance=0cm,font=\small,yshift=1.0cm] {$\sdc$};
  \node[left=of img, node distance=0cm, anchor=center,font=\footnotesize,xshift=0.5cm] {$\Mser{d-1}{3,0}(1)$};
  \node[right=of img, node distance=0cm,font=\scriptsize,xshift=-1.25cm,yshift=1.5cm] {$d=3$};
  \node[right=of img, node distance=0cm,font=\scriptsize,xshift=-1.25cm,yshift=0.6cm] {$d=4$};
  \node[right=of img, node distance=0cm,font=\scriptsize,xshift=-1.25cm,yshift=-0.15cm] {$d=5$};
   \node[right=of img, node distance=0cm,font=\scriptsize,xshift=-1.25cm,yshift=-0.65cm] {$d=6$};
 \end{tikzpicture}
 \end{minipage}
}
\subfloat{
\begin{minipage}[t]{0.5\textwidth}
 \vspace{0.5cm}
\hspace{0pt}
\begin{tikzpicture}
  \node (img)  {\includegraphics[scale=0.275]{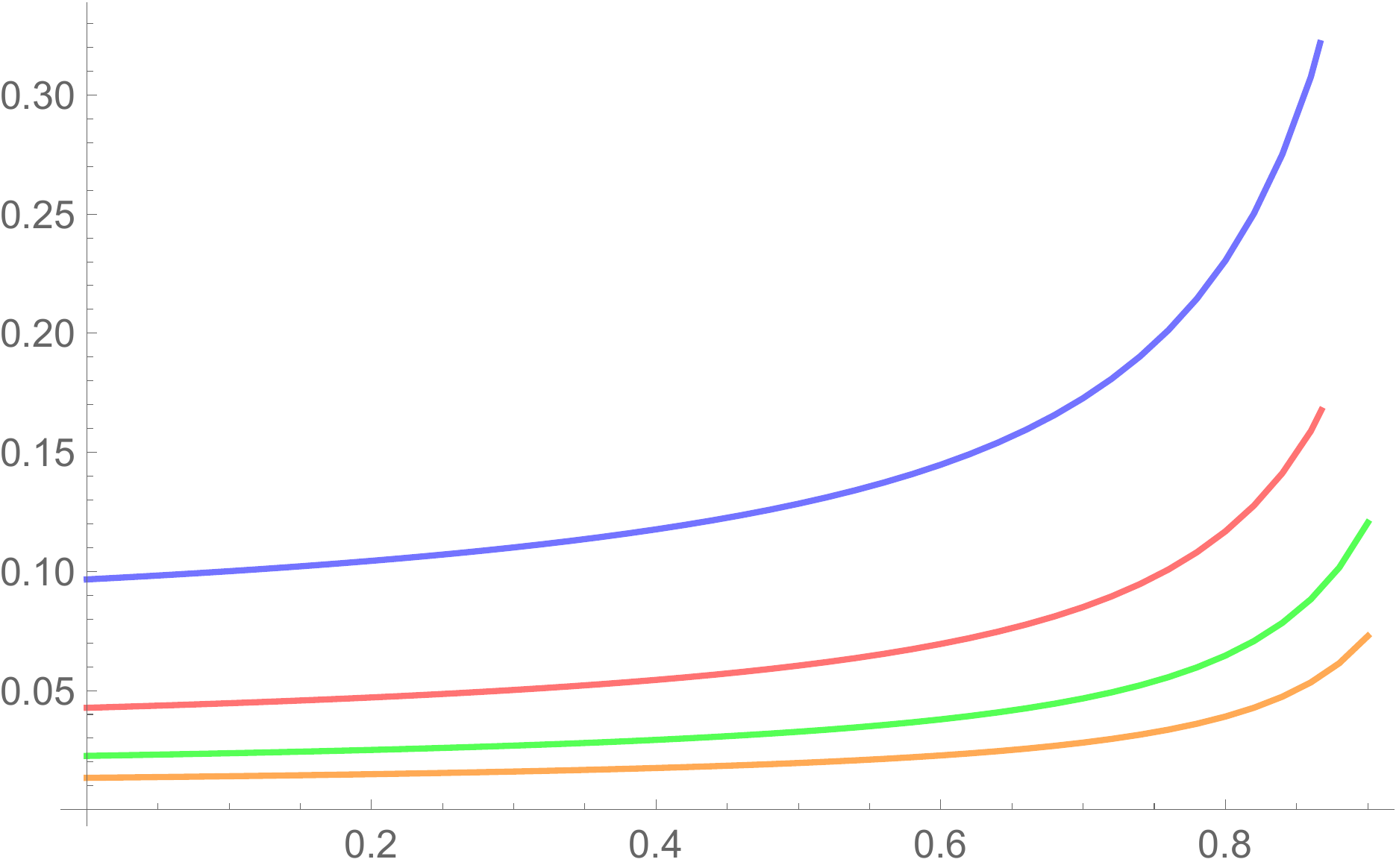}};
  \node[below=of img, node distance=0cm,font=\small,yshift=1.0cm] {$\sdc$};
  \node[left=of img, node distance=0cm, anchor=center,font=\footnotesize,xshift=0.5cm] {$\Mser{d-1}{1,2}(1)$};
  \node[right=of img, node distance=0cm,font=\scriptsize,xshift=-1.25cm,yshift=1.5cm] {$d=3$};
  \node[right=of img, node distance=0cm,font=\scriptsize,xshift=-1.25cm,yshift=0.25cm] {$d=4$};
  \node[right=of img, node distance=0cm,font=\scriptsize,xshift=-1.25cm,yshift=-0.25cm] {$d=5$};
   \node[right=of img, node distance=0cm,font=\scriptsize,xshift=-1.25cm,yshift=-0.75cm] {$d=6$};
 \end{tikzpicture}
 \end{minipage}
 }\\
\subfloat{
\begin{minipage}[t]{0.5\textwidth}
 \vspace{0.5cm}
\hspace{0pt}
\begin{tikzpicture}
  \node (img)  {\includegraphics[scale=0.275]{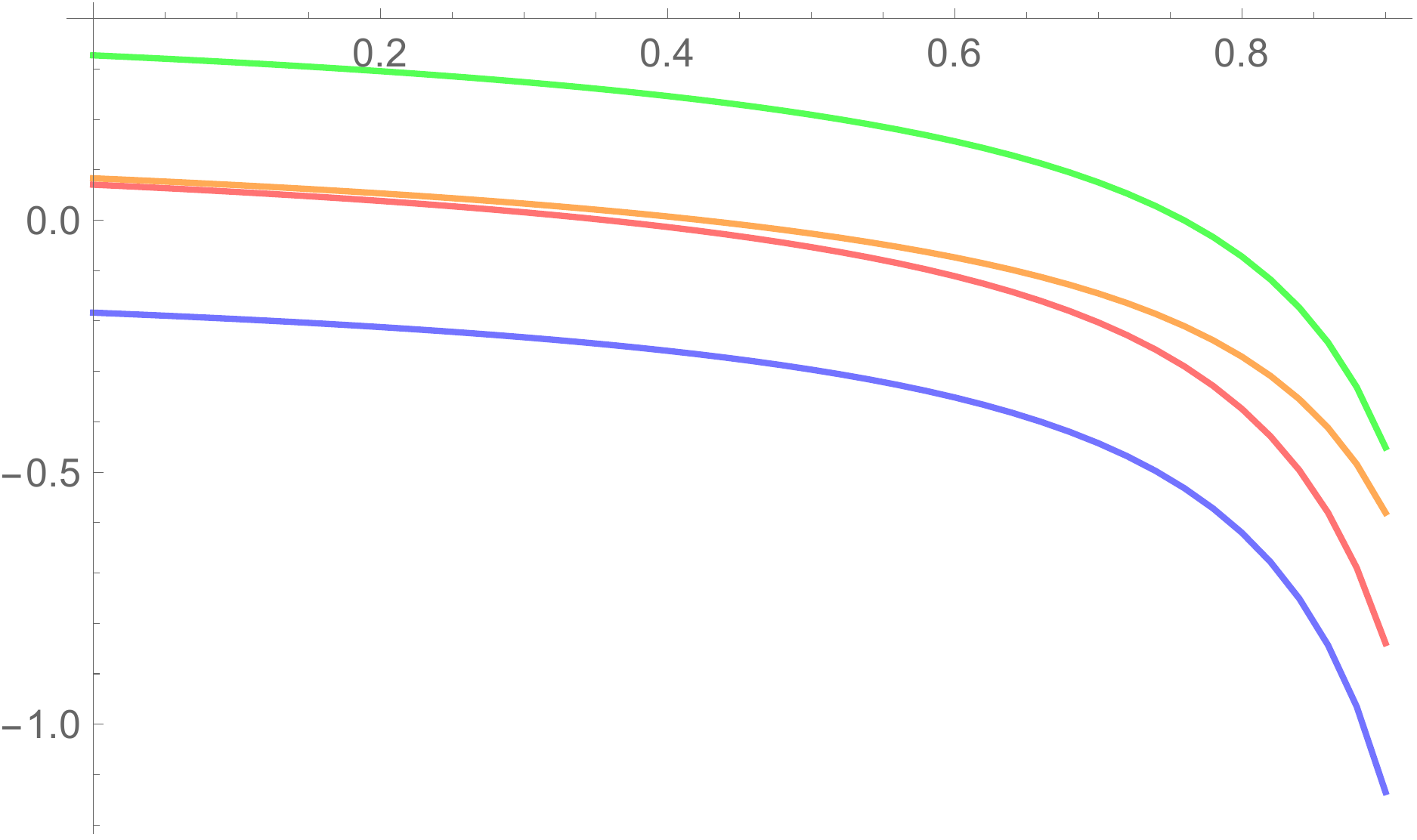}};
  \node[above=of img, node distance=0cm, font=\small,yshift=-1.2cm] {$\sdc$};
  \node[left=of img, node distance=0cm, anchor=center,font=\footnotesize,xshift=0.5cm] {$\Dfn{d-1}{1,2}(1)$};
  \node[right=of img, node distance=0cm,font=\scriptsize,xshift=-1.25cm,yshift=0.1cm] {$d=5$};
  \node[right=of img, node distance=0cm,font=\scriptsize,xshift=-1.25cm,yshift=-0.4cm] {$d=6$};
  \node[right=of img, node distance=0cm,font=\scriptsize,xshift=-1.25cm,yshift=-0.9cm] {$d=4$};
   \node[right=of img, node distance=0cm,font=\scriptsize,xshift=-1.25cm,yshift=-1.4cm] {$d=3$};
 \end{tikzpicture}
 \end{minipage}
 }
\subfloat{
\begin{minipage}[t]{0.5\textwidth}
 \vspace{0.5cm}
\hspace{0pt}
\begin{tikzpicture}
  \node (img)  {\includegraphics[scale=0.275]{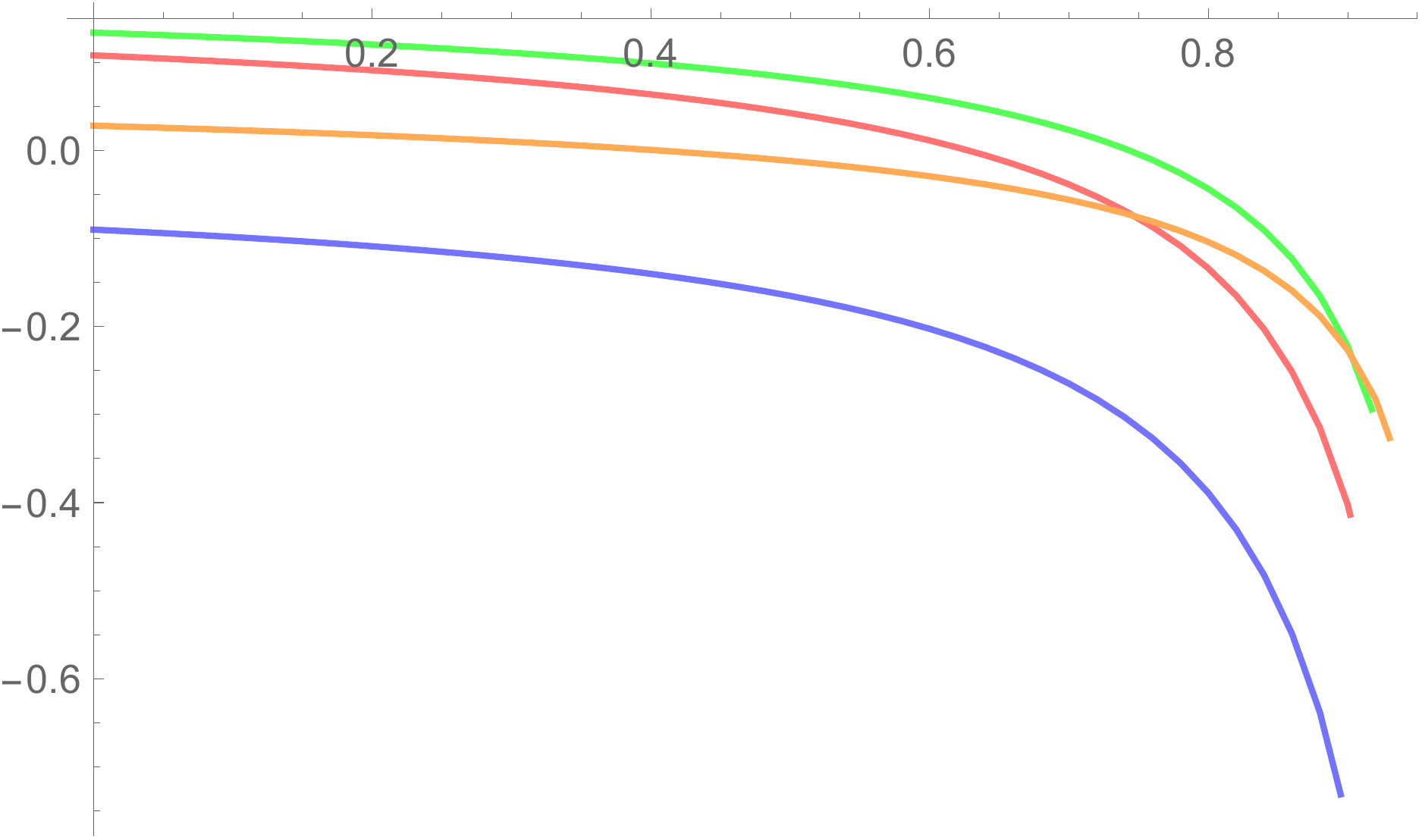}};
  \node[above=of img, node distance=0cm, font=\small,yshift=-1.2cm] {$\sdc$};
  \node[left=of img, node distance=0cm, anchor=center,font=\footnotesize,xshift=0.5cm] {$\Dfn{d-1}{3,0}(1)$};
  \node[right=of img, node distance=0cm,font=\scriptsize,xshift=-1.25cm,yshift=0.4cm] {$d=5$};
  \node[right=of img, node distance=0cm,font=\scriptsize,xshift=-1.25cm,yshift=-0.1cm] {$d=6$};
  \node[right=of img, node distance=0cm,font=\scriptsize,xshift=-1.25cm,yshift=-0.6cm] {$d=4$};
   \node[right=of img, node distance=0cm,font=\scriptsize,xshift=-1.25cm,yshift=-1.4cm] {$d=3$};
 \end{tikzpicture}
 \end{minipage}
 }
\caption{ The charge dependence of the coefficients at the quartic order Green's function of the Markovian tensor graviton polarizations and the non-Markovian momentum diffusion part of vector modes in  various dimensions, supplementing the lower order coefficients depicted in \cref{fig:KinMarT}.}
\label{fig:KinMarT4th}
\end{figure}
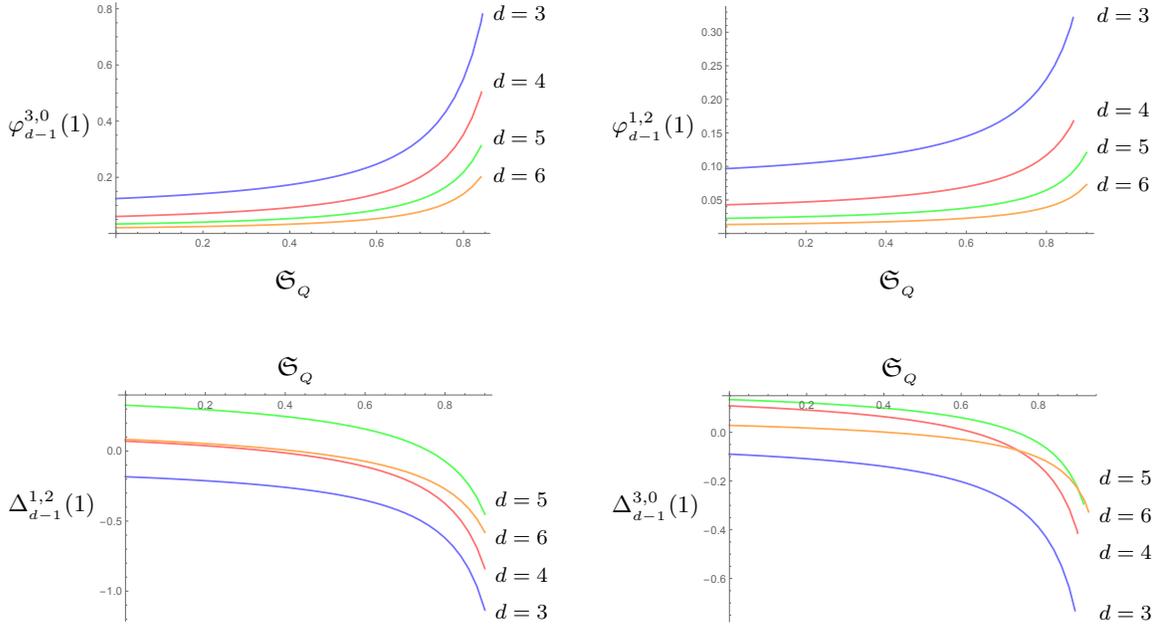

A special case of interest is the minimally coupled massless scalar which corresponds to $\ann = d-1$. For this case we have 
\begin{equation}\label{eq:KinMarkTA}
\begin{split}
\frac{1}{r_+^{d}}\, \Kin{d-1}(\omega, \bk) 
& =
  -i\,\bwt - \frac{\bqt^2}{d-2} + \Dfn{d-1}{2,0}(1)\, \bwt^2 + 2i\,\Mser{d-1}{0,2}(1)\, \bwt\,\bqt^2 -  2i\,\Mser{d-1}{2,0}(1)\, \bwt^3    \\
& \qquad
  - 
    \left[\Mser{d-1}{0,2}(1)-\sen{3-d}^{0,2}(1)\right]\, \frac{\bqt^4}{d-2} - \left[2\, \Mser{d-1}{3,0}(1) - \Dfn{d-1}{3,0}(1)\right]\, \bwt^4 \\
& \qquad
  - 2\left[\Mser{d-1}{1,2}(1) +\frac{\Mser{d-1}{2,0} (1)- \Dfn{d-1}{1,2}(1) }{d-2}  \right]\, \bwt^2\, \bqt^2 \,.
\end{split}
\end{equation}  
This completes the expression given in the main text in \eqref{eq:KinMarkT} to quartic order.

\section{Probe non-Markovian scalars}
\label{sec:probenM}

The non-Markovian fields are distinguished by the fact that their coupling to gravity is damped in the UV region.  These are described by designer scalars with a Markovianity index $\ann \leq  -1$. 
The analysis of \cite{Ghosh:2020lel} indicates that we can work with a probe action written for a field $\sen{-\ann}$ 
\begin{equation}\label{eq:probenMAct}
\begin{split}
S_{-\ann} 
&=
	-\frac{1}{2}\, \int d^{d+1}x \, \sqrt{-g}\, e^{\dils}\, g^{AB}\, \nabla_A\sen{-\ann} \,\nabla_B\sen{-\ann} 
	-\int \, d^dx\, \cpen{-\ann} \,\sen{-\ann} 
	+ S_{-\ann,\text{ct}} \,, \\  
 S_{-\ann,\text{ct}} 
 &= 
 	-\frac{1}{2} \int d^dx\, \sqrt{-\gamma} \, e^{\dils} 
 	\left[ \frac{1}{\ann-1} \,\left(\partial_n\sen{-\ann}\right)^2 +
 		\frac{1}{(\ann-1)^2(\ann-3)} \, \partial_n\sen{-\ann} \Box_\gamma \partial_n \sen{-\ann} 
 	\right] .{}
\end{split}
\end{equation}	
We will work with the convention that the sign of the Markovianity index is explicit (so $\ann$ above is positive and in fact will be taken to be greater than unity) and have written the boundary terms and counterterms in terms of the radial conjugate momentum and the normal derivative \eqref{eq:normalD}.

The main change is that there is now a variational boundary term given by $ \cpen{-\ann} \,\sen{-\ann}$ which implies that the field $\sen{-\ann}$ should be quantized with Neumann boundary conditions.  The reason is that  the solution for $\sen{-\ann}$, which we obtain below, has a solution that grows asymptotically (though there is a codimension-1 locus of normalizable modes), but this is mitigated in $\cpen{-\ann} = -\frac{1}{r^\ann} \, \Dz_+ \sen{-\ann}$. Once this is understood the rest of the counterterms can be fixed by standard procedure and one finds result given in \eqref{eq:probenMAct}.

\paragraph{Solutions in gradient expansion:} To solve the wave equation of the non-Markovian fields we will use a trick following \cite{Ghosh:2020lel} wherein we analytically continue the solutions for $\ann > 1$ to those for $\ann < -1$. This is one reason for introducing the functions $\Dfn{\ann}{m,n}$ in \eqref{eq:DeltaFns} earlier, which in fact help isolate the terms that diverge asymptotically. With this preamble, let us introduce the gradient expansion for the non-Markovian ingoing inverse Green's function: 
\begin{equation}\label{eq:GinexpN}
\begin{split}
\Gin{-\ann}(\ri,\omega, \bk)  
&= 
	\exp\left(\sum_{n,m=1}^\infty \,  (-i)^m \,\bwt^m\, \bqt^{2n}\, \Mser{-\ann}{m,2n}(\ri) \right)  .
\end{split}
\end{equation}	
Solving the wave equation order by order with this ansatz and fixing the arbitrary additive constant to be zero, we find up to the cubic order in derivatives the following: 
\begin{equation}\label{eq:nMgradsol}
\begin{split}
\Mser{-\ann}{1,0}(\ri)
&=
	\Mser{\ann}{1,0}(\ri)+\Dfn{\ann}{2,0}(\ri)  \,, \\ 
\Mser{-\ann}{2,0}(\ri)
&=
	-\Mser{\ann}{2,0} (\ri) -\frac{1}{2}\, \Dfn{\ann}{2,0}(\ri)^2 + \Dfn{\ann}{2,0}(1) \, \Dfn{\ann}{2,0}(\ri) \,,\\ 		
\Mser{-\ann}{0,2}(\ri)
&=
	\frac{1-\ann}{1+\ann} \Mser{\ann}{0,2}(\ri) + \frac{1}{1+\ann}\, \Dfn{\ann}{2,0}(\ri) \,, \\ 
\Mser{-\ann}{3,0}(\ri)
&=
	-\Mser{\ann}{3,0}(\ri) -2\Dfn{\ann}{2,0}(1)\Mser{\ann}{2,0}(\ri)-\Dfn{\ann}{3,0}(\ri)-\Dfn{\ann}{2,0}(1)\Dfn{\ann}{2,0}(\ri)^2+\frac{1}{3}\Dfn{\ann}{2,0}(\ri)^3 \\
&\quad
  +\Dfn{\ann}{2,0}(\ri)\left(2\Mserh{\ann}{2,0}(\ri)+\Dfn{\ann}{2,0}(1)^2\right)
	    \,,\\ 
\Mser{-\ann}{1,2}(\ri)
&= 
	\frac{1-\ann}{1+\ann} \left[ \Mser{\ann}{1,2}(\ri) + 2\, \Dfn{\ann}{2,0}(1)\, \Mser{\ann}{0,2}(\ri)
	 -2\, \Dfn{\ann}{2,0}(\ri)\, \Mserh{\ann}{0,2}(\ri)\right] \\
& \quad 
	-\frac{1}{1+\ann} \left[2 \Dfn{\ann}{1,2}(\ri)  - 2\Dfn{\ann}{2,0}(\ri)   \Dfn{\ann}{2,0}(1)+\Dfn{\ann}{2,0}(\ri)^2\right] .
\end{split}
\end{equation}	
The functions entering at the quartic order are more involved and are in turn given by 
\begin{equation}\label{eq:nMgradsolO4}
\begin{split}	
\Mser{-\ann}{4,0}(\ri)
&= 
-\Mser{\ann}{4,0}(\ri) - \Dfn{\ann}{2,0}(1)\left(2\Mser{\ann}{3,0}(\ri)+\Dfn{\ann}{3,0}(\ri)+2\Dfn{\ann}{2,0}(1)\Mser{\ann}{2,0}(\ri)\right)
	\\
	&\quad
	+\Dfn{\ann}{2,0}(\ri)\left[2\Mserh{\ann}{3,0}(\ri)+\Dfnh{\ann}{3,0}(\ri)-2\left(
	\Dfnh{\ann}{2,0}(\ri)-\Dfn{\ann}{2,0}(1)\right)\Mserh{\ann}{2,0}(\ri)\right]
	\\
	&\quad
	-\frac{1}{4}\left(\Dfnh{\ann}{2,0}(\ri)^4-\Dfnh{\ann}{2,0}(1)^4\right) ,
	   \\ 
\Mser{-\ann}{2,2}(\ri)
&=
 \frac{1-\ann}{1+\ann}\Mser{\ann}{2,2}(\ri) - \left[ \frac{1-\ann}{1+\ann} \Mser{\ann}{1,2}(\ri) - \Mser{-\ann}{1,2}(\ri)\right] \Dfnh{\ann}{2,0}(\ri) \\
 	&\quad 
 	+  \left[ \frac{1-\ann}{1+\ann} \Mser{\ann}{1,2}(1) - \Mserh{-\ann}{1,2}(1)\right] \Dfn{\ann}{2,0}(\ri) + \frac{1}{1+\ann} \left[ \Mser{\ann}{3,0}(\ri) - \Mser{-\ann}{3,0}(\ri) \right] \\
 	&\quad
 	- \frac{1}{3(1+\ann)}\left[ \Dfnh{\ann}{3,0}(\ri)^3 - \Dfn{\ann}{3,0}(1)^3 \right],
	\\
\Mser{-\ann}{0,4}(\ri)
&=
 	-\frac{1}{1+\ann} \ \int_0^\rho		d\rib \frac{\rib^{\ann}}{f(\rib)}\left(\Mserh{2+\ann}{0,2}(\rib)+\frac{\rib^2}{1+\ann}\Dfnh{\ann}{2,0}(\rib)\right)	-\frac{1}{1+\ann}\Dfn{\ann}{2,0}(1)\Mser{2+\ann}{0,2}(\ri)
 	\\
 	&\quad
 	+\frac{1}{1+\ann}\left(\frac{1}{2}\Mser{-\ann}{1,2}(\ri)+\Dfn{\ann}{2,0}(\ri)\Mserh{2+\ann}{0,2}(\ri)-\frac{1}{1+\ann}\Dfn{\ann}{1,2}(\ri)\right)	.	
\end{split}
\end{equation}	

The functions $\Mser{\ann}{m,n}(\ri)$ all vanish at infinity while the divergent terms arise from the $\Dfn{\ann}{m,n} (\ri)$. This makes it easy to extract the asymptotic behaviour of the functions which will be important to isolate the locus where we have a normalizable hydrodynamic mode.

\paragraph{The hydrodynamic locus:} As explained in \cite{Ghosh:2020lel} the inverse Green's function $\Gin{-\ann}(\ri, \omega,\bk)$ can be rewritten to expose the hydrodynamic moduli. One generically finds that the solution factorizes into a normalizable solution related to the Markovian problem with index $\ann$ and a piece that contains the non-normalizable mode. The coefficient of the latter is the dispersion function which defines the codimension-1 locus in $(\omega,\bk)$ where the non-normalizable piece is absent. We write: 
\begin{equation}\label{eq:GinparnM}
\Gin{-\ann}(\ri) = \widetilde{G}^\In_{_{-\ann}}(\ri) \, \exp\left( r_+^{\ann-1}\, \Kin{-\ann} \, \Xi_\text{nn}(\ri,\omega, \bk)\right)  .
\end{equation}	
The precise details of $\widetilde{G}^\In_{_{-\ann}}(\ri) $ and $\Xi_\text{nn}$ will not be relevant for our purposes, but we note that the former can be obtained directly from the corresponding Markovian problem. 
We will however need  the dispersion function $\Kin{-\ann}$ which can be obtained from the gradient expansion functions $\Mser{-\ann}{m,n}$. The result accurate to quartic order in derivatives reads:
\begin{equation}\label{eq:KdispM}
\begin{split}
r_+^{\ann-1}\, \Kin{-\ann}(\omega,\bk)
&=
	-i\,\bwt + \frac{\bqt^2}{1+\ann} - \Dfn{\ann}{2,0}(1)\, \bwt^2 + i\,\bwt^3 \left[\Dfn{\ann}{2,0}(1)^2 -2\, \Mser{\ann}{2,0}(1)\right] \\
&\quad
	-2i\,\frac{\bwt\, \bqt^2}{1+\ann} \left[  \Dfn{\ann}{2,0}(1) + (1-\ann) \, \Mser{\ann}{0,2}(1) \right] \\
&\quad
	- \bwt^4 \left[2\Mser{\ann}{3,0}(1) +4\, \Mser{\ann}{2,0}(1) \, \Dfn{\ann}{2,0}(1) -\Dfn{\ann}{2,0}(1)^3
	+\Dfn{\ann}{3,0}(1) \right] \\
&\quad
	-\frac{\bwt^2\, \bqt^2}{1+\ann} \left[2\, (1-\ann)\left(\Mser{\ann}{1,2}(1) + 2\, \Mser{\ann}{0,2}(1)\, \Dfn{\ann}{2,0}(1)\right) \right.\\
&\left.\qquad \qquad\quad
	+ \left(2\,\Mser{\ann}{2,0}(1) -3\, \Dfn{\ann}{2,0}(1)^2 + 2\, \Dfn{\ann}{1,2}(1)\right)
	\right]	 \\
&\quad 
	+ \frac{\bqt^4}{(1+\ann)^2} 
	\left[(1-\ann)\, \Mser{\ann}{0,2}(1) - (1+\ann) \Mser{\ann+2}{0,2}(1) + \Dfn{\ann}{2,0}(1)
	\right] .
\end{split}
\end{equation}

The vanishing locus of $\Kin{-\ann}$ determines for us the dispersion relation of the associated probe field in the bulk. Solving for $\omega(\bk)$ we find
\begin{equation}\label{eq:dispM}
\bwt = -i\, \frac{\bqt^2}{1+\ann}  +i\,\left[(1-\ann)\, \Mser{\ann}{0,2}(1) + (1+\ann)\, \Mser{\ann+2}{0,2}(1)\right]  \frac{\bqt^4}{(1+\ann)^2}  + \cdots \,.
\end{equation}	
The coefficient of the $\bqt^2$ term fixes the diffusion constant 
\begin{equation}\label{eq:diffusionM}
\mathcal{D}  = \frac{d-(d-2)\,Q^2}{4\pi \,T}\,  \frac{1}{ \abs{\ann}+1}  \,.
\end{equation}	

\paragraph{Comments on the neutral plasma:}
Specializing to a neutral \SAdS{d+1} black hole we find that the dispersion relation \eqref{eq:dispM} can be evaluated explicitly in terms of digamma functions (or equivalently in terms of Harmonic numbers). The general expression accurate to quartic order is 
\begin{equation}\label{eq:Sadsdisp}
\bwt = 
	 -i\, \frac{ \bqt^2 }{1+\ann}  -i\, \frac{\psi\left(\frac{3+\ann}{d}\right) - \psi\left(\frac{1+\ann}{d}\right)}{d\, (1+\ann)}\, \bqt^4 + \cdots \,.
\end{equation}	
Further specializing to the case of momentum diffusion $\abs{\ann} = d-1$ this simplifies considerably. Comparing with the general dispersion relation \eqref{eq:Xlocus} we have for $Q=0$
\begin{equation}\label{eq:Dneutral}
\begin{split}
\mathcal{D}(r_+) &= \frac{1}{d\, r_+}  = \frac{1}{4\pi T}  \,, 
\\  
\mathcal{D}_4(r_+) &= 
\frac{1}{(d\,r_+)^3}\, \text{Har}\left(\frac{2}{d}\right)   =  \text{Har}\left(\frac{2}{d}\right)  \,\frac{1}{(4\pi T)^3} \,.
\end{split}
\end{equation}	
The leading $k^2$ term gives the familiar expression for the diffusion constant $\mathcal{D} = \frac{1}{4\pi\,T}$ which leads to the famous viscosity to entropy density ratio \cite{Policastro:2001yc}. At higher orders, 
it is easy to check that this reproduces the results obtained in literature to this order in specific dimensions.
For example, in $d=4$ and $d=3$ one finds
\begin{equation}\label{eq:momdiff3and4}
\begin{split}
d=4: \qquad 
\omega(k) 
&= 
	-i\, \frac{1}{4\pi T}\, k^2  - i\, \frac{1-\ln 2}{32\, (\pi T)^3}\, k^4 + \cdots \,,\\
d=3: \qquad 
\omega(k) 
&= 
	-i\, \frac{1}{4\pi T}\, k^2  - i\, \frac{9 + \sqrt{3}\pi- 9\ln 3}{384\, (\pi T)^3}\, k^4 + \cdots \,.\\
\end{split}
\end{equation}	
reproducing results in \cite{Baier:2007ix} and \cite{Diles:2019uft}, respectively. This closes a small lacuna in the discussion of \cite[\S 9.4]{Ghosh:2020lel} where the quartic corrections were left undetermined owing to the function $\Mser{-\ann}{0,4}$ being not computed. Since we have this information at hand, we can confirm that the results do continue to work as expected. This also implies that the computation of third order transport data in \cite{Grozdanov:2015kqa} (and also \cite{Diles:2019uft}) is incorrect. It is clear that the quartic order dispersion relies upon getting the terms at order $k^4$ correctly, but these are related to hydrostatic data at the quartic order and undetermined by cubic order transport data alone.

\section{Solution for the vector sector}
\label{sec:XYsolns}
 
We now compile the solutions for the decoupled scalar system parameterizing the vector perturbations in the \eqref{eq:XYfinal} following the gradient expansion strategy used for the Markovian and non-Markovian in \cref{sec:probeM} and \cref{sec:probenM}. We start with our gradient expansion ansatz:
\begin{equation}\label{eq:}
\begin{split}
\Gin{\MX}(\ri,\bwt, \bqt)  
&= 
	\exp\left(\sum_{n,m=1}^\infty \,  (-i)^m \,\bwt^m\, \bqt^{2n}\, \xser{m,2n} (\ri) \right) ,\\
\Gin{\MY}(\ri,\bwt, \bqt)  
&= 
	\exp\left(\sum_{n,m=1}^\infty \,  (-i)^m \,\bwt^m\, \bqt^{2n}\, \yser{m,2n} (\ri) \right) .	
\end{split}
\end{equation}
The strategy for solving for the functions appearing in the gradient expansion $\xser{m,n}$ and $\yser{m,n}$ is analogous to the earlier discussion, especially with the functions appearing in $\Gin{\MX}$ being defined by suitable analytic continuation. In fact, the solutions for $\Gin{\MX}$ are analogous to those for a non-Markovian field of index $\ann  = -(d-1)$ with modifications coming from the potential which gives corrections at order $\bqt^2$ and higher. In particular, $\xser{m,0} (\ri)= \Mser{1-d}{m,0}(\ri)$.  There is a minor complication for $\Gin{\MY}$ since there is an additional factor of $h^2$ in the designer dilaton which is no longer a simple monomial.  We give the general results below without any intermediate details. 

The solution to cubic order is quite straightforwardly related to the non-Markovian designer field, viz., 
\begin{equation}\label{eq:Xgradfns}
\begin{split}
\xser{1,0}(\ri)
&= 
	\Mser{1-d}{1,0}(\ri)  \,,\\
\xser{2,0}(\ri)
&= 
	\Mser{1-d}{2,0}(\ri) \,, \\
\xser{0,2}(\ri)
&= 
	\Mser{1-d}{0,2}(\ri) - \frac{\sdc}{2(d-1)} \Dfn{d-1}{2,0}(\ri) \,,\\
\xser{3,0}(\ri)	
&=
	\Mser{1-d}{3,0}(\ri)  \,, \\ 
\xser{1,2}(\ri)
&= 
	\Mser{1-d}{1,2}(\ri) - \frac{\sdc}{d-1} \Mser{1-d}{2,0}(\ri) \,.
\end{split}
\end{equation}
At quartic order there are some deviations owing to the potential term for $\MX_\ai$. One nevertheless finds a closed form expression:
\begin{equation}\label{eq:XgradfnsO4}
\begin{split}	
\xser{4,0}(\ri)	
&=
	\Mser{1-d}{4,0}(\ri)  \,, 	\\
\xser{2,2}(\ri)
&= 
	\Mser{1-d}{2,2}(\ri) - \frac{\sdc}{d-1}  \left[
	\Mser{1-d}{3,0}(\ri) + \frac{1}{2}\, \Dfn{d-1}{3,0}(\ri) + \frac{1}{6}\, \Dfnh{d-1}{2,0}(\ri)^3
	+ \frac{1}{6}\, \Dfn{d-1}{2,0}(1)^3\right] ,\\		
\xser{0,4}(\ri)
&=
	\Mser{1-d}{0,4}(\ri)  + \frac{\sdc}{4(d-1)\, \bRQ^2}\, \Dfn{d-1}{2,0}(\ri)		
	-\frac{\sdc^2}{4(d-1)^2} \left[\frac{1}{2}\, \Dfn{d-1}{2,0}(\ri)^2 - 2 (d-1)\, \Dfn{\MX}{b}(\ri)\right] \\
& \qquad 
	-\frac{\sdc}{d-1} \left[ \frac{1}{d}\, \Mser{1-d}{2,0}(\ri) + \frac{1}{d}\, \Dfn{d-1}{1,2}(\ri) -\Dfn{d-1}{2,0}(1) \, \Mser{1-d}{0,2}(\ri) + \Dfn{\MX}{a}(\ri)
	\right] .
\end{split}
\end{equation}

We have had to introduce two new functions at the fourth order, which are formally defined in terms of divergent integrals, but can as before, be computed by suitably analytically continuing the expression with $(1-d) \rightarrow (d-1)$.
\begin{equation}\label{eq:DXfns}
\begin{split}
\Dfn{\MX}{a}(\ri)
&= 
	\int_0^\ri\, \frac{d\rib}{f(\rib)} \, \rib^{1-d}\, \int_1^{\rib} \, d\ri'\, \ri'^{d-1}\, \Dfn{d-1}{2,0}(\ri')  \,,\\
&=
	\int_0^\ri\, \frac{d\rib}{f(\rib)} \,\rib^{d-1}\,  \int_1^{\rib} \, d\ri'\, \ri'^{d-1}\, \Dfn{d-1}{2,0}(\ri')  
	+ \int_0^\ri\, d\rib \, f(\rib)\, \Dfn{d-1}{3,0}(\rib) - \frac{\ri^d}{d}\, \Dfn{d-1}{2,0}(1)^2 \,, \\
& \qquad
	+ \; 2\, \Dfn{d-1}{2,0}(1)\, \int_0^\ri	 \, d\rib \, \rib^{d-1}\, \Dfn{d-1}{2,0}(\rib) - \Dfn{d-1}{2,0}(\ri) \, \int_1^\ri\, d\rib \, \rib^{d-1}\, \Dfn{d-1}{2,0}(\rib) - f(\ri) \Dfn{d-1}{3,0}(\ri) \,, \\
\Dfn{\MX}{b}(\ri)
&= 
	\int_0^\ri\, \frac{d\rib}{f(\rib)} \, \rib^{1-d} \,\int_1^{\rib} \, d\ri'\, \ri'^{2d-3}\, \Dfn{d-1}{2,0}(\ri')  \\
&= 
	\int_0^\ri\, \frac{d\rib}{f(\rib)} \,\rib^{d-1}\,  \int_1^{\rib} \, d\ri'\, \ri'^{2d-3}\, \Dfn{d-1}{2,0}(\ri')  	
	+ \int_0^\ri\, d\rib \,\rib^{2d-3}\, \Dfn{d-1}{2,0}(\rib)^2\\
&\qquad
	 -\; \Dfn{d-1}{2,0}(\ri) \, \int_1^\ri\, d\rib \,\rib^{2d-3}\, \Dfn{d-1}{2,0}(\rib) \,.
\end{split}
\end{equation}

The solutions for the Markovian mode $\MY$ can be obtained by dressing the solutions for the Markovian problem with $\ann = d-3$ with suitable factors of $h(\ri)$ and accounting for the potential involving $\bqt^2$. We can write the general solution in terms of the integral transform introduced in \eqref{eq:MItransform} . The additional factor of $h^2$ however implies that we also need a second integral transform with kernel 
$\frac{1}{fh^2}$, i.e.,
\begin{equation}\label{eq:MItransformH}
\mathfrak{H}\big[\mathfrak{g}\big](\ri) \equiv \int_0^\ri\, \frac{d\rib}{f(\rib)\, h(\rib)^2}\, \mathfrak{g}(\rib) , \ \hspace{0.5cm} 
\hat{\mathfrak{H}}\big[\mathfrak{g}\big](\ri) \equiv \int_1^\ri\, \frac{d\rib}{f(\rib) \, h(\rib)^2}\, \mathfrak{g}(\rib) \,.
\end{equation}	

With these definitions we may then write the solution up to the cubic order for $\MY_\ai$ as
\begin{equation}\label{eq:Ygradfns}
\begin{split}
\yser{1,0}(\ri)
&=
	\mathfrak{T}[1] - h(1)^2\, \mathfrak{H}\left[\ri^{d-3}\right] ,\\
\yser{2,0}(\ri)
&=
	- h(1)^2\, \mathfrak{H} \left[ \ri^{d-3}\, \Dfnh{\MY}{2,0}(\ri) \right]  ,\\
\yser{0,2}(\ri)
&=
		-\frac{1}{d-4} \, \mathfrak{H}\left[\ri\left(1-\ri^{d-4}\right)  \right]  + \sdc\, \Dfn{\MY}{0,2}(\ri) \,,\\
\yser{3,0}(\ri)
&=
	2\, h(1)^2\, \mathfrak{H}\left[\ri^{d-3}\,  \yserh{2,0}(\ri)\right]\,, \\
\yser{1,2}(\ri)
&=
	2\, h(1)^2\,  \mathfrak{H}\left[\ri^{d-3}\, \yserh{0,2}(\ri)  \right]  . 
\end{split}
\end{equation}
At the quartic order one finds
\begin{equation}\label{eq:YgradfnsO4}
\begin{split}
\yser{4,0}(\ri)
&=
	2\, h(1)^2 \, \mathfrak{H}\left[ \ri^{d-3}\, \left( \yserh{3,0}(\ri) 
		+\frac{1}{2}\, \Dfnh{\MY}{3,0}(\ri)\right)\right] ,\\
\yser{2,2}(\ri)
&=
	2\, h(1)^2 \,\mathfrak{H}\left[  \ri^{d-3} 
	\left(\yserh{1,2}(\ri) + \frac{1}{d-4} \left(\Dfnh{\MY}{1,2}(\ri) - \frac{1}{h(1)^2}\, \yserh{2,0}(\ri)\right)
	+ \sdc \, \Dfnh{\MY}{2,2}(\ri)\right)
	\right] ,\\
\yser{0,4}(\ri)
&=
	- \mathfrak{H} \left[ \ri^{d-3}  \Dfnh{\MY}{0,4}(\ri)\right] \,. 
\end{split}
\end{equation}
There are several intermediate functions introduced above which convert all the functions to satisfy first order ordinary differential equations. These are up to the cubic order nicely expressed as integral transforms 
\begin{equation}\label{eq:DYfns}
\begin{split}
 \Dfnh{\MY}{2,0}(\ri)
 &= 
	h(1)^2 \hat{\mathfrak{H}}\left[ \ri^{d-3}\right] 
	- \frac{1}{h(1)^2}\, \hat{\mathfrak{T}}\left[\frac{h(\ri)^2}{\ri^{d-3}}\right] , \\
 \Dfnh{\MY}{3,0}(\ri)
 &= 
 - h(1)^2 \, \hat{\mathfrak{H}}\left[ \ri^{d-3}\,\Dfnh{\MY}{2,0}(\ri)^2  \right] , \\ 
 \Dfn{\MY}{0,2}(\ri)
 &= 
 	 \frac{1}{2\,d(d-1)} \, \mathfrak{H}\left[\ri^{d-3}  \left( d \left(\ri^2\, h(\ri)^2 - h(1)^2\right) -2\left(\ri^2\, h(\ri) - h(1)\right) + (d^2-2) \, (1-\ri^2) 
  \right)\right] ,\\  
 \Dfnh{\MY}{1,2}(\ri)
 &= 
 	-\hat{\mathfrak{H}}\left[\ri\,\Dfnh{\MY}{2,0}(\ri)  \right] .
\end{split}
\end{equation}
At the quartic order owing to the contribution from the potential term the expressions are a bit more complicated, but can be brought to a simple form using the functions defined above
\begin{equation}\label{eq:DYfnsO4}
\begin{split}
 \Dfnh{\MY}{2,2}(\ri)
 &= 
  \int_1^\ri \, d\rib\,  \Dfnh{\MY}{2,0}(\rib)\, \dv{\rib}\Dfn{\MY}{0,2}(\rib)   \,,  \\
 \Dfnh{\MY}{0,4}(\ri)
 &=
 	\int_1^{\ri}\, d\rib \,  h(\rib)^2
 	\left[\frac{f(\rib)}{ \rib^{d-3} } \, \left(\dv{\yser{0,2}(\rib)}{\rib}\right)^2 + 
 	\frac{\sdc}{2\, \bRQ^2}\, \rib\right]   .
\end{split}
\end{equation}

 While these solutions are valid in general $d>4$, the case of $d=4$ needs to be dealt with care as this is a marginal case in our analysis. As can be seen from the occurrence of factors of $\frac{1}{d-4}$ in the above expressions there are divergences owing to the presence of a logarithmic mode. The issue only arises for $\yser{m,n}(\ri)$ with $n\neq 0$ and  can all be suitably accounted for with care. To the order we are working in one finds modifications to the following two functions  
\begin{equation}\label{eq:Yfns4d}
\begin{split}
\yser{0,2}(\ri)
& = 
	\mathfrak{H}\left[\ri \, \log \ri\right]+ \sdc\,  \Dfn{\MY}{0,2}(\ri)  \,,\\
\Dfnh{\MY}{1,2}(\ri)
& = 
	- \hat{\mathfrak{H}}\left[\ri \, \left(1-\log \rib\right) \Dfn{\MY}{2,0}(\rib)	\right] .
\end{split}
\end{equation}	

The asymptotic behaviour of the functions in \eqref{eq:Ygradfns} and \eqref{eq:DYfns} can be determined as before. We record them below as they are relevant for extracting the physical Green's functions of the boundary currents. 
\begin{equation}
\begin{split}
\yser{1,0}(\ri) 
		&= \ri - \frac{h(1)^2}{d-2}\ri^{d-2} \,, \\
\yser{0,2}(\ri) 
		&=  -\frac{\ri^2}{2(d-4)} + \frac{\ri^{d-2}}{(d-4)(d-2)} +\frac{d^2-d h(1)^2 -2\sdc}{2d(d-1)(d-2)}\sdc \ri^{d-2} -\frac{\sdc}{2(d-4)}\ri^{d} \,, \\
\yser{2,0}(\ri)
		&= -\frac{\ri^2}{2(d-4)} +\frac{h(1)^2 \Dfn{\MY}{2,0}(1)}{d-2}\ri^{d-2} - \frac{(d-2)\sdc}{d(d-4)}\ri^{d}\,, \\
\yser{3,0}(\ri)
		&= -\frac{2h(1)^2\yser{2,0}(1)}{d-2}\ri^{d-2} - \frac{h(1)^2}{d(d-4)}\ri^d\,,\\
\yser{1,2}(\ri)
		&=  -\frac{2h(1)^2\yser{0,2}(1)}{d-2}\ri^{d-2} - \frac{h(1)^2}{d(d-4)}\ri^d\,,\\
\yser{4,0}(\ri)
		&= \frac{\ri^4}{4(d-4)^2(d-6)} -\frac{h(1)^2\left(\Dfn{\MY}{3,0}(1)+2\yser{3,0}(1)\right)}{d-2}\ri^{d-2} + \frac{\Dfn{\MY}{2,0}(1)h(1)^2}{d(d-4)}\ri^{d}\,,\\
\yser{2,2}(\ri)
		&= \frac{\ri^4}{2(d-4)^2(d-6)} +\frac{1}{d(d-4)}\left(\frac{1}{d-4}+\frac{\sdc}{2}+\frac{\sdc^2}{d}-\frac{\sdc^3}{2(d-1)}+h(1)^2\Dfn{\MY}{2,0}(1)\right)\ri^{d}\,,\\
			&+\frac{2\yser{2,0}(1)-2h(1)^2\left(\Dfn{\MY}{1,2}(1)+(d-4)\left(\yser{1,2}(1)+\sdc\Dfn{\MY}{2,2}(1)\right)\right)}{(d-4)(d-2)}\ri^{d-2}\,,\\
\yser{0,4}(\ri)
		&= \frac{\ri^{4}}{2(d-6)(d-4)^2} + \frac{\Dfn{\MY}{0,4}(1)}{d-2}\ri^{d-2}\\
		&+\left(\frac{1}{d(d-4)}\left(\frac{1}{d-4}+\frac{\sdc}{2}+\frac{\sdc^2}{d}-\frac{\sdc^3}{2(d-1)}\right) - \frac{\sdc}{4d\bRQ^2}\right)\ri^{d} \,.
\end{split}
\end{equation}
%

\begin{figure}[h]
\centering
\subfloat{
\begin{minipage}[t]{0.5\textwidth}
\vspace{0.5cm}
\hspace{0pt}
\begin{tikzpicture}
  \node (img)  {\includegraphics[scale=0.275]{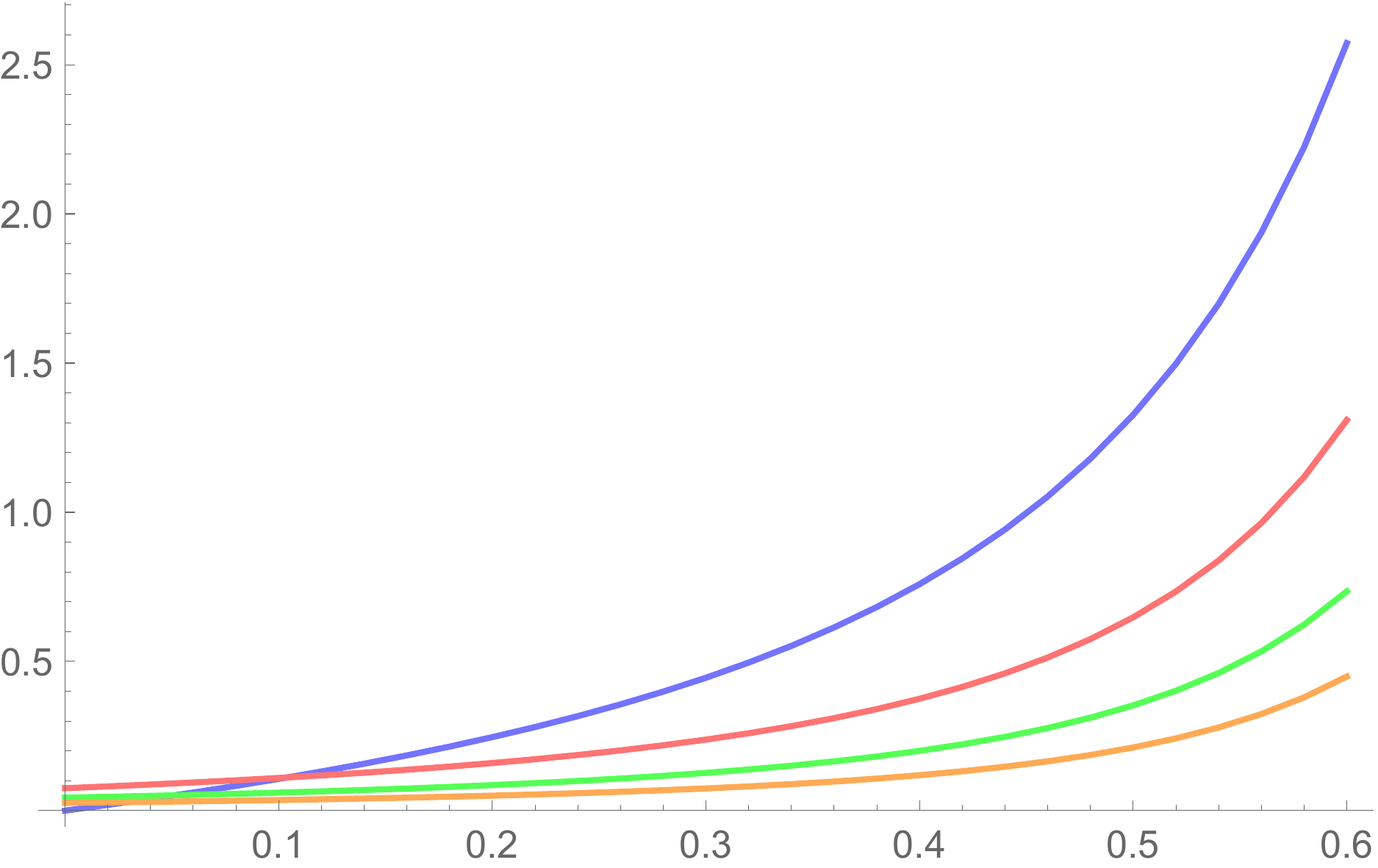}};
  \node[below=of img, node distance=0cm,font=\small,yshift=1.0cm] {$\sdc$};
  \node[left=of img, node distance=0cm, anchor=center,font=\footnotesize,xshift=0.5cm] {$\yser{3,0}(1)$};
  \node[right=of img, node distance=0cm,font=\scriptsize,xshift=-1.25cm,yshift=1.5cm] {$d=3$};
  \node[right=of img, node distance=0cm,font=\scriptsize,xshift=-1.25cm,yshift=0.25cm] {$d=4$};
  \node[right=of img, node distance=0cm,font=\scriptsize,xshift=-1.25cm,yshift=-0.35cm] {$d=5$};
   \node[right=of img, node distance=0cm,font=\scriptsize,xshift=-1.25cm,yshift=-0.85cm] {$d=6$};
 \end{tikzpicture}
 \end{minipage}
 }
 \subfloat{
 \begin{minipage}[t]{0.5\textwidth}
 \vspace{0.5cm}
\hspace{0pt}
\begin{tikzpicture}
  \node (img)  {\includegraphics[scale=0.275]{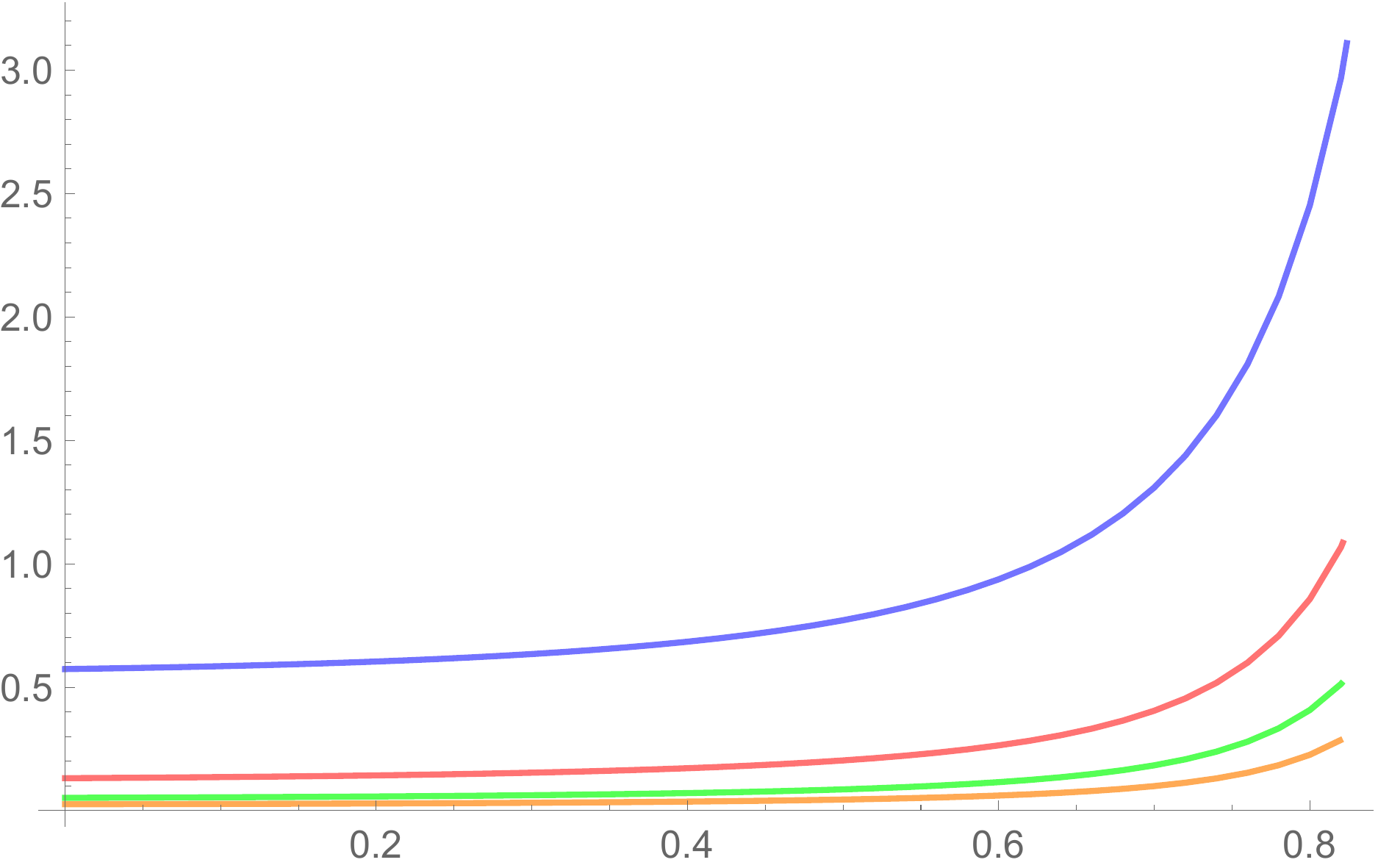}};
  \node[below=of img, node distance=0cm,font=\small,yshift=1.0cm] {$\sdc$};
  \node[left=of img, node distance=0cm, anchor=center,font=\footnotesize,xshift=0.5cm] {$\yser{1,2}(1)$};
  \node[right=of img, node distance=0cm,font=\scriptsize,xshift=-1.25cm,yshift=1.5cm] {$d=3$};
  \node[right=of img, node distance=0cm,font=\scriptsize,xshift=-1.25cm,yshift=-0.25cm] {$d=4$};
  \node[right=of img, node distance=0cm,font=\scriptsize,xshift=-1.25cm,yshift=-0.8cm] {$d=5$};
   \node[right=of img, node distance=0cm,font=\scriptsize,xshift=-1.25cm,yshift=-1.15cm] {$d=6$};
 \end{tikzpicture}
 \end{minipage}
 } \\
 \subfloat{
 \begin{minipage}[t]{0.5\textwidth}
 \vspace{0.5cm}
\hspace{0pt}
\begin{tikzpicture}
  \node (img)  {\includegraphics[scale=0.275]{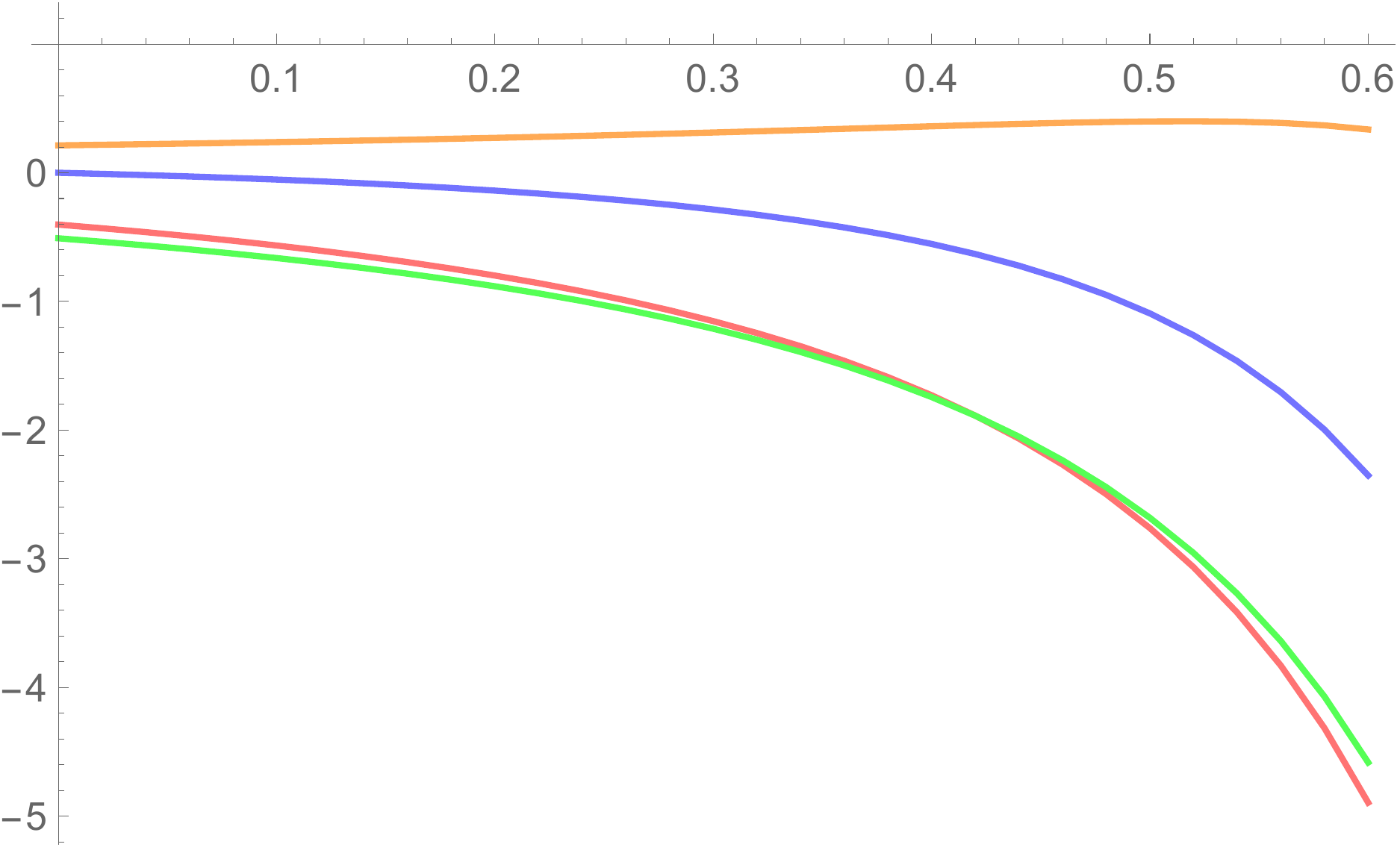}};
  \node[above=of img, node distance=0cm, font=\small,yshift=-1.2cm] {$\sdc$};
  \node[left=of img, node distance=0cm, anchor=center,font=\footnotesize,xshift=0.5cm] {$\Dfn{\MY}{1,2}(1)$};
  \node[right=of img, node distance=0cm,font=\scriptsize,xshift=-1.25cm,yshift=1.1cm] {$d=6$};
  \node[right=of img, node distance=0cm,font=\scriptsize,xshift=-1.25cm,yshift=-0.3cm] {$d=3$};
  \node[right=of img, node distance=0cm,font=\scriptsize,xshift=-1.25cm,yshift=-1.1cm] {$d=5$};
   \node[right=of img, node distance=0cm,font=\scriptsize,xshift=-1.25cm,yshift=-1.5cm] {$d=4$};
 \end{tikzpicture}
 \end{minipage}
}
 \subfloat{
 \begin{minipage}[t]{0.5\textwidth}
 \vspace{0.5cm}
\hspace{0pt}
\begin{tikzpicture}
  \node (img)  {\includegraphics[scale=0.275]{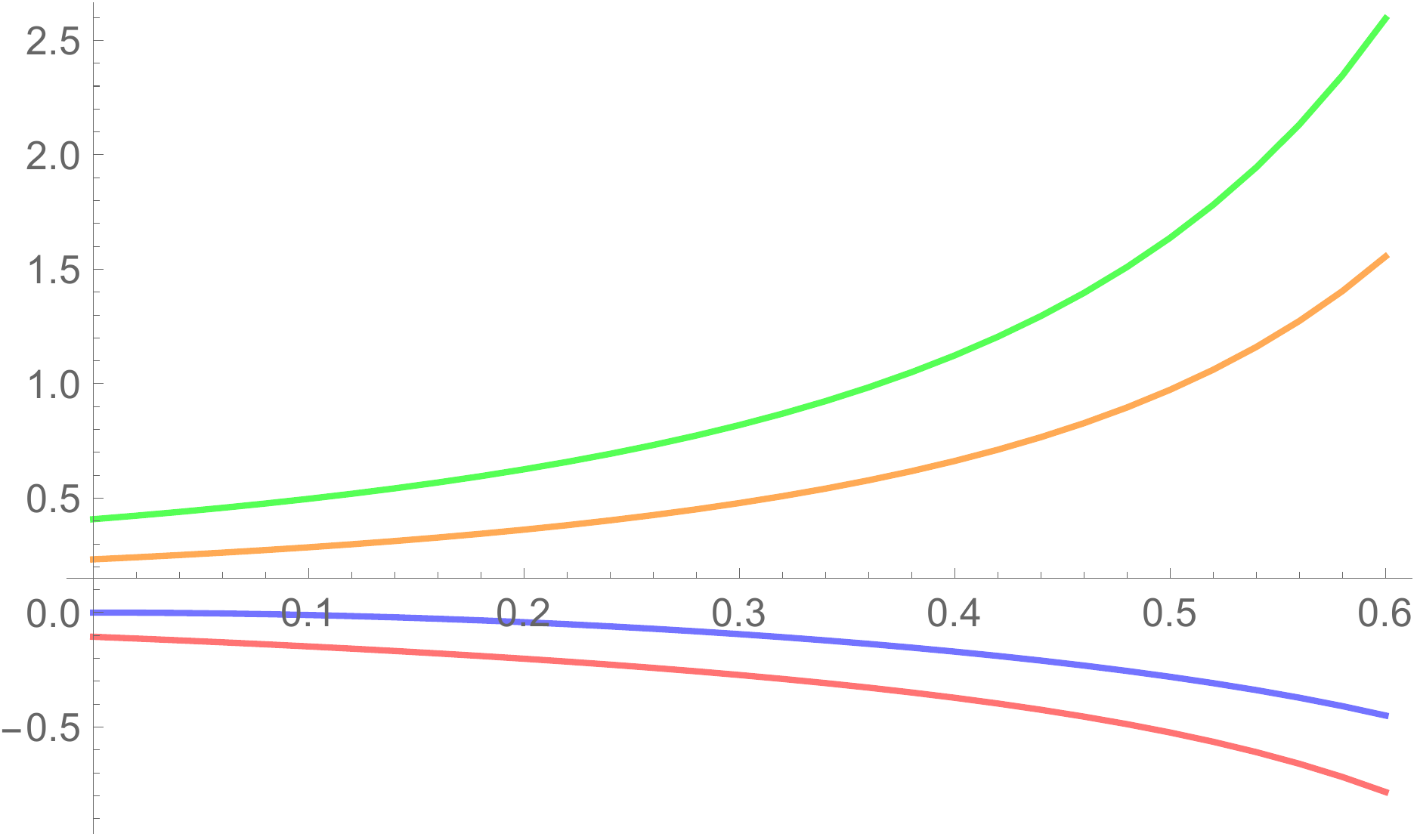}};
  \node[right=of img, node distance=0cm, font=\small,xshift=-1.25cm,yshift=-0.5cm] {$\sdc$};
  \node[left=of img, node distance=0cm, anchor=center,font=\footnotesize,xshift=0.5cm] {$\Dfn{\MY}{3,0}(1)$};
  \node[right=of img, node distance=0cm,font=\scriptsize,xshift=-1.25cm,yshift=1.6cm] {$d=5$};
  \node[right=of img, node distance=0cm,font=\scriptsize,xshift=-1.25cm,yshift=0.8cm] {$d=6$};
  \node[right=of img, node distance=0cm,font=\scriptsize,xshift=-1.25cm,yshift=-1.1cm] {$d=3$};
   \node[right=of img, node distance=0cm,font=\scriptsize,xshift=-1.25cm,yshift=-1.4cm] {$d=4$};
 \end{tikzpicture}
 \end{minipage}
}\\
 \subfloat{
 \begin{minipage}[t]{0.5\textwidth}
 \vspace{0.5cm}
\hspace{0pt}
\begin{tikzpicture}
  \node (img)  {\includegraphics[scale=0.275]{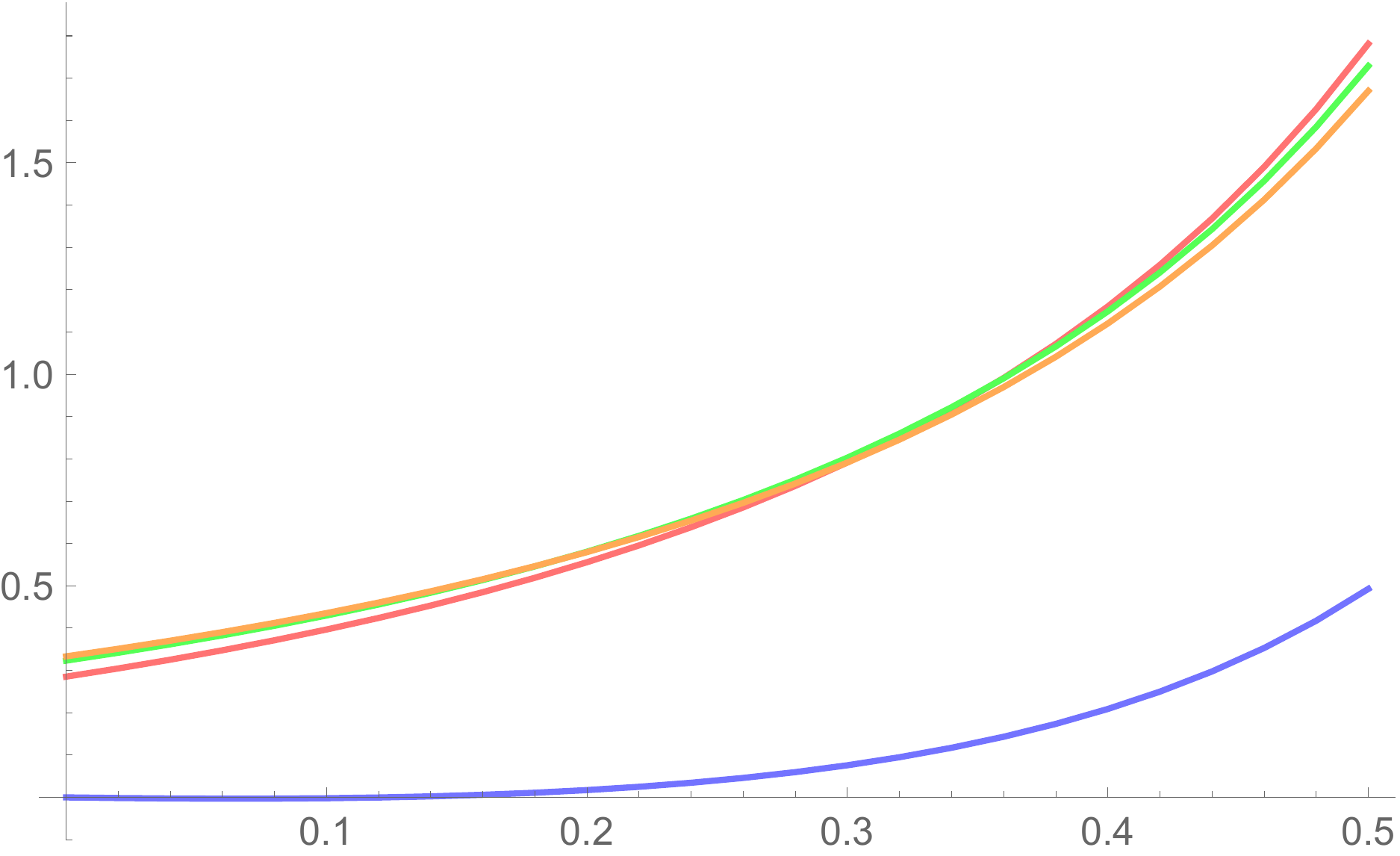}};
  \node[below=of img, node distance=0cm,font=\small,yshift=1.0cm] {$\sdc$};
  \node[left=of img, node distance=0cm, anchor=center,font=\footnotesize,xshift=0.5cm] {$\Dfn{\MY}{2,2}(1)$};
  \node[right=of img, node distance=0cm,font=\scriptsize,xshift=-1.25cm,yshift=1.65cm] {$d=4$};
  \node[right=of img, node distance=0cm,font=\scriptsize,xshift=-1.25cm,yshift=1.4cm] {$d=5$};
  \node[right=of img, node distance=0cm,font=\scriptsize,xshift=-1.25cm,yshift=1.15cm] {$d=6$};
   \node[right=of img, node distance=0cm,font=\scriptsize,xshift=-1.25cm,yshift=-0.5cm] {$d=3$};
 \end{tikzpicture}
 \end{minipage}
}
 \subfloat{
 \begin{minipage}[t]{0.5\textwidth}
 \vspace{0.5cm}
\hspace{0pt}
\begin{tikzpicture}
  \node (img)  {\includegraphics[scale=0.275]{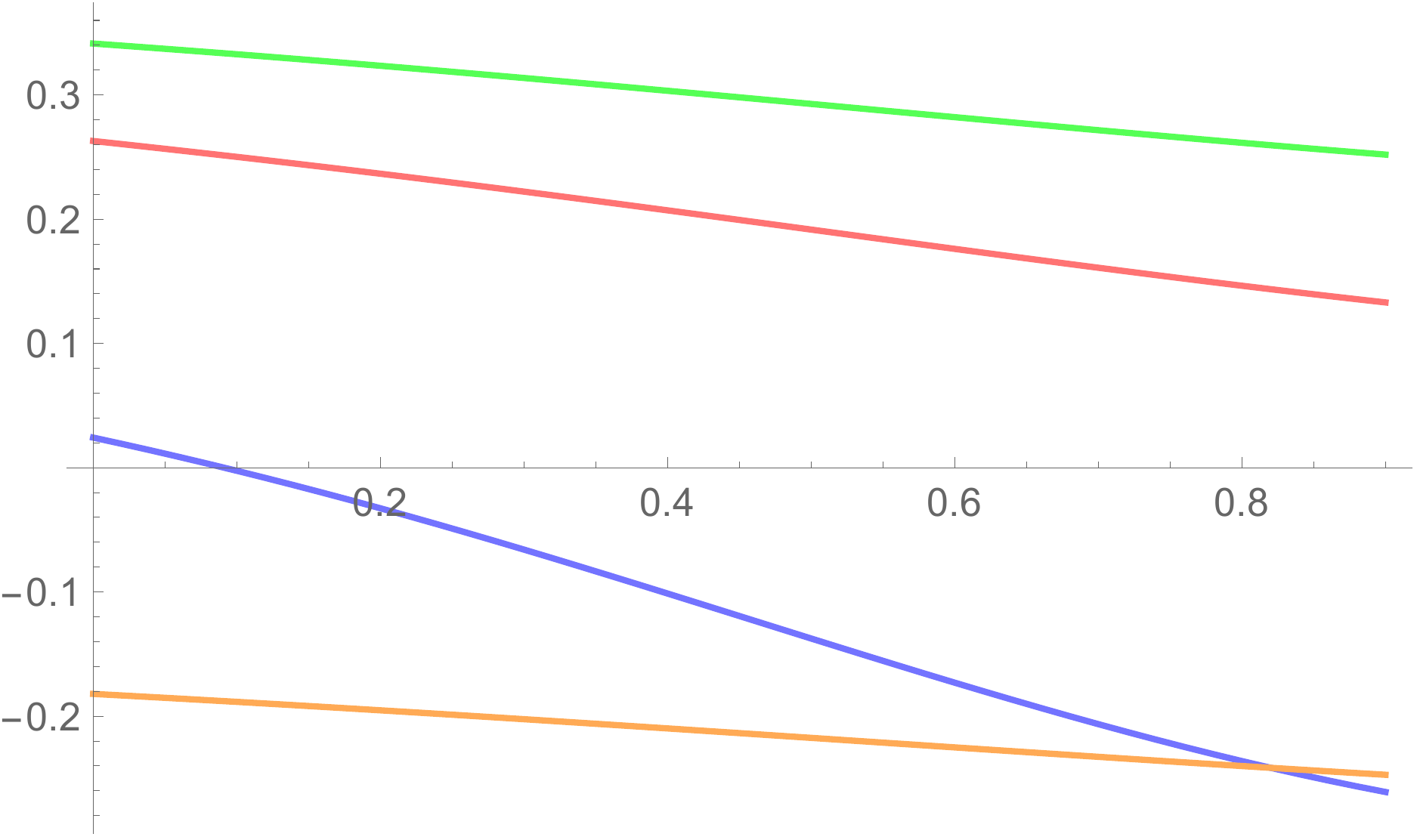}};
  \node[right=of img, node distance=0cm, font=\small,xshift=-1.25cm,yshift=-0.15cm] {$\sdc$};
  \node[left=of img, node distance=0cm, anchor=center,font=\footnotesize,xshift=0.5cm] {$\Dfn{\MY}{0,4}(1)$};
  \node[right=of img, node distance=0cm,font=\scriptsize,xshift=-1.25cm,yshift=1.1cm] {$d=5$};
  \node[right=of img, node distance=0cm,font=\scriptsize,xshift=-1.25cm,yshift=0.5cm] {$d=4$};
  \node[right=of img, node distance=0cm,font=\scriptsize,xshift=-1.25cm,yshift=-1.1cm] {$d=6$};
   \node[right=of img, node distance=0cm,font=\scriptsize,xshift=-1.25cm,yshift=-1.4cm] {$d=3$};
 \end{tikzpicture}
 \end{minipage}
}
\caption{ The charge dependence of parameters determining the coefficients at the quartic order Green's function of the Markovian vector  polarizations  encoded in $\MY_\ai$ for various dimensions, supplementing the lower order coefficients depicted in \cref{fig:KinMarT}. }
\label{fig:KinMarY4th}
\end{figure}
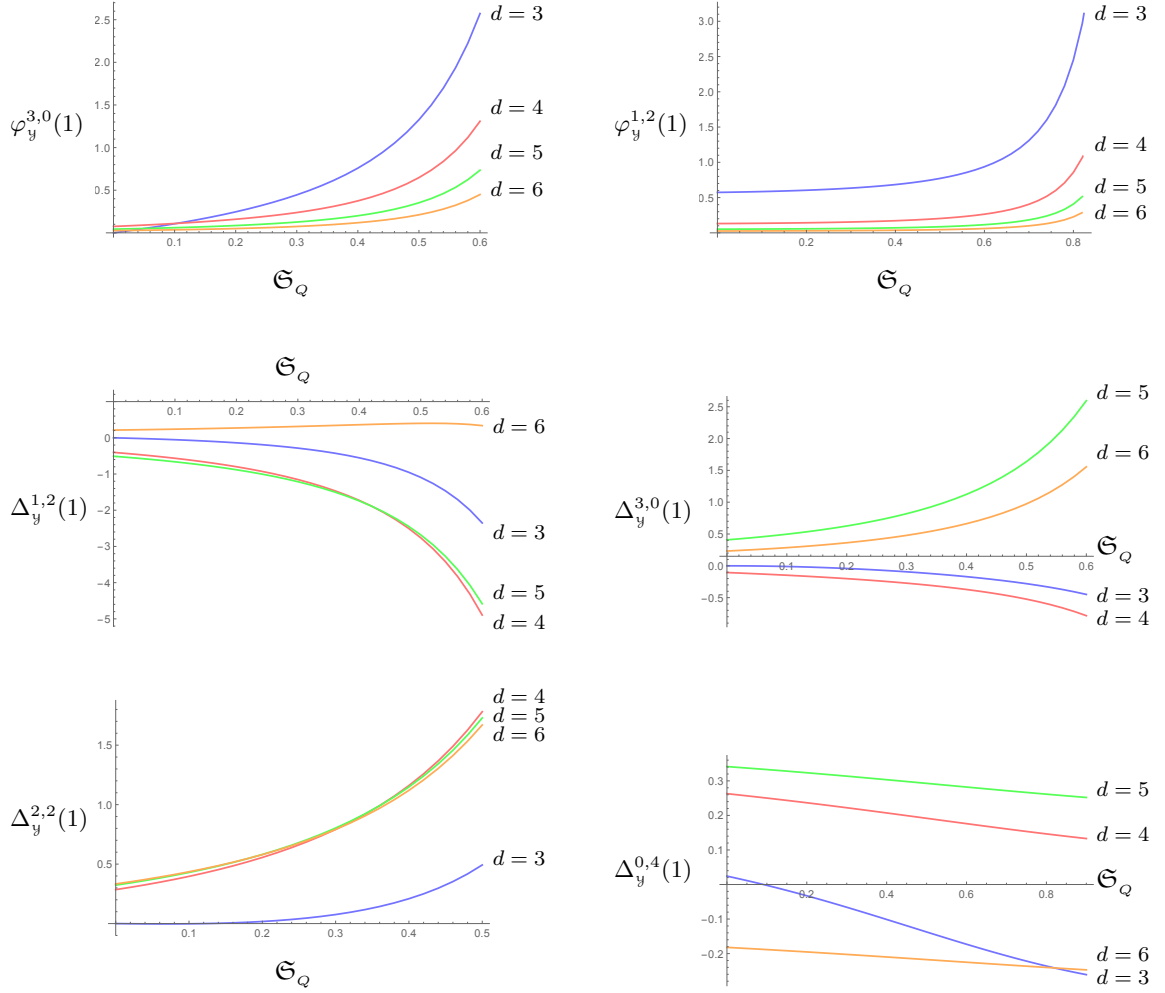

The physical information we need from these solutions is the boundary retarded Green's function for $\MY_\ai$ and the inverse Green's function for $\MX_\ai$. Given the bulk propagator, using the asymptotic expansions of the solutions determined above one finds the Green's function for $\MY_\ai$ to be 
\begin{equation}\label{eq:KinYA}
\begin{split}
\Kin{\MY}(\omega,\bk) 
&=
  r_+^{d-2}\,\bigg\{ -i\,  (1-\sdc)^2 \,\bwt 
  - \left[\frac{1}{d-4} + \frac{\sdc}{2} + \frac{\sdc^2}{d} - \frac{\sdc^3}{2(d-1)}\right] \bqt^2 \\
&\qquad 
   + (1-\sdc)^2 \left[ \Dfn{\MY}{2,0}(1) \, \bwt^2  - 2i\, \yser{0,2}(1)\,  \bwt \bqt^2 + 2i \,\yser{2,0}(1)\, \bwt^3 \right]\\
 &\qquad 
  +  (1-\sdc)^2\left[2\,\yser{3,0}(1) + \Dfn{\MY}{3,0}(1)\right] \, \bwt^4   - \Dfn{\MY}{0,4}(1)\bqt^4 \\
 &\qquad 
  - 2 (1-\sdc)^2\,\left[ \yser{1,2}(1) + \sdc\, \Dfn{\MY}{2,2}(1)
  - \frac{1}{d-4}\left( \frac{\yser{2,0}(1)}{(1-\sdc)^2}  -  \Dfn{\MY}{1,2}(1) \right) \right]
    \bwt^2\, \bqt^2\bigg\} \,.
\end{split}
\end{equation}  
A similar computation for $\MX_\ai$ results in
\small{
\begin{equation}\label{eq:KinXA}
\begin{split}
&\frac{1}{r_+^{2-d}} \, 
\Kin{\MX}(\omega,\bk) \\
&=
  -i\, \bwt 
  + \left[\frac{1}{d} - \frac{\sdc}{2(d-1)} \right] \bqt^2  - \Dfn{d-1}{2,0}(1) \, \bwt^2 \\
&\qquad 
  + i\, \bwt^3 \left[\Dfn{d-1}{2,0}(1)^2  -2\Mser{d-1}{2,0}(1) \right]  
  + i\bwt\,\bqt^2 \left[ \frac{2(d-2)}{d}\, \Mser{d-1}{0,2}(1) 
  + \left(\frac{\sdc}{d-1}-\frac{2}{d}\right) \Dfn{d-1}{2,0}(1)
  \right]\\
 &\qquad  
  + \bwt^4 \left[ \Dfn{d-1}{2,0}(1)^3 - 2 \Mser{d-1}{3,0}(1) - \Dfn{d-1}{3,0}(1) - 4 \Mser{d-1}{2,0}(1) \Dfn{d-1}{2,0}(1) \right] \\
  &\qquad + \bwt^2\bqt^2 \left[ \frac{\sdc}{2(d-1)} \left( 3 \Dfn{d-1}{2,0}(1)^2 - 4 \Mser{d-1}{2,0}(1) \right)  \right. \\
  &\qquad 
  \left. + \frac{1}{d}\left( 2 (d-2) \Mser{d-1}{1,2}(1) + 2 \Mser{d-1}{2,0}(1) + 2 \Dfn{d-1}{1,2}(1) + 4(d-2) \Mser{d-1}{0,2}(1) \Dfn{d-1}{2,0}(1) - 3 \Dfn{d-1}{2,0}(1)^2  \right) \right] \\
  &\qquad 
  + \bqt^4 \left[ \frac{\sdc}{d-2} \left( \frac{2\, \sdc}{d(3d-2)} + \frac{d-2}{4(d-1)\bRQ^2} - \frac{d+2}{2d^2} \right) - \frac{1}{d^2} \left( (d-2) \Mser{d-1}{0,2}(1) + d \Mser{d+1}{0,2}(1) - \Dfn{d-1}{2,0}(1) \right) \right].
\end{split}
\end{equation}}
\normalsize

These expressions enter our Wilsonian influence phase and thus determine the current correlation functions. We have given here the results up to fourth order in gradients updating the equations presented in \eqref{eq:KinY} and \eqref{eq:KinX}.

\section{Boundary observables}
\label{sec:bdyobs}

Having discussed the repackaging of Einstein-Maxwell dynamics, at the quadratic level, into a set of designer scalar fields $\tGR_\bi$, $\MX_\ai$ and $\MY_\ai$, we now turn to computing boundary observables. Of particular interest to us are is the on-shell action obtained in order by order in a boundary gradient expansion, as well as the expression for the boundary conserved currents. These directly determine the physical observables of the theory reported in \cref{sec:cplasmadata,sec:bdycurrents}.

\subsection{On-shell action}
\label{sec:osaction}

Once one has the solution to the bulk equations of motion, we should derive the on-shell action, parameterized either in terms of the sources of the Markovian fields or the expectation values of the non-Markovian operators. The analysis for the tensor polarizations is straightforward and parallels the neutral fluid discussion in \cite{Ghosh:2020lel}. We will therefore focus on explaining the salient features in deriving the result for the $\MX-\MY$ system parameterizing the vector polarizations.

Our starting point in the action \eqref{eq:SXYdecoupleA} which upon imposing the equations of motion leads to a boundary term on the grSK geometry: 
\begin{equation}
S^{\text{V}}_{\text{EM,bulk}} + S^{\text{V}}_{\text{bdy}} 
=S^\text{V}_\skR - S^\text{V}_\skL\,,
\end{equation}
with 
\begin{align}\label{eq:on-shell}
\begin{split}
S^\text{V}_\skR
&= 	
	 \lim_{r\to \infty-i0}\int_k \sum_{\ai=1}^{N_V} \, k^2\, 
	\left[ -(\BQT^2 + 2) \,\,\cpen{\MX}^\ai \,\MX_{\ai} + \frac{\BQT^2}{2\, \RQ^{2(d-2)}} \, 
	 \cpen{\MY}^\ai \,\MY_{\ai}   \right. \\
&\qquad 
	+ \frac{d-2}{\bRQ^2} \, \frac{f}{r^{d-2}} \,\bqt^2\left(\MX_{\ai}^2-2\,\MX_{\ai}\, \MY_{\ai} \right) 
	+  \frac{d-2}{2\,\RQ^{d-2}}\, 
	 f \, h\, \BQT^2 \left(\frac{h}{1-h}\, \BQT^2 - 1 \right) \MY_{\ai}^2  \\ 
&\qquad 
	\left. + \frac{\BQT^2 (\BQT^2 + 2) }{\RQ^{d-2}} \left( \frac{1}{h \,r^{d-2}} \,\MX_{\ai} \, \cpen{\MY}^\ai + h \,r^{d-2}\, \MY_{\ai} \,\cpen{\MX}^\ai \right)  \right],
\end{split}
\end{align}
and similarly at the left boundary. This action is divergent and should be regulated by  the counterterm action \eqref{eq:LctermsXY}. We have chosen to parameterize the counterterm action so that it contributes no finite part, and thus one can analyze the asymptotic behaviour of the various fields to extract the regulated on-shell action from \eqref{eq:on-shell} directly.

Using the definition of the conjugate momenta \eqref{eq:XYconjmom}, we can derive to leading order in large $r$
\begin{align}\label{eq:XpiYasy}
	\lim_{r\to\infty\pm i0} \MX_\ai 
	= -\frac{1}{d-2}\,r^{d-2} \,  \cpen{\MX}^\ai \bigg|_{\skL,\skR} \,, \qquad \lim_{r\to\infty\pm i 0} \cpen{\MY}^\ai = \frac{\omega^2 - k^2}{d-4} \, r^{d-4} \,\MY_\ai \bigg|_{\skL,\skR}\,.
\end{align}
As expected, because $\MX_\ai$ is non-Markovian with Markovian index $\ann = 1-d$,  it diverges as $r^{|\ann|-1} = r^{d-2}$ near the boundary while its conjugate momentum $\cpen{\MX}^\ai$ approaches a finite value. On the other hand, $\MY_\ai$ is Markovian with Markovian index $\ann = d-3$, so $\MY_\ai$ is finite near the boundary while its conjugate momentum $\cpen{\MY}^\ai$ diverges as 
$r^{d-4}$. This prompts us to introduce sources for the fields $\MX_\ai$ and  $\MY_\ai$ which we do with the  following boundary conditions:
\begin{align}\label{eq:BC}
	\lim_{r\to\infty\pm i 0 } \cpen{\MX}^\ai = -\XsJ^\ai_{\skL/\skR}  \,, 
	\qquad \lim_{r \to \infty \pm i 0 } \MY_\ai = \YsQ^\ai_{\skL/\skR} \,.
\end{align}
From the boundary condition for $\cpen{\MX}^\ai$ and the asymptotic behavior of the solution we deduce:
\begin{equation}\label{eq:XsourceRL}
\begin{split}
 \XsJ^\ai_\skR  
 &= 
 	\Kin{\MX} \, \left[ (\nB+1)\, \XP^\ai_\skR - \nB\, \XP^\ai_\skL\right]  - \nB \Krev{\MX} \left[ \XP^\ai_\skR - \XP^\ai_\skL\right] , \\
   \XsJ^\ai_\skL  
 &= 
 	\Kin{\MX} \, \left[ (\nB+1)\, \XP^\ai_\skR - \nB\, \XP^\ai_\skL\right]  -( \nB+1) \Krev{\MX} \left[ \XP^\ai_\skR - \XP^\ai_\skL\right] .
\end{split}
\end{equation}
To extract the on-shell action from \eqref{eq:on-shell}, we need to extract the finite terms in the action. This is complicated by the fact some of the coefficients of the fields $\MX_\ai, \MY_\ai, \cpen{\MX}^\ai$, and $\cpen{\MY}^\ai$ have explicit $r$ dependence in the numerator, so there are finite terms arising from subleading terms of the large $r$ expansion of the various fields. In particular, it is useful to recall from \eqref{eq:KinMark} that since $\MY$ is a Markovian scalar, the finite piece of $\cpen{\MY}^\ai$ and the $r^{-(d-2)}$ subleading piece of $\MY_\ai$ are
\begin{align}\label{eq:renormY}
\lim_{r\to\infty \pm i 0} \cpen{\MY}^\ai \bigg|_{\text{ren}} = -  \Kin{\MY}(\omega,\bk)\, \YsQ^\ai_{\skL/\skR} \,, 
\qquad  
\lim_{r \to \infty \pm i 0} r^{d-2} \MY_\ai \bigg|_{\text{ren}} = -\frac{1}{d-2} \Kin{\MY}(\omega, \bk)\, \YsQ^\ai_{\skL/\skR} \, .
\end{align}

Using the above expressions, we proceed to obtain the finite on-shell action from \eqref{eq:on-shell} term by term. For each term, we first use \eqref{eq:XpiYasy} to write either $\MX_\ai$ in terms of $\cpen{\MX}^\ai$ or $\cpen{\MY}^\ai$ in terms of $\MY_\ai$. We can then replace $\cpen{\MX}^\ai$ and $\MY_\ai$ by their boundary values via \eqref{eq:BC}, and finally use \eqref{eq:renormY} to extract out the finite piece of the term.

As an example, consider the $\MX_\ai \cpen{\MY}^\ai$ term from the last line of \eqref{eq:on-shell} . We observe that we can replace $h$ in the denominator by unity since by \eqref{eq:XpiYasy}, all subleading terms vanish near the boundary. Using \eqref{eq:XpiYasy} to write $\MX_\ai$ in terms of $\cpen{\MX}^\ai$, we obtain (retaining only the radial dependent terms)
\begin{align}\label{eq:XpiYterm}
	\frac{1}{ h\,r^{d-2}} \MX_\ai \cpen{\MY}^\ai = -\frac{1}{d-2} \, \cpen{\MX}^\ai \cpen{\MY}^\ai.
\end{align}
We can now replace $\cpen{\MX}^\ai$ with its source given in \eqref{eq:BC}, and the only term that is not divergent (and hence canceled by a counterterm) or vanishing at the boundary comes from the finite piece of $\cpen{\MY}^\ai$ given in \eqref{eq:renormY}.

Performing this procedure for each term and simplifying, we obtain that the on-shell action is given by
\begin{equation}\label{eq:onshellfinal}
\begin{split}
 S^{\text{V}}_{\text{on-shell}} 
 &= 
 -\, \int_k \;  \sum_{\ai=1}^{N_V}  \,k^2  
 \left\{ \NY(\BQT) \, (\YsQ^\ai_\skR -\YsQ^\ai_\skL)^\dag \,
  \Kint{\MY} \left[(\nB+1)\YsQ^\ai_\skR - \nB \,\YsQ^\ai_\skL \right]\right.\\
  &\left. \hspace{3cm}
  - \NX(\BQT) \, (\XP^\ai_\skR  - \XP^\ai_\skL)^\dag \,
    \Kin{\MX} \left[(\nB+1)\,\XP^\ai_\skR- \nB\,\XP^\ai_\skL \right]
    \right\} .
\end{split}
\end{equation}	
We have used here the ingoing propagator $\Kint{\MY}(\omega,\bk)$ which is the ingoing propagator for the field $\MY_\ai$ with an additional contact term defined in \eqref{eq:KintY}.

Upon converting to the average-difference basis we obtain the expression \eqref{eq:WIFEin} quoted in the main text. Note that the on shell action for the $\MX_\ai - \MY_\ai$ system arising from the Einstein-Maxwell dynamics is not canonically normalized owing to the momentum dependent factors $\NX$ and $\NY$. We have found it convenient to analyze the boundary observables for this set of modes, treated as probe fields in the \RNAdS{d+1} background, without these normalization factors. Indeed, our choice of boundary conditions to  identify the sources and vevs for the bulk $\MX_\ai$ and $\MY_\ai$ in \eqref{eq:BC},  \eqref{eq:nMVop}, and \eqref{eq:renormY}, respectively. We will account for these factors when we compute the boundary current observables as they are of course present in the Einstein-Maxwell system parameterized in terms of $\MX_\ai$ and $\MY_\ai$.

\subsection{Conserved currents from the ingoing solution}
\label{sec:TJ}

The conserved boundary charge current is given in terms of asymptotic behaviour of the bulk Maxwell field. One has 
\begin{equation}\label{eq:Jcftgen}
\begin{split}
\Jcft_v 
&= 
	- \lim_{r\to \infty}\, r^{d-1}\, \left[F_{rv} -\frac{1}{d-4}\, \frac{1}{r^3\,\sqrt{f}}\, \partial_i\, F_{vi} + \cdots \right] , \\
\Jcft_i
&=
	-\lim_{r\to\infty} \, r^{d-3} \left[r^2 f\, F_{ri} + F_{vi} - \frac{1}{d-4}\, \frac{1}{r\,\sqrt{f}} \, \left(\partial_v F_{vi} 
	-f\, \partial_j F_{ji}\right) + \cdots\right] .
\end{split}
\end{equation}	
We have indicated here the leading order counterterm contribution given in \eqref{eq:SEMct} which suffice for $d\leq 6$; there are higher order terms necessary beyond that. Evaluating this on our linear perturbations \eqref{eq:perturb} one finds in Fourier domain:
\begin{equation}\label{eq:J01}
\begin{split}
\Jcft_\mu 
&= 
	 J^\text{Ideal}_\mu + c_\text{eff}\, \widehat{J}_\mu  \,,\\
\widehat{J}_\mu
&=
		-\lim_{r\to\infty} \, \sum_{\ai=1}^{N_v} \, 
		\VV_i^\ai 
		 \left[ r^{d-3} \, \Dz_+\vMax_\ai +  \frac{1}{d-4}\, r^{d-4}\, \sqrt{f} \, \left(\frac{\omega^2}{f}-k^2\right) \vMax_\ai + \cdots\right] ,\\
&=
		\lim_{r\to\infty} \, \sum_{\ai=1}^{N_v} \, 
		\VV_i^\ai \frac{k^2}{(d-2) \,\mu \,r_+^{d-2} }
		 \left[ r^{d-3} \, \Dz_+Y_\ai +  \frac{1}{d-4}\, r^{d-4}\, \sqrt{f} \, \left(\frac{\omega^2}{f}-k^2\right) Y_\ai + \cdots\right] .		 
\end{split}
\end{equation}	
where we used the field redefinition \eqref{eq:Debyeinspire} in deriving the third line.  The ideal contribution to the current picks out the background charge density since in the inertial frame of the fluid $u^\mu = (\partial_v)^\mu$ and is given in \eqref{eq:idealfluid}.  

The boundary stress tensor includes contributions from the quasilocal Brown-York tensor and counterterms given in \eqref{eq:SEMct}. To quartic order it reads \cite{deHaro:2000vlm}:
\begin{equation}\label{eq:Tcftgen}
\begin{split}
\sqrt{-\gcft}\, \Tcft_{\mu\nu}
&=
	c_\text{eff} \, \lim_{r\to \infty} \, \frac{2}{r^2}\, \sqrt{-\gamma}  \left[ K\, \gamma_{\mu\nu}  -  K_{\mu\nu} -  (d-1)\, \gamma_{\mu\nu} + \frac{1}{d-2}\, \tensor[^\gamma]{G}{_{\mu\nu}} \right. \\
& \left. \quad
	+\frac{1}{(d-2)^2\,(d-4)} \left(\tensor[^\gamma]{\nabla}{^2}\, \tensor[^\gamma]{R}{_{\mu\nu}}
	+ 2\, \tensor[^\gamma]{R}{_{\mu\rho\nu\sigma}}\, \tensor[^\gamma]{R}{^{\rho\sigma}} 
	\right. \right.\\
& \left. \left. \quad	
	+ \frac{1}{2\,(d-1)}\, \left[-
		(d-2)\,\tensor[^\gamma]{\nabla}{_\mu}\, \tensor[^\gamma]{\nabla}{_\nu} \tensor[^\gamma]{R}{}  - 
		 d\, \tensor[^\gamma]{R}{} \, \tensor[^\gamma]{R}{_{\mu\nu}}
		  \right]
	\right. \right. \\
& \left. \left. \quad
	-\frac{1}{2} \,\gamma_{\mu\nu} \left(
		 \tensor[^\gamma]{R}{_{\rho\sigma}}\,  \tensor[^\gamma]{R}{^{\rho\sigma}}  -\frac{d}{4(d-1)}\,  \tensor[^\gamma]{R}{^2} + \frac{1}{d-1}\, \tensor[^\gamma]{\nabla}{^2}\, \tensor[^\gamma]{R}{}
		\right)
	\right)\right] .
\end{split}
\end{equation}
Evaluating on our perturbation ansatz we find a zeroth order ideal fluid contribution and along with a  contribution from the graviton fluctuations, viz.,  
\begin{equation}\label{eq:T01}
\Tcft_{\mu\nu} 
 =  T^\text{Ideal}_{\mu\nu} + c_\text{eff}\, \widehat{T}_{\mu\nu} \,.
\end{equation}	
The ideal piece is the familiar hydrodynamic stress tensor as in \eqref{eq:idealfluid} with the change that the induced metric is no longer flat, viz.,  
\begin{equation}\label{eq:Tideal}
T^\text{Ideal}_{\mu\nu} 
 = c_\text{eff} \, r_+^d (1+Q^2) \left(\gamma_{\mu\nu} + d\, u_\mu\, u_\nu\right)  ,
\end{equation}	
we have used the conformal fluid equation of state $\varepsilon = (d-1)\, p$. 

The contribution from the perturbations is easy to evaluate in terms of the fields $X_\ai$ and $Y_\ai$ first introduced in \cref{sec:Vcoupled}. The Fourier domain stress-tensor works out to be 
\begin{equation}\label{eq:Tcftcoupled}
\begin{split}
(\widehat{T}_{vi})_{\skL/\skR}
&= 
 - \sum_{\ai=1}^{N_V} \, k^2\, \VV_i^\ai  \lim_{r\to \infty\pm i 0} \left\{
 X_\ai + 2\, Y_\ai - \frac{1}{d-2}\, \frac{1}{r\sqrt{f}}\, \Dz_+ X_\ai  + \cdots \right\} ,\\ 
(\widehat{T}_{ij})_{\skL/\skR}
&= 
 - \sum_{\bi=1}^{N_T} \, \TT_{ij}^\bi  \lim_{r\to \infty\pm i 0} \left\{
 r^{d-1}\, \Dz_+ \tGR_\bi + \frac{k^2 \, f- \omega^2}{d-2}\, \frac{r^{d-2}}{\sqrt{f}} 
 \left[1+ \frac{k^2\, f- \omega^2}{(d-2)\,(d-4)\, r^2 f} \right] \tGR_\bi  + \cdots \right\} \\
& + 
	\sum_{\ai=1}^{N_V} \, \omega \left(k_i\, \VV_j^\ai + k_j \VV_i^\ai\right)
	 \lim_{r\to \infty\pm i 0} \left\{
	 	X_\ai - \frac{1}{d-2}\, \frac{1}{r\sqrt{f}} \left[1- \frac{k^2\, f-\omega^2}{(d-2)\,(d-4)\, r^2\, \sqrt{f}} \right] 
	 	\Dz_+ X_\ai+\cdots
	\right\} .
\end{split}
\end{equation}

To make contact with our physical parameterization we need to rewrite the currents \eqref{eq:J01} and \eqref{eq:Tcftcoupled} in terms of the fields $\MX_\ai$ and $\MY_\ai$ and the conjugates. This is straightforward to do using \eqref{eq:XYredef} and \eqref{eq:DzXYmap}. For the Maxwell current \eqref{eq:J01} we get by direct substitution
\begin{align}\label{eq:JcftXYoriginal}
\begin{split}
	(\widehat{J}_i)_{\skL/\skR}
&=
		\lim_{r\to\infty\pm i 0} \, \sum_{\ai=1}^{N_v} \, 
		\VV_i^\ai \frac{k^2}{(d-2)\, \mu \, r_+^{d-2} }
		 \bigg\{  - r^{d-2}\, \cpen{\MX}^\ai - \frac{1}{h}\cpen{\MY}^\ai  \\
&\qquad\qquad		 
		 + \left[ \frac{1}{(d-4)r^2} \sqrt f \left( \frac{\omega^2}{f} - k^2 \right) - (d-2) f  \right] \RQ^{d-2} \MX_\ai  \\
&\qquad\qquad
		+  \left[ \frac{r^{d-4}}{d-4}\sqrt f h \left( \frac{\omega^2}{f} - k^2 \right) + (d-2) \RQ^{d-2} f \right] \MY_\ai + \cdots \bigg\} .	
\end{split}
\end{align}

As in the case of the on-shell action the computation can be organized to either obtain the operator expression for the  currents in terms of the boundary operators $\XOp^\ai$ and $\YOp^\ai$. Equivalently, one can compute the expectation value of the currents in the presence of the boundary sources $\XsJ^\ai$ and $\YsQ^\ai$ defined in \eqref{eq:BC}. To obtain either form,  we perform an expansion in large $r$ and discard all terms that are either divergent  or subleading. The latter vanish on the boundary, while the former are canceled by the boundary counterterms (contained in the ellipses). Utilizing the leading asymptotic behavior of $\MX_\ai$ and $\cpen{\MY}^\ai$ from \eqref{eq:XpiYasy}, the boundary conditions for the fields \eqref{eq:BC}, and the finite piece of both $\cpen{\MY}^\ai$ (given in \eqref{eq:renormY}) and $\MX_\ai$, we find the desired expressions after a bit of algebra.

Let us begin with the current $\widehat{J}_i$. There are two contributions in \eqref{eq:JcftXYoriginal} which are additive coming from the non-Markovian and the Markovian degrees of freedom, respectively. Carrying out the asymptotic analysis current operator can be shown to be 
\begin{equation}\label{eq:currentPA}
\left(\frac{\mu}{\sdc} \, \hat{J}_i\right)_{\skL/\skR}
= - \sum_{\ai=1}^{N_V}\, k^2\, \VV_i ^\ai\, \left[  \XOp^\ai + \frac{1}{(d-2)\, \RQ^{d-2}}\YOp^\ai - \YsQ^\ai \right]_{\skL/\skR}.
\end{equation}	

Similarly, beginning with \eqref{eq:Tcftcoupled} and performing the analogous computation for the vector polarization of the perturbed stress tensor leads to 
\begin{equation}\label{eq:TcftPA}
\begin{split}
(\widehat{T}_{vi} )_{\skL/\skR}
&=  
	-\sum_{\ai = 1}^{N_V}\, k^2\, \VV_i^\ai \left[
		 \left(\frac{\BQT^2}{(d-2)\, \RQ^{d-2}}  \YOp^\ai - (\BQT^2+2) 	\XOp^\ai  \right)_{\skL/\skR} 
		 - \BQT^2 \,  \YsQ^\ai_{\skL/\skR} \right.\\
& \left. \qquad \qquad \qquad \qquad  \qquad 
	+\;	2\left( \frac{\RQ^{d-2}}{d-2} \, \XsJ^\ai_{\skL/\skR}  + \YsQ^\ai_{\skL/\skR}\right) 
	 \right]	 ,	 \\
(\widehat{T}_{ij}^\text{vec})_{\skL/\skR}
&=
	\sum_{\ai = 1}^{N_V}\,\omega \left(k_i \,\VV_j^\ai + k_j \VV_i^\ai \right)
	\left[\frac{\BQT^2}{(d-2)\, \RQ^{d-2}} \, \YOp^\ai -(\BQT^2+2)  \, \XOp^\ai - \BQT^2\, \YsQ^\ai
 	\right]_{\skL/\skR} .
 \end{split}
\end{equation}
where the vector piece of $\widehat{T}_{ij}$ is the second line of $\widehat{T}_{ij}$ in \eqref{eq:Tcftcoupled}. 
Note that while the shear-stress part of the energy-momentum tensor is simply expressed in terms of the operators $\XOp^\ai$ and $\YOp^\ai$ the momentum-flux current $\widehat{T}_{vi}$ also has contribution from the sources of these operators., which one can check is proportional to the charge current source $\left(\vMax^\ai\right)_\infty$ obtained in  \eqref{eq:sources}.

It is easy to isolate the contributions to the currents from the non-Markovian and Markovian sectors,  respectively. One can check that the linear combinations $\XT_i$ and $\YJ_i$ defined in \eqref{eq:CTJdef} serve to achieve this decoupling of the modes. Note that the $vi$ component of the stress tensor by virtue of the above is not independent, but rather is determined in terms of the currents $\mathcal{T}_i$ and $\mathcal{J}_i$. 

Finally, we can write down the expression for the one-point function of the currents as a functional of the background sources $\XsJ^\ai$ and $\YsQ^\ai$. For the charge current one finds the expectation value to be 
\begin{equation}\label{eq:JcftSKad}
\begin{split}
\expval{\left(\frac{\mu}{\sdc}\, \widehat{J}_i\right)_a}
&= 
\sum_{\ai=1}^{N_V}\,  \VV_i \, k^2\,  
  \bigg\{\frac{\Kint{\MY}}{(d-2)\, \RQ^{d-2}}  \, \YsQ_a^\ai -\frac{1}{\Kin{\MX}}\, \XsJ^\ai_a   \\
& \qquad \qquad \qquad 
  + \left(\nB+\frac{1}{2}\right)  
  \left( \frac{\Kint{\MY} -\Krevt{\MY}}{(d-2)\, \RQ^{d-2}} \,  
     \YsQ_d^\ai+\frac{\Kin{\MX} -\Krev{\MX} }{ \Kin{\MX} \, \Krev{\MX}   } \XsJ^\ai_d  \right)
  \bigg\} , \\
\expval{\left(\frac{\mu}{\sdc}\, \widehat{J}_i\right)_d}
&=  
	\sum_{\ai=1}^{N_V}\,  \VV_i \,k^2\,  
  \bigg\{ \frac{\Krevt{\MY}}{(d-2)\, \RQ^{d-2}}\, \YsQ_d^\ai  - \frac{1}{\Krev{\MX}} \, \XsJ^\ai_d 
  \bigg\} .
\end{split}
\end{equation}  
where we used \eqref{eq:KintY}.
A similar exercise for shear-stress part of the energy-momentum tensor leads to 
\begin{equation}\label{eq:TijCFTSKad}
\begin{split}
\expval{\left(\hat{T}_{ij}^\text{vec}\right)_a }
&= 
 -  \sum_{\ai = 1}^{N_V}\,\omega \left(k_i \,\VV_j^\ai + k_j \VV_i^\ai \right)  
   \bigg\{ 
  	\BQT^2 \,  \frac{\Kint{\MY}}{(d-2)\, \RQ^{d-2}}\, \YsQ_a^\ai  
  		+ \frac{\BQT^2+2}{\Kin{\MX}} \, \XsJ^\ai_a
          \\
& \qquad \qquad 
    +\left(\nB+\frac{1}{2}\right) 
    \left(\BQT^2 \frac{\Kint{\MY} -\Krevt{\MY}}{(d-2)\,\RQ^{d-2}} \YsQ_d^\ai
    	-(\BQT^2+2)  \, \frac{\Kin{\MX}-\Krev{\MX}}{\Kin{\MX}\,\Krev{\MX}}
    	 \,  \XsJ^\ai_d    \right) 
    \bigg\} \,,
  \\
\expval{\left(\hat{T}_{ij}^\text{vec}\right)_d} 
&= 
 -  \sum_{\ai = 1}^{N_V}\,\omega \left(k_i \,\VV_j^\ai + k_j \VV_i^\ai \right)  \bigg\{ 
 	\BQT^2 \,   \frac{\Krevt{\MY}}{(d-2)\, \RQ^{d-2}}\, \YsQ_d^\ai   
    + \frac{\BQT^2+2}{\Krev{\MX}}\, \XsJ^{\ai}_d  
   \bigg\}  .
\end{split}
\end{equation}

For the spatio-temporal component of the energy-momentum tensor,  there are contributions from both the  sources and the hydrodynamic operators. The latter are computed in terms of the sources as usual in terms of the Schwinger-Keldysh Green's functions. However, the background  contributions to the current remain untouched when computing expectation values. To wit, 
\begin{equation}\label{eq:TviCFTSKad}
\begin{split}
\expval{\left(\hat{T}_{vi}\right)_a }   
&= 
 - \sum_{\ai = 1}^{N_V}\, k^2\, \VV_i^\ai \bigg\{ 
   	\left(2-\BQT^2\, \frac{\Kint{\MY}}{(d-2)\, \RQ^{d-2}}\right) \, \YsQ_a^\ai  + 
   \left(	\frac{2\,\RQ^{d-2}}{d-2} - \frac{\BQT^2+2}{\Kin{\MX}} \right) \XsJ^\ai_a\\
&\qquad
	+ \left(\nB + \frac{1}{2}\right) \left(
		 \BQT^2 \, \frac{\Krevt{\MY} -\Kint{\MY}}{(d-2)\, \RQ^{d-2}} \, \YsQ_d^\ai
		 +(\BQT^2+2)  \, \frac{\Kin{\MX}-\Krev{\MX}}{\Kin{\MX}\,\Krev{\MX}}
    	 \,  \XsJ^\ai_d
		 \right)   \bigg\} \,,\\
\expval{\left(\hat{T}_{vi}\right)_d }   
&= 
   -\sum_{\ai = 1}^{N_V}\, k^2\, \VV_i^\ai \bigg\{  
   	\left(2-\BQT^2\, \frac{\Krevt{\MY}}{(d-2)\, \RQ^{d-2}}\right) \, \YsQ_d^\ai 
   	+\left(	\frac{2\,\RQ^{d-2}}{d-2}- \frac{\BQT^2+2}{\Krev{\MX}}\right) \XsJ^{\ai}_d
   	   \bigg\}		 .
\end{split}
\end{equation}
The source terms in the above expression represent the polarization of the energy-momentum tensor  from the source for the background charge current given in  \eqref{eq:sources}.  This term gives a contact term contribution to the two-point function of the charge current and the energy-momentum tensor proportional to the background charge density (when we use the third expression in \eqref{eq:TJvariations}), which we have refrained from writing in \eqref{eq:JTspectral}.


\input{gravsk-charge-refs}

\end{document}

%% file: gravsk-macros.tex
\newcommand{\RNAdS}[1]{Reissner-Nordstr\"om-AdS$_{#1}$}
\newcommand{\lieD}{\pounds} 
\newcommand{\RQ}{r_{_{\!Q}}}    
\newcommand{\bRQ}{\mathfrak{C}}   
\newcommand{\sdc}{\mathfrak{S}_{_Q}}  

\newcommand{\ann}{\mathscr{M}}   
\newcommand{\sen}[1]{\varphi_{_{#1}}} 
\newcommand{\cpen}[1]{\pi_{_{#1}}} 

\newcommand{\bwt}{\mathfrak{w}}   
\newcommand{\bqt}{\mathfrak{q}}    
\newcommand{\bk}{\vb{k}}  
\newcommand{\bx}{\vb{x}}   
\newcommand{\dils}{\chi_s}  

\newcommand{\Mser}[2]{\varphi_{_{#1}}^{#2}}  
\newcommand{\Mserh}[2]{\hat{\varphi}_{_{#1}}^{#2}} 
\newcommand{\Dfn}[2]{\Delta_{_{#1}}^{#2}} 
\newcommand{\Dfnh}[2]{\hat{\Delta}_{_{#1}}^{#2}} 
\newcommand{\xser}[1]{\varphi_{_{\MX}}^{#1}}  
\newcommand{\yser}[1]{\varphi_{_{\MY}}^{#1}}  
\newcommand{\yserh}[1]{\hat{\varphi}_{_{\MY}}^{#1}}  

\newcommand{\ctor}{\zeta}  
\newcommand{\ri}{\varrho}   
\newcommand{\rib}{u}    
\newcommand{\Dz}{\mathbb{D}}  
\newcommand{\Dann}{\mathfrak{D}_\ann}  

\newcommand{\In}{\text{\tiny{in}}}     
\newcommand{\Rev}{\text{\tiny{rev}}}  

\newcommand{\nB}{n_{_B}}   
\newcommand{\Gin}[1]{G_{_{#1}}^\In}     
\newcommand{\Kin}[1]{K_{_{#1}}^\In}      
\newcommand{\Kint}[1]{\widetilde{K}_{_{#1}}^\In}      
\newcommand{\Grev}[1]{G_{_#1}^\Rev}   
\newcommand{\Krev}[1]{K_{_{#1}}^\Rev}    
\newcommand{\Krevt}[1]{\widetilde{K}_{_{#1}}^\Rev}    

 \newcommand{\MX}{\mathscr{X}}   
 \newcommand{\MY}{\mathscr{Y}}   
\newcommand{\dilY}{\chi_{_\MY}}   
\newcommand{\dilX}{\chi_{_\MX}}   

 \newcommand{\BQT}{\mathfrak{p}}    

\newcommand{\TT}{\mathbb{T}}   
\newcommand{\VV}{\mathbb{V}}  
\newcommand{\ScS}{\mathbb{S}} 
\newcommand{\ai}{\alpha}   
\newcommand{\bi}{\sigma}  

\newcommand{\EEin}{\mathcal{E}^{\text{\tiny{Ein}}}}   
\newcommand{\EMax}{\mathcal{E}^{\text{\tiny{Max}}}}  

\newcommand{\tGR}{\Phi}   
\newcommand{\vGR}{\Psi}   
  
\newcommand{\vMax}{\Xi}     

\newcommand{\Vm}{\mathscr{A}}      
\newcommand{\AGR}{\mathscr{A}}  
\newcommand{\FGR}{\mathscr{F}}   

\newcommand{\Jcft}{J^\text{\tiny{CFT}}}   
\newcommand{\Tcft}{T^{\text{\tiny{CFT}}}}   
\newcommand{\gcft}{g^{\text{\tiny{CFT}}}}   

\newcommand{\XT}{\breve{\mathcal{T}}}  
\newcommand{\YJ}{\mathcal{J}} 

\newcommand{\JMarP}{\bm{\gamma}}  

\newcommand{\YsQ}{\bm{\alpha} }  
\newcommand{\XsJ}{\breve{\bm{\chi}}}  

\newcommand{\XP}{\breve{\mathcal{P}}}  
\newcommand{\XOp}{\breve{\mathcal{O}}_{_{\!\MX}}}  
\newcommand{\YOp}{\mathcal{O}_{_{\!\MY}}} 

\newcommand{\NX}{\mathsf{N}_{_\MX}}    
\newcommand{\NY}{\mathsf{N}_{_\MY}}     

\newcommand{\JMar}{J}  

\newcommand{\JMarQ}{\mathcal{A}} 
\newcommand{\JMarQF}{\mathcal{F}}  

\newcommand{\JnMar}{\breve{J}}  
\newcommand{\snMar}{ \breve{\Phi}}   
\newcommand{\OnMar}{\breve{\mathcal{O}}}  
\newcommand{\skR}{\text{\tiny R}}  
\newcommand{\skL}{\text{\tiny L}}  

\newcommand{\psGR}{\Pi}
\newcommand{\pJGR}{P}



%% file: gravsk-charge-refs.tex
\providecommand{\href}[2]{#2}\begingroup\raggedright\endgroup